\documentclass[aps,prd,amsmath,amssymb,nofootinbib,tightenlines,floatfix,superscriptaddress,preprint,longbibliography]{revtex4-1}
\usepackage{bm}
\usepackage{hyperref}
\usepackage{latexsym}
\usepackage[pdftex]{graphicx,color}

\newcommand{\be}{\begin{equation}}
\newcommand{\ee}{\end{equation}}
\newcommand{\bea}{\begin{eqnarray}}
\newcommand{\eea}{\end{eqnarray}}

\begin{document}

\title{\Large LATTICE GAUGE THEORIES AT THE ENERGY FRONTIER}
%\author{}

\author{Thomas Appelquist, Richard Brower, Simon Catterall, George Fleming, Joel Giedt, Anna Hasenfratz, Julius Kuti, Ethan Neil, 
and David Schaich}
\noaffiliation
\collaboration{USQCD Collaboration}
\noaffiliation

\date{February 11, 2013}

\begin{figure}[!b]
\begin{minipage}[!b]{0.95\linewidth}%

\vskip -11in
{\large\bf Abstract}\\

\begin{flushleft}

{\footnotesize 
This White Paper has been prepared as a planning document for the Division of High Energy Physics of the U. S. Department of Energy.
Recent progress in lattice-based studies of physics beyond the standard model is summarized, and major current goals of USQCD research in this area are presented. Challenges and opportunities associated with the recently discovered 126 GeV Higgs-like particle are highlighted.
Computational resources needed for reaching important goals are described.
The document was finalized on February 11, 2013 with references 
that are not aimed to be complete, or account for an accurate historical record of the field.}
\end{flushleft}

\end{minipage}
\end{figure}

%%% SS: use of ``\rule'' is a kluge to make more space
%%%     between the date and the contents

%%% SS: subsubsections not included in Contents since
%%%     they make them too long to fit on the title page

\renewcommand{\abstractname}{\rule{1in}{0pt}\vphantom\strut \\[.5in]
\bf CONTENTS}
\begin{abstract}
\begin{itemize}

\item[]{\bf \ref{sec:exec}. EXECUTIVE SUMMARY}
\item[]{\bf \ref{sec:intro}. INTRODUCTION}
\item[]{\bf \ref{sec:roadmap}. LATTICE BSM RESEARCH}
\item[]{\bf\ref{subsec:results}. Highlights of Recent Results and Future Goals}
\item[]{\bf \ref{sec:dilaton}. THE LIGHT HIGGS AND THE DILATON NEAR CONFORMALITY}
\begin{itemize}
\item[]{\bf \ref{subsec:scalar}. The $0^{++}$ Scalar Spectrum in BSM Lattice Models}
\item[]{\bf\ref{subsec:pcdc}. The PCDC Dilaton Relation on the Lattice}
\end{itemize}
\item[]{\bf \ref{sec:pngb}. HIGGS AS A PSEUDO-GOLDSTONE BOSON}
\begin{itemize}
\item[]{\bf \ref{subsec:minimal}. Minimal PNGB Model}
\item[]{\bf\ref{subsec:lattice}. Lattice Formulation}
\end{itemize}
\item[]{\bf \ref{sec:susy}. STUDIES OF SUPERSYMMETRIC THEORIES ON THE LATTICE}
\begin{itemize}
\item[]{\bf\ref{subsec:N4}. {\cal N}=4 SYM}
\item[]{\bf\ref{subsec:susyQCD}. Dynamical SUSY Breaking and Super QCD}
\end{itemize}
\item[]{\bf \ref{sec:methods}. METHODS AND PHENOMENOLOGICAL APPLICATIONS}
\begin{itemize}
\item[]{\bf\ref{subsec:coupling}. Running Coupling and ${\bf \beta}$-function}
\item[]{\bf\ref{subsec:dirac}. Chiral Symmetry Breaking, Dirac Spectrum, and Anomalous Mass Dimension}
\item[]{\bf\ref{subsec:condensate}. Chiral Perturbation Theory and Condensate Enhancement}
\item[]{\bf\ref{subsec:s-parameter}. S-parameter}
\item[]{\bf\ref{subsec:ww}. WW Scattering on the Lattice}
\item[]{\bf\ref{subsec:dark}. Composite Dark Matter}
\end{itemize}
\item[]{\bf \ref{sec:NewMethods}. NEW METHODS FOR BSM LATTICE FIELD THEORY}
\begin{itemize}
\item[]{\bf\ref{subsec:multiscale}. Traditional Multi-scale Tools}
\item[]{\bf\ref{subsec:radial}. Radial Quantization}
\end{itemize}
\item[]{\bf \ref{sec:resources}. RESOURCES FOR STUDIES AT THE ENERGY FRONTIER}
\begin{itemize}
\item[]{\bf\ref{subsec:scalarResource}. Scalar Spectrum and ${\rm \chi SB}$ Project}
\item[]{\bf\ref{subsec:pngbResource}. PNGB SU(2) Color Higgs Project}
\item[]{\bf\ref{subsec:susyResource}. SUSY ${\cal N}=1$ Yang-Mills Project}
\end{itemize}

\end{itemize}
\end{abstract}

\maketitle
\thispagestyle{empty}

\section{Executive Summary}
\label{sec:exec}

This report on { \em Lattice Gauge Theory at the Energy Frontier} is a
companion document to three other reports: { \em Lattice QCD at the
Intensity Frontier}, { \em Lattice QCD for Cold Nuclear Physics}, and
{ \em Computational Challenges in QCD Thermodynamics}. Together they
address prospects in the next five years to advance lattice field
theory calculations to support the high-energy physics and nuclear
physics programs.

Lattice based studies of  strongly coupled gauge theories proposed
for the description of physics beyond the standard
model (BSM) have accelerated in recent years. New numerical and
theoretical tools have enabled physicists to address a broad range of
theories that could, for example, lie at the heart of the dynamics of the Higgs
effective theory of electroweak symmetry breaking. With the dramatic
discovery at the LHC of a 126 GeV Higgs-like particle, lattice BSM research
is moving to the study of mechanisms for the origin of a Higgs scalar field
that are consistent with this discovery. Rather than accepting an elementary
Higgs boson with no new physics at accessible scales, lattice BSM research
explores its possible origin from TeV-scale physics involving compositeness and new symmetries.
Studying these theories requires non-perturbative lattice methods, which  build
directly on substantial USQCD lattice-BSM accomplishments of recent years:

\begin{itemize}
\renewcommand{\labelitemi}{$\circ$}

\item Investigations of strongly coupled BSM gauge theories identified conformal or near conformal behavior, 
demonstrating that the anomalous mass dimensions and chiral condensates are  enhanced near conformality, with interesting
implications for model building.

\item Electroweak precision experimental constraints were compared with numerical estimates of the S-parameter, 
W-W scattering, and the composite spectra.  In particular in contrast with naive estimates,  these studies demonstrate that the S-paramenter  in near-conformal theories  may be reduced in better agreement with experimental constraints.

\item Investigations of ${\cal N} = 1$ supersymmetric Yang Mills theory (gauge bosons and gauginos) produced estimates of the  gluino condensate and string tension in these theories.

\end{itemize}

Building on these significant and computationally demanding accomplishments,
the USQCD BSM  program
has developed three major directions for future lattice BSM research
with well-defined calculational goals:

\begin{itemize}
\renewcommand{\labelitemi}{$\bullet$}

\item To determine whether a composite dilaton-like particle or light Higgs can emerge in near-conformal quantum field theories. 

\item  To investigate strongly coupled theories with a composite Higgs as a pseudo-Goldstone boson.

\item To investigate the nature of  $\mathcal{N}=1$ SUSY breaking with matter multiplets and ${\cal N} =4$ conformal
SUSY as a test bed for AdS/CFT theoretical conjectures. 

\end{itemize}

For each, we describe challenges and prospects with an
emphasis on the broad range of phenomenological investigations underway
and the development of new lattice-field-theory methods targeted at
BSM research. We identify the computational resources
needed to make  major contributions to our understanding
of gauge theories and to apply  the theoretical framework to
illuminate experiments at the energy frontier.

\vfill\newpage

\section{Introduction}
\label{sec:intro}

From its inception in the 1970’s, lattice gauge theory has been
employed with great success to study QCD, the one strongly coupled
gauge theory known to describe the real world.  The remarkable
progress of recent years includes increasingly precise computations of
hadron structure, finite-temperature behavior, and the matrix elements
crucial for precision tests of the standard model (SM).

During this same period, the many questions posed by the standard
model have led theorists to propose and study a vast array of other
strongly coupled gauge theories.  They have included QCD-like theories,
supersymmetric (SUSY) gauge theories, conformal field theories, and
others. These beyond-standard-model (BSM) theories could play a role
in extending the standard model to ever higher energies, and perhaps
lead to an understanding of the many parameters of the standard
model. Theorists have envisioned these theories to describe a range of
phenomena including electroweak symmetry breaking, SUSY breaking, and
the generation of the flavor hierarchies of the standard model. While
not yet making definitive contact with experimental measurement, this
work has given birth to many ideas for exploring SUSY and non-SUSY
gauge field theories. Approaches such as degree-of-freedom based
constraints on renormalization-group flow, duality and AdS/CFT
constraints on SUSY gauge theories, and symmetry-based constraints on
conformal field theories have provided important new insights.  

In recent years, lattice simulations have taken a prominent place
among these tools. They have been brought to bear on a variety of
strongly coupled gauge field theories, exploring QCD-like theories
extensively and laying the groundwork for similar studies of SUSY
gauge theories. For certain QCD-like theories, dramatic quantitative
changes have been observed as the number of light fermions 
is increased, or their color representation is changed,
showing a much smaller relative scale for vacuum chiral 
symmetry breaking.   In fact, these same studies raise 
the strong possibility of the emergence of conformal 
symmetry as the fermion content is increased.  Rather than 
confining and spontaneously breaking chiral symmetries as 
in QCD, these theories may exhibit an exact or approximate 
infrared conformal symmetry.  The nature of this transition, 
including the spectrum of particle states and other properties, 
is the focus of intensive, ongoing research.
%For certain QCD-like theories, it has been shown
%that as the number of light fermions is increased, or their color representation is changed, 
%a qualitatively new kind of infrared behavior emerges. Rather than
%confining and spontaneously breaking chiral symmetries as in QCD,
%these theories can exhibit an exact or approximate infrared conformal
%symmetry.  The nature of this transition, including the spectrum of
%particle states and other properties, is the focus of intensive,
%ongoing research. 
If these theories play a role in electroweak
breaking, then these transitional features could have immediate
measurable consequences. Lattice simulation of SUSY gauge theories
will be especially exciting, making contact with other powerful
approaches to these theories.

Lattice-based work on BSM theories is already being powerfully shaped
by the dramatic discovery at the Large Hadron Collider (LHC) of a 126
GeV Higgs-like particle.
%An immediate question is whether a strongly-coupled gauge theory can
%produce a light composite Higgs boson, the lightness being enabled by
%it being an approximate Nambu-Goldstone boson associated with the
%breaking of an approximate conformal or global symmetry and separated from heavy resonances on the TeV scale. 
An immediate question is whether a strongly-coupled gauge theory can
produce a  composite Higgs boson, which is light relative  to all the heavier resonances on the TeV scale. This
parametric light scalar maybe  an  approximate  Nambu-Goldstone boson  associated either with the
breaking of an  approximate conformal symmetry or  with the spontaneous breaking of a global symmetry.
Both of these possibilities are receiving much recent attention.

As lattice-based studies of BSM theories have blossomed in recent
years, a whole set of simulation and analysis tools have been
developed, sharpened, and shared. They are widely applicable, and will
no doubt continue to evolve and enable BSM studies as new
possibilities and directions are identified. All of this progress and
promise for the future depends critically on the availability of
large-scale computational resources.  It is an investment that is beginning to pay off, promising
to deepen our understanding of gauge field theories and 
substantially increasing our ability to interpret experiments 
at the energy frontier.
%It is an investment that has paid
%off handsomely, deepening our understanding of gauge field theories
%and offering much promise for future elucidation of experiments at the
%energy frontier.

\section{Lattice BSM Research}
\label{sec:roadmap}

The recent discovery of the Higgs-like resonance at 126 GeV by the
CMS~\cite{CMS:2012gu} and ATLAS~\cite{ATLAS:2012gk} experiments at the
Large Hadron Collider (LHC) provides the first insight into the origin
of electroweak symmetry breaking (EWSB) in the standard model.  The
minimal realization of EWSB is implemented by introducing an ${\rm
  SU(2)}$ doublet scalar Higgs field whose vacuum expectation value,
${\rm v = 246~GeV}$, sets the electroweak scale, or VEV.  This simple
``Mexican-hat" solution should be regarded as a parametrization rather
than a dynamical explanation of EWSB. In particular, the mass-squared
parameter of the light Higgs has to be finely tuned, leading to the
well-known hierarchy problem.  Searching for a deeper dynamical
explanation, and resolving the shortcomings of the minimal standard
model with its elementary Higgs doublet, the USQCD BSM program has
developed three research directions. One employs strongly coupled
gauge theories near
conformality~\cite{Weinberg:1979bn,Susskind:1978ms,Dimopoulos:1979es,
  Eichten:1979ah,Farhi:1980xs,Holdom:1984sk,Appelquist:1987fc,Yamawaki:1985zg,Caswell:1974gg,Banks:1981nn,
  Marciano:1980zf,Kogut:1984sb,Appelquist:2003hn,Sannino:2004qp,Dietrich:2005jn,Luty:2004ye,Dietrich:2006cm,Kurachi:2006ej,
  Ellis:2012hz,Low:2012rj,Elander:2012fk,Bardeen:1985sm,Holdom:1986ub,Miransky:1996pd,Goldberger:2007zk,Appelquist:2010gy,
  Grinstein:2011dq,Antipin:2011aa,Hashimoto:2010nw,Matsuzaki:2012xx,Matsuzaki:2012vc},
another envisions the new particle as a light pseudo-Goldstone boson
(PNGB) in the spirit of little Higgs
scenarios~\cite{ArkaniHamed:2001nc,ArkaniHamed:2002qx,ArkaniHamed:2002qy,Kaplan:1983fs,
  Kaplan:1983sm,Georgi:1984af,Thaler:2005kr,Schmaltz:2008vd}, and the
third begins to explore SUSY gauge theories. In each of these three
research directions, new degrees of freedom are expected at the TeV
scale.

\begin{figure}[h!]
\begin{center}
\begin{tabular}{cc}
\includegraphics[width=0.43\textwidth]{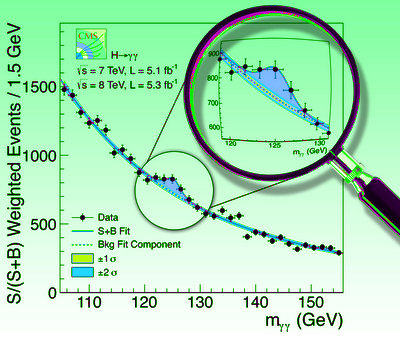} \hskip 1 cm
\includegraphics[width=0.52\textwidth]{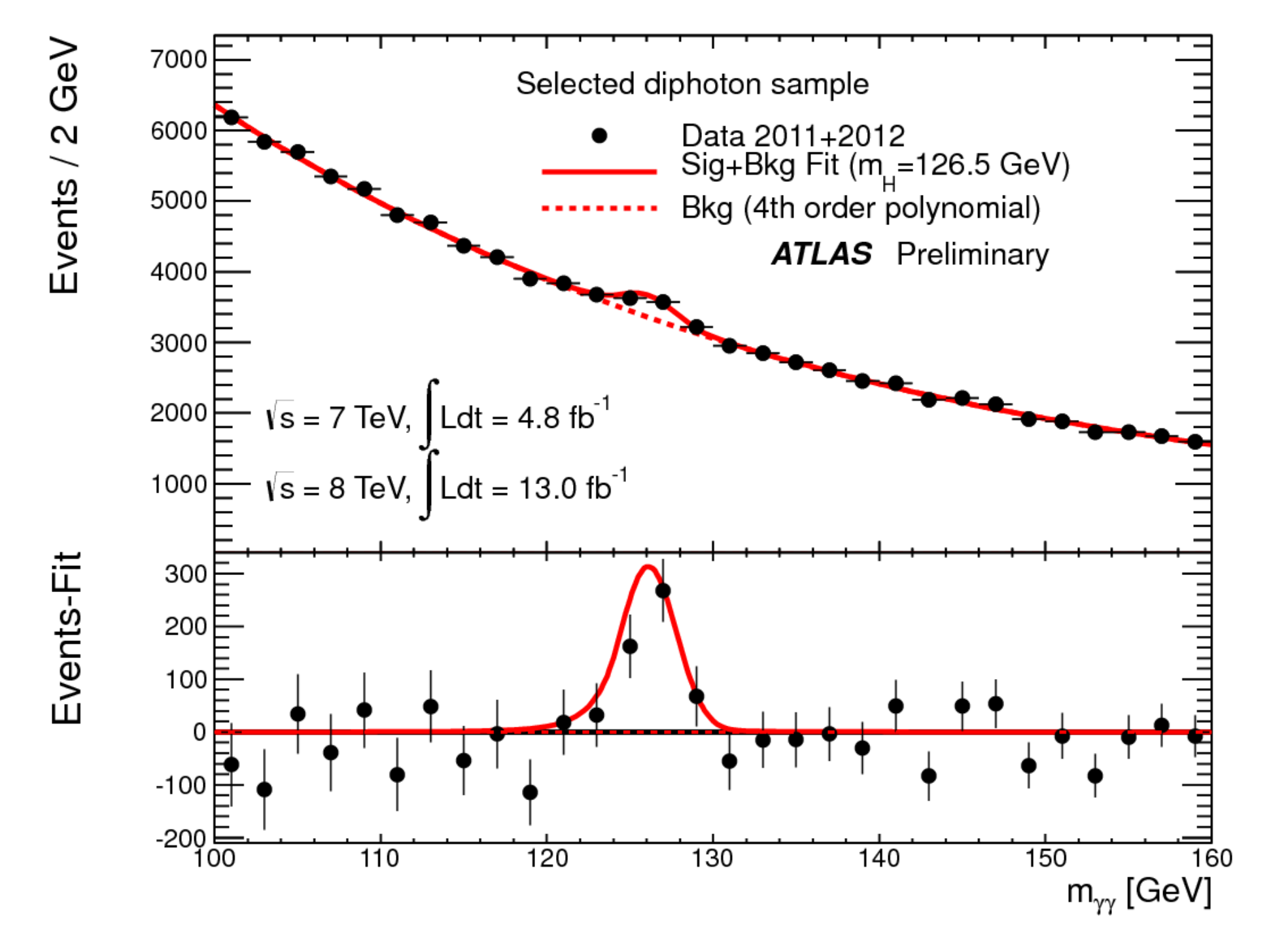}
\end{tabular}
\end{center}
%\vskip -0.25in
\caption{\label{fig:Higgs}
The plots illustrate the Higgs discovery of the CMS and ATLAS collaborations in the
prominent two-photon decay channel. On the right, ATLAS data is shown
for the invariant mass distribution of diphoton candidates for the combined
 $\sqrt{s}$ = 7 TeV and $\sqrt{s}$ = 8 TeV data samples. On the left, the two-photon
 signal of the CMS collaboration is shown.}
\end{figure}

The dramatic LHC discovery presents a challenge for four-dimensional BSM gauge theories where the electroweak symmetry is broken
via strong dynamics. Can such theories lead to a naturally light and narrow Higgs-like state in the scalar spectrum
with ${\rm 0^{++}}$ quantum numbers? An approach to the viability of this class of models, the main focus of the first two of the
three current BSM directions, envisions strong dynamics that could produce a Higgs-like light and narrow scalar composite state with
${\rm 0^{++}}$ quantum numbers. One possibility is that an underlying strongly coupled gauge theory could produce this state
as a dilaton, the PNGB of conformal symmetry breaking. The minimal standard model itself, with a light Higgs particle, has
this approximate symmetry at the classical level. The landscape of strongly interacting gauge theories also provides
intriguing chiral symmetry-breaking patterns in which the light state can exist as a PNGB of an approximate  expanded
global symmetry, similar in spirit to the little Higgs scenarios. In addition, the landscape of SUSY gauge theories includes many
attractive features, among them natural flat directions which can make plausible the dilaton interpretation of the 126 GeV particle.

%Each of our three BSM research directions is supported by an expanding USQCD BSM lattice tool set together with phenomenological
%approaches of general validity, such as chiral perturbation theory. In this white paper, we identify computational resources needed
%during the next five years for important representative projects of the three research directions.

\subsection{Highlights of Recent Results and Future Goals}
\label{subsec:results}

Rather than accepting an elementary
Higgs boson with no new physics at accessible scales, lattice BSM research
explores its possible origin from TeV-scale physics involving compositeness and new symmetries.
Studying these theories requires non-perturbative lattice methods, which  build
directly on substantial lattice-BSM accomplishments of recent years:

\begin{itemize}
\renewcommand{\labelitemi}{$-$}

\item Investigations of strongly coupled BSM gauge theories identified conformal or near conformal behavior, 
demonstrating that the anomalous mass dimensions and chiral condensates are  enhanced near conformality, with interesting
implications for model building.

\item Electroweak precision experimental constraints were compared with numerical estimates of the S-parameter, 
W-W scattering, and the composite spectra.  In particular in contrast with naive estimates,  these studies demonstrate that the S-paramenter  in near-conformal theories  may be reduced in better agreement with experimental constraints.

\item Investigations of ${\cal N} = 1$ supersymmetric Yang Mills theory (gauge bosons and gauginos) produced estimates of the  gluino condensate and string tension in these theories.

\end{itemize}

Building on these significant and computationally demanding accomplishments,
the USQCD BSM  program
has developed three major directions for future lattice BSM research
with well-defined calculational goals:

\begin{itemize}
\renewcommand{\labelitemi}{$-$}

\item To determine whether a composite dilaton-like particle or light Higgs can emerge in near-conformal quantum field theories. 

\item  To investigate strongly coupled theories with a composite Higgs as a pseudo-Goldstone boson.

\item To investigate the nature of  $\mathcal{N}=1$ SUSY breaking with matter multiplets and ${\cal N} =4$ conformal
SUSY as a test bed for AdS/CFT theoretical conjectures. 

\end{itemize}

For each we describe challenges and prospects with an
emphasis on the broad range of phenomenological investigations underway
and the development of new lattice-field-theory methods targeted at
BSM research. We identify the computational resources
needed to make  major contributions to our understanding
of gauge theories and to prepare the theoretical framework to
illuminate experiments at the energy frontier.

Each of our three BSM research directions is supported by an expanding USQCD BSM lattice tool set, 
including techniques and phenomenological approaches (such as chiral perturbation theory) which are useful for all three directions.  
In this white paper, we first present in sections IV-VI the physics motivation and current status of each of the major research directions: 
Higgs as a dilaton, Higgs as a PNGB, and supersymmetric theories.  In section VII we give an overview of phenomenological applications, 
generally using methods developed in QCD simulations.  Section VIII presents new methods under development specifically for the 
study of BSM theories on the lattice, while section IX identifies computational resources needed during the next five years for important 
representative projects of the three research directions.

\section{The light Higgs and the dilaton  near conformality}
\label{sec:dilaton}

It is a truth universally acknowledged that at high energies particle masses become unimportant.
In the absence of electroweak symmetry breaking, the interactions of standard-model gauge
bosons and fermions show approximate conformal symmetry down to the QCD scale.
In this sense, electroweak symmetry breaking in the standard model appears together with the dynamical breaking of scale invariance.
This opens up the possibility that the Higgs mode and the dilaton mode, the pseudo-Goldstone boson of
spontaneously broken scale invariance, are perhaps intimately related.
The important properties of the
standard-model Higgs boson are basically determined by the approximate conformal invariance in the
limit when the Higgs potential is turned off. In this case the Higgs VEV is arbitrary (it is a flat direction),
and its value at ${\rm v = 246~GeV}$ will spontaneously break the approximate conformal symmetry and the electroweak symmetry.
In this limit the Higgs particle can be identified with the massless dilaton  of conformal symmetry breaking~\cite{Goldberger:2007zk} at the scale ${\rm f_\sigma = v}$. Higgs properties in this weak coupling scenario are  associated with approximate conformal invariance.
%When coupled to a small number of massless fermions $N_f$, an $SU(N_c)$ Yang-Mills gauge theory such as QCD exhibits the well-known properties of confinement %and spontaneous chiral symmetry breaking. However, as $N_f$ is increased, or the fermion representation changed, 
%we expect to find a critical value $N_f^c$ at which the infrared properties of the theory undergo a transition, due to the appearance of an infrared fixed point; this occurs in %perturbation theory for large enough $N_f$ \cite{Caswell:1974gg,Banks:1981nn}.  Theories with $N_f > N_f^c$ are said to lie in the ``conformal window".  The precise %value of $N_f^c$ and the nature of the transition itself can only be determined non-perturbatively, requiring the use of lattice simulation for definitive study.

The possibility of approximate conformal symmetry and its breaking in strong gauge
dynamics motivates intriguing BSM possibilities close to the conformal window.
%Some gauge theories with strong dynamics are expected to be  close to the conformal window with their approximate
%conformal symmetry broken at a scale separated perhaps from the onset of the chiral condensate in the infrared.
The onset of the chiral condensate with electroweak symmetry breaking and the spontaneously broken conformal invariance with
its dilaton might create a scenario consistent with the standard-model Higgs mechanism  and the
light Higgs impostor, in a natural setting and with minimal tuning~\cite{Appelquist:2010gy,Hashimoto:2010nw,Matsuzaki:2012vc} .

The knowledge of the phase diagram of nearly conformal gauge theories is important as the number of colors $N_c$, 
number of fermion flavors $N_f$ , and the fermion representation $R$ of the color gauge group are varied 
in theory space. For fixed $N_c$ and $R$ the theory is in the chirally broken phase for low $N_f$ and asymptotic 
freedom is maintained with a negative $\beta$-function. On the other hand, if $N_f$ is large enough, 
the $\beta$-function is positive for
all couplings, and the theory is trivial. The conformal window is defined by the range of $N_f$ for which 
the $\beta$-function has an infrared fixed point, where the theory is in fact conformal. 

Close to the conformal window, BSM gauge theories with strong dynamics have a broad appeal.
If the strongly coupled BSM gauge model is  very close to the conformal window
with a small but nonvanishing $\beta$-function, a necessary
condition is satisfied for spontaneous  breaking of scale invariance and generating the light PNGB dilaton state.
Since these models  exhibit chiral symmetry breaking ($\chi{\rm SB}$)
a Goldstone pion spectrum is generated and when coupled to the electroweak sector, the onset of electroweak
symmetry breaking with the Higgs mechanism is realized.
The very small beta function (walking) and $\chi{\rm SB}$ are
not sufficient to guarantee a light dilaton state if scale symmetry breaking and $\chi{\rm SB}$
are entangled in a complicated way.

It is far from clear that the dilaton mechanism is naturally realized in strong dynamics.
However,  a light Higgs-like scalar might still be expected to emerge near the conformal window as a composite scalar state
with $0^{++}$ quantum numbers, not necessarily with dilaton interpretation.
This scalar state has to be light and would  not be required to exhibit exactly the observed ${\rm 126~GeV}$
mass. The dynamical Higgs mass ${\rm M^{0^{++}}_H}$ from composite strong dynamics is expected to be shifted
by electroweak loop corrections, most likely dominated by the large negative mass shift from the top quark loop~\cite{Foadi:2012bb}.
The Higgs mass emerging from strong gauge dynamics cannot be very heavy, but  an ${\rm M^{0^{++}}_H}$ in the range of
several hundred GeV, before electroweak corrections shift it downward, should not be dismissed for viable model building.

A high priority goal for non-perturbative BSM lattice studies of the USQCD collaboration
is to determine the scalar spectrum
of strong gauge dynamics with Higgs quantum numbers, close to the conformal window.
This presents an enormous challenge in BSM lattice simulations, and the first pilot studies are just beginning
to emerge, outlining a roadmap toward this goal. Once the scalar spectrum is established, the possibility of an intriguing connection
with the dilaton mechanism can be further investigated with the potential of important phenomenological consequences.

\subsection{The  $\mathbf {0^{++}}$ Scalar Spectrum in BSM Lattice Models}
\label{subsec:scalar}
The generic features and the required resources for the scalar spectrum to include a state as light as the 
experimentally observed Higgs particle  are illustrated with
a pilot study of a frequently discussed
BSM gauge theory with twelve fermion flavors  
in the fundamental  representation
of the SU(3) color gauge group. With ${\rm N_f=12}$ being close to the critical flavor number of the conformal edge,
the model has attracted a great deal 
of attention in the lattice community~\cite{Appelquist:2007hu,Fodor:2009wk,Fodor:2011tw,Fodor:2011tu,Fodor:2012et,Lin:2012iw,
Deuzeman:2009mh,Deuzeman:2011pa,Hasenfratz:2009ea,Hasenfratz:2010fi,Cheng:2011ic,Jin:2009mc,
Jin:2012dw,Aoki:2012eq}, and off-lattice as well.
Although the precise position of the model with respect to the conformal window remains unresolved,
the required lattice technology to capture the scalar spectrum in BSM lattice models is quite robust in a model independent way.
In Section 8 the resource estimate will be presented for another frequently discussed strongly interacting
gauge theory with a fermion flavor doublet in the two-index symmetric (sextet) representation
of the SU(3) color gauge group~\cite{Fodor:2012ty,Kogut:2010cz,Kogut:2011ty}. 
Close to the conformal edge, the scalar spectrum of this model
is being explored for the minimal realization of the composite Higgs mechanism.

In BSM models, close to the conformal window, a low mass ${\rm 0^{++}}$ glueball is expected to mix with the 
composite scalar state of the fermion-antifermion pair. With the ultimate goal of reaching to the chiral limit of massless fermions,
the mixing in the scalar spectrum has to include Goldstone pairs with vacuum quantum numbers and exotic states made of
two fermions and two antifermions with ${\rm 0^{++}}$  quantum numbers. Realistic studies require a 3-channel solution, even if exotica are excluded from
the analysis. The pilot study presented here for future planning is restricted to the single channel problem using 
scalar correlators which are built from connected and disconnected loops of fermion propagators~\cite{Fodor:2013xxx,Rinaldi:2012xxx}. 

\begin{figure}[h!]
\begin{center}
\begin{tabular}{cc}
\includegraphics[width=0.5\textwidth]{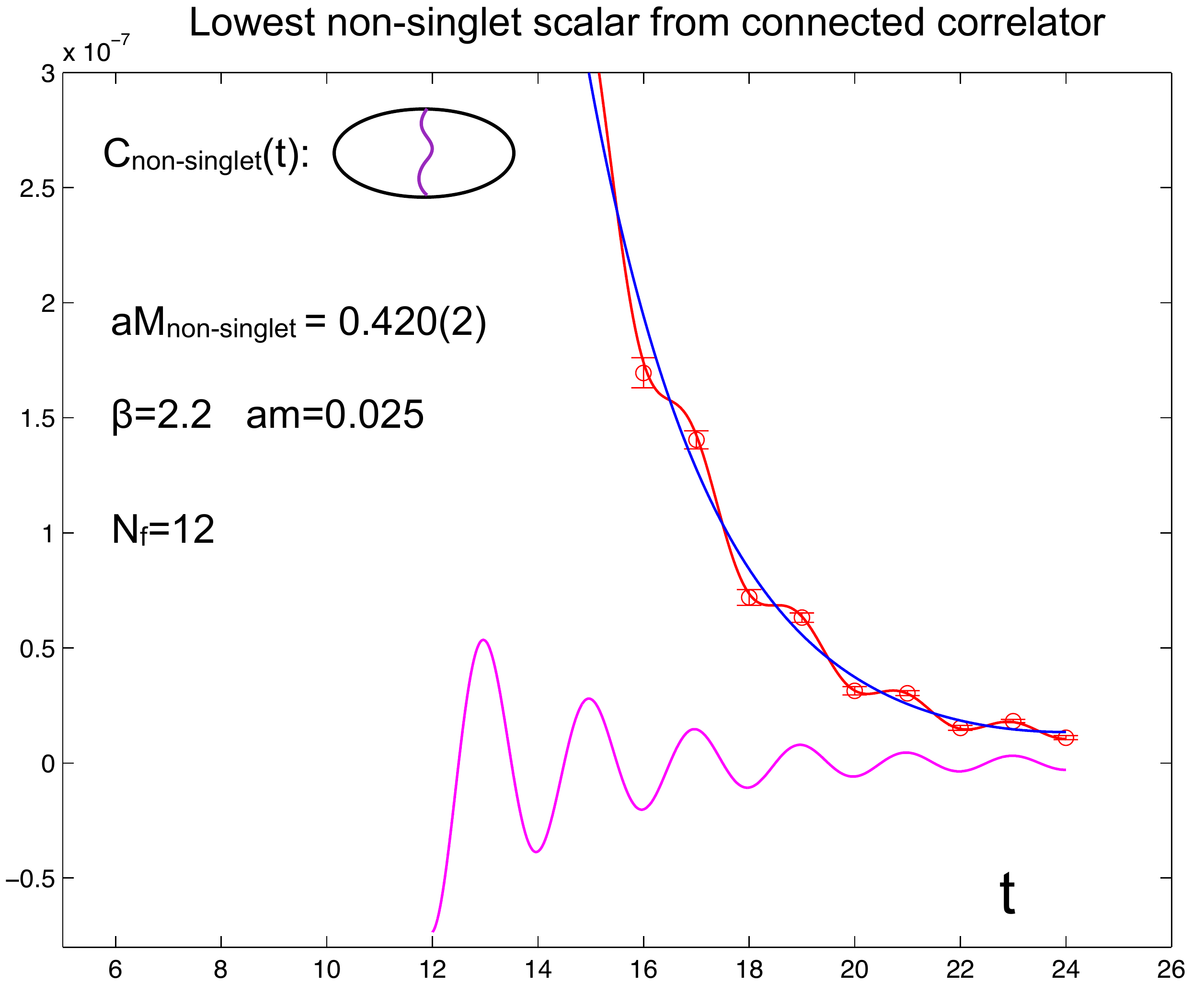} 
\includegraphics[width=0.5\textwidth]{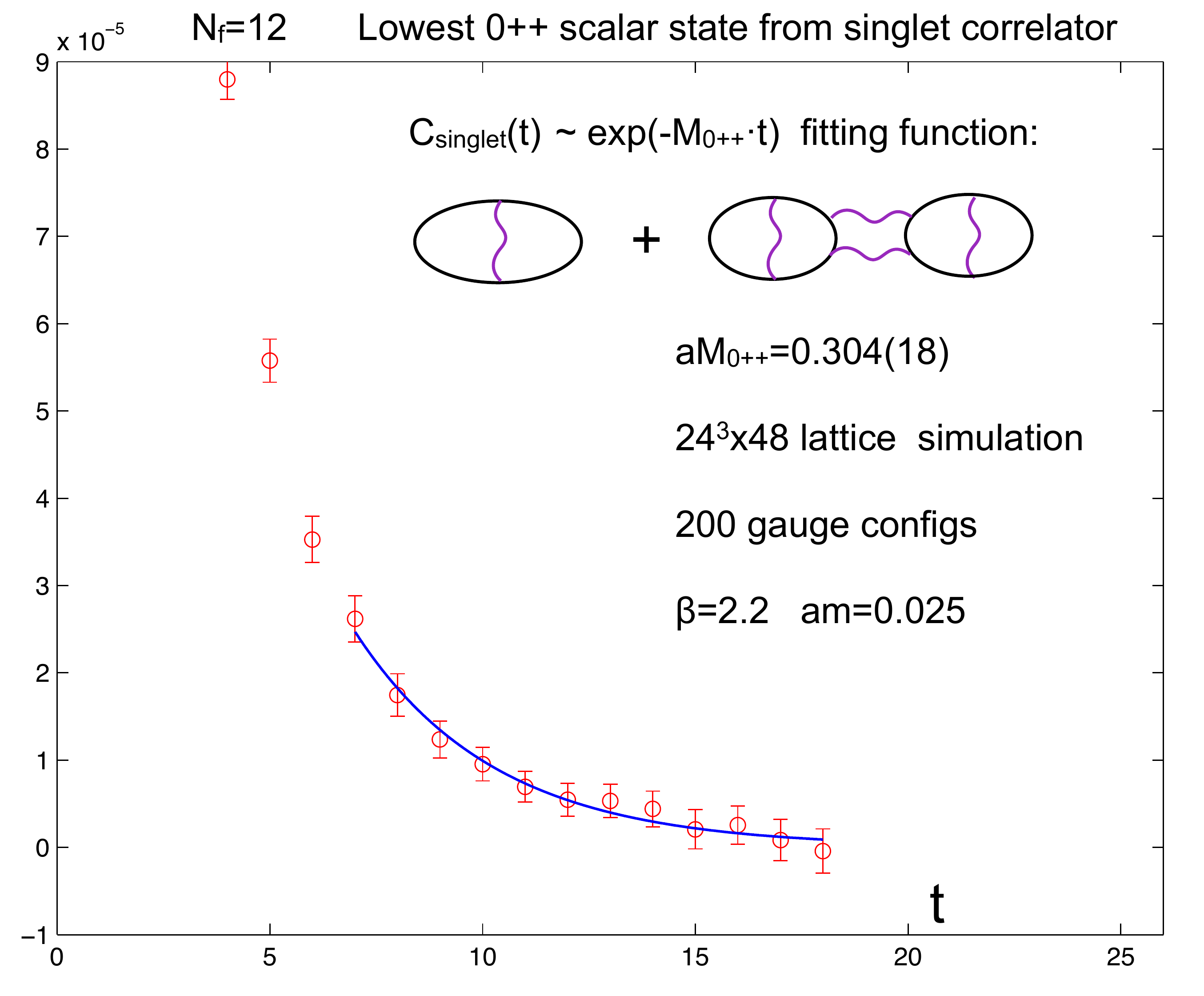} 
\end{tabular}
\end{center}
\vskip -0.25in
\caption{\label{fig:scalar} 
The fermion loops were evaluated using
stochastic methods with full time dilution and $48$ noise vectors on each gauge configuration~\cite{Fodor:2013xxx}.
The correlator ${\rm C_{conn}(t)}$ on the left plot and the correlator ${\rm C_{singlet}=C_{conn}+C_{disc}(t)}$ 
on the right plot were assembled from the stochastic fermion propagators. The left side plot shows the mass
of the lowest non-singlet scalar (blue exponential fit). 
The plot also displays the oscillating pseudo-scalar parity partner (magenta) 
and the full correlator (red) fitting the data. 
On the right side plot, with larger errors
in the limited pilot study, the scalar singlet mass is considerably downshifted (blue exponential) and
the presence of a pseudo-scalar parity partner is not detectable. 
The conventional $\beta=6/g^2$ lattice gauge coupling, setting the lattice spacing $a$, is shown in addition 
to the finite fermion mass $am$ of the simulation.
}
\end{figure}

 The staggered lattice fermion formulation is deployed in the pilot study to demonstrate feasibility with control of $\chi{\rm SB}$ and
 serves as a lower bound for the required resources. Domain wall fermions would be 10-20 times more demanding.
 The Symanzik improved tree level gauge action is used with stout smeared
 gauge links to minimize lattice cut-off effects in the study.
A staggered operator which creates a state that lies
in the spin-taste representation $\Gamma_{S}\otimes \Gamma_{T}$
also couples to one lying in the $\gamma_{4} \gamma_{5} \Gamma_{S}
\otimes \gamma_{4}
\gamma_{5} \Gamma_{T}$ representation.
Thus a staggered meson correlator has the general form
$$
{\rm C(t) = \sum_n \bigl[A_n e^{-m_n(\Gamma_{S}\otimes
\Gamma_{T})t}
+ (-1)^{t} B_n
e^{-m_n(\gamma_{4} \gamma_{5} \Gamma_{S} \otimes \gamma_{4} \gamma_{5}
    \Gamma_{T})t}}\bigr]
$$
with oscillating contributions from 
parity partner states. For the scalar meson
($\Gamma_{S}\otimes \Gamma_{T}= 1\otimes 1$), the parity partner
is $\gamma_{4} \gamma_{5} \otimes \gamma_{4} \gamma_{5}$ which corresponds to
one of the pseudoscalars in the analysis. 
%For the non-singlet scalar this is one
%of the (taste-split) pions and so is a significant low lying
%contribution which must be included in the fits.
%
For flavour singlet mesons, the correlator is of the form
$
{\rm C(t) = C_{conn}(t) + C_{disc}(t)}
$
where ${\rm C_{conn}(t)}$ is the correlator coupled to the non-singlet meson state
and ${\rm C_{disc}(t)}$ is the contribution of disconnected fermion loops in the annihilation diagram. 
% correlator suitably
%corrected ( by $1/4$) to account for the extra $4$ tastes that
%can contribute as compared with the connected correlator.
%In this pilot study $n=3$ and the three fermion masses are degenerate representing 12 flavors in the staggered approach.
%There is no oscillating contribution to ${\rm D_{conn}(t)}$ since, in this case,
%the parity partner would be taste non-singlet.
%
Figure~\ref{fig:scalar} on the left shows the propagation of the lowest flavor-nonsinglet state 
together with its oscillating parity partner, as determined by ${\rm C_{conn}(t)}$. The singlet scalar
mass, the Higgs particle of the strongly coupled gauge model, is determined from the flavor singlet correlator
${\rm C(t)}$ that includes the annihilation diagram on the right side of Figure~\ref{fig:scalar}.
The result shows a significant downward shift of the Higgs mass from the annihilation diagram 
of the scalar singlet correlator, as expected close to the conformal edge. Very large scale simulations
are needed with substantial resources to confirm the emerging picture of this pilot study.

Once the technology of scalar spectra is under control,
the USQCD BSM groups plan to investigate the
dynamics of strongly coupled BSM gauge theories that could be responsible for producing a light dilaton-like state.
This possibility remains consistent with the  preliminary values of the observed decay modes
of the Higgs-like resonance discovered at the LHC. Using symmetry arguments,
the dilaton couplings are generically ${\rm v/f_\sigma }$ suppressed compared to standard model  Higgs couplings.
Fermion couplings are also modified by the anomalous dimensions of the SM fermions.
Couplings to massless gauge bosons are loop induced and are determined by the $\beta$-function coefficients
of the composites.  The
coupling to the W and Z can be made realistic and consistent with electroweak precision
tests only if the scale of conformal breaking is close to the scale of  electroweak symmetry breaking
with ${\rm v/f_\sigma \approx 1}$.
%This is a dynamical challenge which requires non-perturbative lattice
%calculations to investigate the plausibility of a light dilaton scenario
%in strongly coupled BSM gauge theories without fine tuning.

\subsection{The  PCDC Dilaton Relation on the Lattice}
\label{subsec:pcdc}

The USQCD BSM groups plan to deploy
non-perturbative lattice methods
to explore the implications of the partially
conserved dilatation current (PCDC) relation when applied to four-dimensional BSM gauge models.
The PCDC relation connects the dilaton mass $m_\sigma$ and its decay amplitude $f_\sigma$ with
the non-perturbative gluon condensate.
For discussion of the PCDC relation constraining
the properties of the dilaton, standard arguments,
like in~\cite{Appelquist:2010gy,Hashimoto:2010nw,Matsuzaki:2012vc} can be followed.

In strongly interacting gauge theories, like the sextet model under active USQCD BSM investigation~\cite{Fodor:2012ty,DeGrand:2012yq}, a dilatation current
${\mathcal D}^\mu=\Theta^{\mu\nu}x_\nu$
can be defined from the symmetric energy-momentum tensor $\Theta^{\mu\nu}$. Although the massless theory is scale invariant on the classical
level, from the scale anomaly the dilatation current has a non-vanishing divergence,
$
\partial_\mu \mathcal {D}^\mu = \Theta_\mu^\mu=\frac{\beta (\alpha)}{4\alpha}G^a_{\mu\nu}G^{a\mu\nu} \, .
$
While $\alpha(\mu)$ and $G^a_{\mu\nu}G^{a\mu\nu} $ depend on the renormalization scale $\mu$, the trace of the
energy-momentum tensor is scheme independent after renormalization. In BSM models, the massless fermions are in the
fundamental representation, or in the two-index symmetric representation of the SU(N) color gauge group in most studies.

If we assume that the perturbative parts of the composite operators $G^a_{\mu\nu}G^{a\mu\nu}$ and $\Theta_\mu^\mu$
are removed, only the non-perturbative (NP) infrared parts have to be considered
in $\partial_\mu \mathcal {D}^\mu $.
The dilaton mass $m_\sigma$ and its decay amplitude  $f_\sigma$ are defined by the vacuum to dilaton matrix element
of $\partial_\mu \mathcal {D}^\mu $,
$
 \langle 0| \partial_\mu\mathcal{D}^\mu(x)|\sigma(p)\rangle = f_\sigma m^2_\sigma  e^{-ipx}\, .
$
The important partially conserved dilatation current (PCDC) relation,
$$
m^2_\sigma\simeq - \frac{4}{f^2_\sigma }\langle 0|\Bigl[\Theta^\mu_\mu (0)\Bigr]_{NP}|0\rangle \, ,
$$
then connects $m_\sigma$ and $f_\sigma$ to the nonperturbative gluon condensate close
to the conformal window~\cite{Appelquist:2010gy}.

Predictions for $m_\sigma$ when close to the conformal window depend on the behavior
of  $f_\sigma$ and the non-perturbative gluon condensate $G^a_{\mu\nu}G^{a\mu\nu}|_{NP}$.
There are two
different expectations about the limit of the gluon condensate to $f_\sigma$ ratio when the conformal window is approached.
In one interpretation,  the right-hand side of the PCDC relation is predicted to approach zero in the limit, so that the dilaton mass
$m_\sigma^2 \simeq (N^c_f -N_f)\cdot\Lambda^2$
would parametrically vanish when the conformal limit is reached.
The $\Lambda$ scale is defined where the running coupling becomes strong to trigger
$\chi{\rm SB}$. The formal parameter $N^c_f -N_f$ with the non-physical (fractional) critical number of fermions  vanishes
when the conformal phase is reached~\cite{Appelquist:2010gy}.
In an alternate interpretation the right-hand side ratio of the PCDC remains finite in the conformal limit and a
residual dilaton mass is expected when scaled with  $f_\sigma \simeq \Lambda$ ~\cite{Hashimoto:2010nw,Matsuzaki:2012vc}.

It is important to note that there is no guarantee, even with a very small $\beta$-function 
near the conformal window, for
the realization of  a light enough dilaton to act as the new Higgs-like particle. 
Realistic BSM models have not been built with
parametric tuning  close to the conformal window. 
For example, the sextet model is at some intrinsically determined position
near the conformal window and
only non-perturbative  lattice calculations can explore the physical properties of the scalar particle.

%\subsection{The gluon condensate}
The lattice determination of the non-perturbative gluon condensate is important in exploring the consequences of the PCDC relation.
Power divergences are severe in the
calculation of the lattice gluon condensate,
because the operator $\alpha G^a_{\mu\nu}G^{a\mu\nu}$ has quartic UV divergences.
If the gluon condensate is computed on the lattice from the
expectation value of the plaquette operator $U_P$ 
the severe UV divergences make the separation of the finite NP part difficult,
or even suspect on theoretical grounds.
If it turns out that the meaningful separation of the gluon condensate 
into perturbative and non-perturbative parts exists,
as suggested by some recent developments,
it will be very important to undertake these investigations in BSM gauge models with full fermion dynamics.
One of the most interesting recent developments relates the energy from gradient flow
to $G^a_{\mu\nu}G^{a\mu\nu}|_{NP}$ on lattice configurations
at small flow times~\cite{Luscher:2010iy,Luscher:2011bx}. 
The USQCD BSM groups plan to investigate this important alternative approach.

\section{Higgs as a pseudo-Goldstone boson}
\label{sec:pngb}
With the discovery of a new particle at 126 GeV with properties
so-far consistent with the standard model Higgs boson, it is natural
to explore strong dynamics where this  Higgs  is a scalar
pseudo-Nambu-Goldstone boson (PNGB).
%The global symmetries that give
%raise  to a composite  Higgs bound state in a confining gauge theory  will also remove the dangerous quadratic
%mass divergence of the elementary Higgs, solving the problem of ``naturalness'',  with
%the Higgs mass lifted to electroweak scale by symmetry breaking terms induce by
%coupling to the standard model.
This PNGB Higgs mechanism~\cite{Peskin:1980gc}
%,  has a long history with the early introduction of
%the technicolor ideas by Weinberg~\cite{Weinberg:1979bn}, and Susskind~\cite{Susskind:1978ms} and more recently it
plays a crucial role in little Higgs
~\cite{Schmaltz:2005ky,Katz:2003sn} and minimal conformal technicolor~\cite{Galloway:2010bp} models.
In little Higgs theory, an interplay of global symmetries is devised in an effective
Lagrangian  to cancel the quadratic divergences for the Higgs mass to one loop. This provides a
weakly coupled effective theory with little fine tuning up to energies
on order of 10 TeV. A large range of phenomenologically interesting models
continues to be explored as weakly coupled extensions of the standard model
with PNGB Higgs scalars.

Lattice field theory can explore realizations of this approach, starting
with an ultraviolet complete theory for strongly interacting gauge
theories with fermions in real or pseudo-real representations of the
gauge group giving rise to Goldstone scalars. There are no ultraviolet
divergences in this strongly interacting sector, and in the chiral limit the Higgs scalar is
a member of set of Goldstone Bosons that includes the triplet of
``techni-pions'', required to give masses to the W and Z. Subsequently the
mass of this Higgs is naturally light with its mass induced by small
couplings to the weak sector. A central challenge to support this
scenario for models based on effective phenomenological Lagrangian is
to use lattice field theory to demonstrate that viable UV complete theories
exist and to understand the impact of extending the standard
model with this strong sector replacing the weakly coupled elementary
Higgs.

\subsection{Minimal PNGB Model}
\label{subsec:minimal}
The minimal PNBG Higgs model
%, sometime referred as minimal conformal technicolor~\cite{Galloway:2010bp} ,
consists of a $SU(2)$ color gauge
theory with $N_f =2$ fundamental massless fermions.  Additional sterile
flavors with $N_f >2$ can be added ~\cite{Galloway:2010bp}
to drive the theory close to or into the conformal window.
Because of the pseudo-real properties of $SU(2)$ color group, the conventional
$SU(N_f)_L \times SU(N_f)_R$ vector-axial symmetry becomes a larger
$SU( 2 N_f)$ flavor symmetry combining the $2 N_f$ left/right 2-component
chiral spinors. Most-attractive-channel arguments suggest that $SU(2 N_f)$ will break
dynamically to $Sp(2N_f )$. If explicit masses are given to $N_f -2$
flavors, the remaining 2 massless flavors yield the $SU( 4)/Sp(4)$ coset
with 5 Goldstone Bosons: the  isotriplet pseudo-scalars (or ``techni-pions'' ) to give mass to the W, Z, and
two isosinglet scalars. With additional explicit breaking, these two  extra Goldstone Bosons become massive.
The lighter is a candidate for a composite Higgs and the heavier is a possible
candidate for dark matter.

%It is exciting goal for
%lattice field theory to establish the existence of such theories that
%are not in contradiction with precision electro-weak experiments.

For the lattice formulation, it is convenient to use the standard
four-component Dirac spinors for the vector-like theory. It can be shown
that the most general mass matrix in the $SU(4)$ symmetric two-component
chiral representation is equivalent to the up and down quark masses
of the Dirac four- component theory, up to $SU(4)$ transformations plus a
theta  term to absorb the complex phase. (See Kogut et
al \cite{Kogut:2000ek} for the basic formalism required for this
demonstration.) As in QCD one fixes the vacuum theta angle to zero to avoid CP
violation. 
Standard lattice methods are suited to investigate this model at this
stage.  As a first step, lattice
calculations for $N_f =2$ have been performed~\cite{Lewis:2011zb} using
Wilson fermions that give non-perturbative support to the breaking pattern,
$SU(4) \rightarrow Sp(4)$, favored by the most attractive channel argument in the
chiral limit.

However this is still not the correct vacuum alignment. With the
fermion mass parameters, all 5 GBs are given a common mass. To lift the
two scalars keeping the isotriplet pseudo-scalars  massless, one must
appeal to the weak interactions coupling the Higgs composite to the
top quark. As demonstrated by the use of the chiral Lagrangian~\cite{Galloway:2010bp,Katz:2003sn}
in the 5-Goldstone sector, the top quark loop destabilizes the vacuum
alignment in $Sp(4)$ shifting to a new vacuum preserving the 3
pseudoscalars ( $\pi_T$)  but lifting the 2 scalars as required. The CP =1 scalar is
identified as a composite Higgs ($\bf h$), and the CP = -1 scalar ($\bf A$)
is a possible candidate for  dark matter.  The new alignment
includes an alignment angle $\theta$  determined by the ratio of the top quark
loop and the mass matrix that fixes the mass of the Higgs relative to
the additional scalar. The alignment angle interpolates between a
weakly coupled effective low energy theory and a strongly coupled
composite Higgs depending on the ratio of the Higgs vacuum
expectation value ($\bf v$) to the chiral Lagrangian coupling parameter
$\bf f $. For $\bf v/f = \sin \theta$ small, with the weak scale set by $\bf v$ = 246 GeV,
after the ``techni-pions'' are ``eaten'' by  W and Z,
the ``dark matter'' scalar ($m^2_A = m^2_h/\sin^2 \theta$) along with all other
composites are heavy at the new strong scale set by
$\bf f$. The theory approximates the standard model following the
methods of the little Higgs approach. For large $\bf v/f = O(1) $,
the model is  typical of strong coupling technicolor  with the massive
PNGB scalar merging with a rich technicolor composites. In between
there is a range of finely tuned theories  worthy of careful study.

It is important to confirm rigorously the above scenario as the
correct description of the low energy effective theory in the context
of {\em ab initio} non-perturbative lattice calculation. To include the effect of
the top quark into the lattice simulations, one must add the four
Fermi term induced by the top quark loop. The four Fermi term is then
recast as a quadratic term coupled to a Gaussian auxiliary field.
It is  an open question whether or not the four-fermion operator requires
an explicit cut-off or whether it becomes a relevant operator due
to a large anomalous dimension. A central challenge is to use
lattice field theory to demonstrate that UV complete theories are
viable with PNGB scalars consistent with current experimental
constraints.

\subsection{Lattice Formulation}
\label{subsec:lattice}
In the continuum there is a rather large range of models proposed for
strongly interacting gauge theories with a candidate PNGB for the
Higgs. However lattice formulations and numerical simulations for each new
theory require a substantial investment in software and algorithmic
development, so it is reasonable to start with the simplest example.
Fortunately there is already considerable experience
with this  two-color, two-flavor lattice gauge theory example.
In fact lattice gauge theories with SU(2) color have been studied extensively as
toy models for QCD at
finite chemical potential because, unlike SU(3) QCD, the chemical
potential  does not introduce a sign problem coming from the
Fermion determinant. It is well known that these sign problems, when they
occur, usually make lattice studies extremely difficult if not impossible.

A similar difficulty might present itself for this study.  For the
PNGB application, we must consider two new mass-like extensions to the
the lattice action.  One is the introduction of
mass term for the diquark and the other is the four Fermi term
induced by the top-quark loop. Remarkably, each has been studied to
some degree.  For example the diquark condensate is analyzed in Kogut et
al ~\cite{Kogut:1999iv}, and Hands et al  have
demonstrated ~\cite{Hands:2000ei,Hands:2007uc} that adding the diquark
term to the Wilson or staggered action does not cause a sign
problem. In the case of the four-Fermi term, this must be recast via
a Lagrange multiplier as local bilinear mass like term. Again,  a similar term has
already been considered as an chiral extension for QCD ( $\chi
QCD$) ~\cite{Brower:1995vf}.  In $\chi QCD$, it was realized that the
(mild) sign problem for SU(3) color is absent for SU(2) color.
All these observations rely on the special properties of the
pseudo-real nature of the SU(2) group.   Still, there is more to
consider in this application when one includes
simultaneously  both the diquark term and the $\chi QCD$-like extension.

When both the diquark and bilinear for the four Fermi term are included together, the measure is now a
Pfaffian, or the square root of the fermionic determinant. Remarkably, it  can still be proven that for this  specific example of
the four-Fermi coupling of $\chi QCD$, the fermionic
measure for the Wilson action is again positive
definite.    Another example has been considered recently for staggered fermions by 
Catterall~\cite{Catterall:2012vu}. Clearly the  extent of parameter space devoid of a severe sign problem
needs further study.  For example, the possibility that the breaking of custodial symmetry reintroduces
a sign problem has not been fully explored. Most likely for some models, one will have to
learn how to perturb away from examples with positive definite measure with methods
similar to those used for QCD with small chemical potentials. Nonetheless,  it appears that
substantial progress can be made studying the vacuum alignment in strong coupling gauge theories
by conventional lattice Monte Carlo methods.

The next issue is the choice of action: staggered, Wilson or domain
wall? Our first choice is to start with the Wilson action while we
investigate further possible advantages for the least expensive
staggered case and more expensive domain wall case. The Wilson
term  breaks chiral symmetry but fortunately it aligns  the vacuum
consistent with the breaking of $SU(4)$ to $Sp(4)$.  Research is underway to
see if domain wall fermions at finite lattice spacing have, as expected,
the full enhanced $SU(4)$ symmetry at infinite wall separation. If
this is proven to hold, clearly the domain wall implementation, although
more costly, does have theoretical advantages worth considering.
As has been to be the case for lattice QCD research, we anticipate that each action has advantages for
specific questions.

\section{Studies of Supersymmetric Theories on the Lattice}
\label{sec:susy}
%\subsection{Introduction}

One natural solution to the problems with the Standard Model Higgs is to incorporate supersymmetry.
The Higgs is naturally light in such theories since it is paired with a fermonic superpartner whose mass is
protected by chiral symmetries.

Supersymmetric theories have been extensively studied over many decades now both in the context of LHC
phenomenology and string theory.
However, for many years, significant barriers prevented serious study of such theories on the lattice.
Recently this has changed; a series of studies has shown that $\mathcal{N}=1$ super Yang-Mills can be studied
successfully using domain wall fermions~\cite{Endres:2009yp} and using these ideas one can even contemplate lattice study of super
QCD theories.

A second recent development has been the construction of lattice theories which retain some exact
supersymmetry at non zero lattice spacing. 
The prime example of such a theory is $\mathcal{N}=4$ super Yang-Mills. One of the longstanding barriers to
studying supersymmetric theories on lattices is the
need to
tune the bare couplings for large numbers of
supersymmetry breaking operators to achieve a supersymmetric
continuum limit. The exact supersymmetry present in the ${\cal N}=4$
lattice theory we are studying saves the theory from this fine tuning problem
and renders it possible for the first time to use lattice simulations to
study the non-perturbative structure of this theory~\cite{Catterall:2012yq} - which lies at
the heart of the AdS/CFT correspondence and hence to
quantum gravity.

As part of its BSM physics program USQCD is engaged in extensive studies of supersymmetric lattice theories. 
Currently this is concentrated in two major directions: studies of ${\cal N}=4$ super Yang-Mills using new 
formulations which possess exact supersymmetry and a long range effort to
determine the non-perturbative features of super QCD using domain wall fermions. 

The work on ${\cal N}=4$ is exciting from a variety of
perpectives; it potentially allows us to test and extend the AdS/CFT correspondence beyond the
leading supergravity approximation and hence to
learn about quantum gravity. But, in addition, it serves as a useful arena in which we
can hope to develop the tools and techniques necessary to extract physics from four dimensional
conformal field theories in general. In this way it
is complementary to USQCD's effort to
determine the conformal window and physical content of non-supersymmetric near conformal gauge theories. Indeed, in light of the recent discovery of a light Higgs at the LHC, it is important to
understand whether a light Higgs might arise as a pseudo dilaton associated with spontaneous
breaking of conformal symmetry. In this context, the flat directions generically encountered in
supersymmetric theories, and those seen in ${\cal N}=4$ in particular,  may play an important role since they allow for just such a 
spontaneous breaking phenomenon to occur.

The work in super QCD is important for gaining a detailed understanding of the dynamical breaking of
supersymmetry. This has important phenonenological implications since
non-perturbative supersymmetry breaking in some high scale hidden sector generically feeds down to
determine the structure of soft breaking terms in any low scale 
supersymmetric theory such as the MSSM. Indeed the values of many of these soft parameters are generically determined by just a handful of non-perturbative
quantities in the high scale theory. An ab initio lattice calculation of such quantities in the  high scale
theory can thus provide valuable constraints on models of BSM physics. 
While super QCD is known to have no exact susy breaking
vacua it does possess many metastable vacua whose lifetimes can be sufficiently large that
they can play this role \cite{Intriligator:2006dd}. To
facilitate such studies USQCD plans to extend its existing domain wall codes focused on ${\cal N}=1$ super Yang-Mills to allow for the inclusion of the chiral multiplets needed to explore super QCD.

\subsection{${\cal N}=4$ SYM}
\label{subsec:N4}

In recent years a new approach to the problem of putting supersymmetric
field theories on the lattice has been developed based on
orbifolding and topological twisting - see the recent
Physics Report \cite{Catterall:2009it}. The central advantage of the new formulations is that they
retain some exact supersymmetry for non-zero lattice spacing. 

In the past year members of USQCD have been using these
ideas to push forward a
program to conduct a non-perturbative study of ${\cal N}=4$ super Yang-Mills theory on the lattice.
The lattice formulation that is being studied corresponds to a discretization of a topologically
twisted form of the ${\cal N}=4$ action. The constraints inherent in the construction pick out
uniquely both the form of action, the distribution of fermions and bosons over the
lattice, the structure of gauge covariant difference operators and even the type of lattice - in
this case the so-called $A_4^*$ lattice (a lattice whose fundamental basis vectors correspond to
the fundamental weight lattice of $SU(5)$).
This approach has the crucial advantage of leading to a lattice action which
possesses a single exact supersymmetry at non zero lattice spacing. This theory may then
be studied using the
standard tools and algorithms of lattice QCD. The existence
of a supersymmetric lattice formulation should dramatically reduce the usual fine tuning
problem associated with supersymmetric lattice theories and makes 
it possible, for the first time, to attempt
a serious non-perturbative study of the theory. 

The availability of a supersymmetric lattice 
construction for this theory is clearly very exciting from the point of 
view of exploring the possible
holographic connections between gauge theories and 
string/gravitational theories. For example, there is growing interest to relate
the decay constant of the holographic techni-dilaton
to important physics of the 126 GeV Higgs-like boson~\cite{Elander:2012fk}. 
For the initial work on supersymmetric lattice contructions, 
see \cite{Catterall:2010jh,Catterall:2010fx,Catterall:2010ng,Catterall:2010gf,Catterall:2009xn,Catterall:2008yz}. 
Lattice constructions
allow new strategies to be employed such as strong coupling
expansions and Monte Carlo simulation, which may lead to new insights
and tools for
extracting non-perturbative information. 
Preliminary
results on the phase diagram of ${\cal N}=4$ super Yang-Mills have been reported \cite{Catterall:2011pd}.
Initial
results point to the existence of a single phase corresponding to a deconfined conformal field
theory in the infrared for any value of the bare gauge coupling. Work is ongoing to develop a parallel
code to extend this work to larger lattices.

In addition to the development of simulation code it is important to
understand in detail how much tuning of the bare lattice parameters will be needed
to recover full supersymmetry as the lattice spacing is sent to zero. Substantial analytic
progress 
on this question has been obtained recently. A perturbative calculation has been completed
and published~\cite{Catterall:2012yq} which contains several remarkable results; perhaps the most interesting of these is a calculation which shows 
that the effective potential of the lattice theory vanishes to all orders in perturbation theory in complete analogy to the continuum theory.
The topological character of the twisted supersymmetry
which is realized in the lattice theory plays a crucial role in this result. This result
prohibits
the appearance of any scalar mass terms as a result of
quantum corrections. In addition, the authors were able to
show that no fine tuning was needed at one loop to target the usual ${\cal N}=4$ theory -- a single
wave function renormalization was the only correction needed to absorb all divergences of
the lattice theory. Indeed, the authors give general arguments that the beta function of the lattice theory 
vanishes at one loop - affording the only known example of a non-trivial
four dimensional lattice
theory which is finite at 1-loop!

Beyond one loop there is potential for logarithmic running of
various dimensionless couplings in the lattice
action and numerical methods will likely be needed to proceed
further. 
Work in this direction has already
started with the elucidation of
the action of all the additional twisted supersymmetries on the lattice fields. This analytic
work must now be incorporated into the simulations to allow for numerical tests of various
broken supersymmetric Ward identities.

In addition, a study has recently been completed on the question of whether the lattice
theory exhibits a sign problem. The Pfaffian representing the
effect of the fermion loops is generically complex in Euclidean space
rendering the use of Monte Carlo methods potentially
problematic. Members of USQCD have been examining this issue in some detail in the
dimensional reduction of this model to two dimensions. The 
results reported in \cite{Catterall:2011aa} are encouraging; the
observed fluctuations in the Pfaffian phase 
as measured in the phase quenched Monte Carlo
ensembles used for simulation are actually rather small allowing for
the possibility of the use of reweighting techniques to compute expectation values. Indeed for small lattices the phase is sufficiently
close to zero for gauge group $U(2)$ that it can be neglected in comparison with
current statistical errors. Furthermore, there exist formal arguments based on the structure of
the lattice moduli space and the topological character of the lattice partition function that
also suggest that the sign problem is absent in this theory.

Finally there are efforts currently underway to study a variety of two point correlation functions in the theory.
In principle on a sufficiently large lattice these are expected to exhibit power law rather than
exponential behavior consistent with the conformal nature of the continuum theory. Some
specific predictions for certain operators exist in the continuum literature; eg for the
Maldecena loop or for the anomalous dimensions for various scalar operators like the
Konishi and supergravity multiplet. 

Unlike the study of
the conformal window in non supersymmetric lattice gauge theories the exact supersymmetry
prohibits any mass terms for the fermions and so the finite box size and lattice spacing constitute the
only features that break conformality even at strong coupling. Furthermore, some of the leading
cut-off effects stemming from tadpole terms in lattice perturbation theory are absent in the
${\cal N}=4$ action since the gauge fields are valued in the algebra and not the group. It remains to
be seen if these differences reduce lattice artifacts to levels where the power laws can be differentiated
from exponentials on realistic lattice sizes. If not other approaches such as those based on exploiting
the conformal mapping $R^4\to S^3\times R$ may need to be employed to make reliable
measurements of quantities like anomalous dimensions.

Since the discovery of a light Higgs-like state at the LHC there there has been renewed interest in
the possibility that the Higgs is a (pseudo) dilaton associated with spontaneous breaking of
conformal invariance. Such a state is particularly simple to realize in a supersymmetric theory
with flat directions such as ${\cal N}=4$ Yang-Mills - the dilaton corresponding to
translations along such a flat direction. The ${\cal N}=4$ lattice action we are investigating
possesses such flat directions and furthermore these flat directions are
stable against quantum corrections due to the exact lattice supersymmetry. 
To realize the spontaneously broken state and the corresponding dilaton in the
lattice theory then requires that a non-zero vacuum expectation value of one or more scalar fields
be fixed
and the Monte Carlo procedure modifed so that no integration over
this mode is carried out. This procedure can be carried out in a straightforward manner and will
be investigated in the near future.

\subsection{Dynamical SUSY Breaking and Super QCD}
\label{subsec:susyQCD}

It is straightforward to ``supersymmetrize'' the usual theories of particle
physics; the Minimal Supersymmetric Standard Model (MSSM) is perhaps the most studied extension
of the Standard Model of particle physics. In such models the Higgs
is naturally light since it is accompanied by a fermionic partner whose
mass is protected by chiral symmetries.
A great deal of effort has been expended on predicting signals for SUSY/MSSM at LHC.

Of course the low energy world we inhabit is manifestly not supersymmetric; so a key component
of any realistic theory of Beyond Standard Model physics must provide a mechanism for spontaneous supersymmetry
breaking. In general a variety of no go theorems ensure that any such symmetry breaking must
be non-perturbative in nature and hence inaccessible to most continuum calculations.
As a consequence this breaking is usually handled by explicit soft breaking terms in the
low energy theory. In general there are very many of
these terms and their corresponding couplings (so-called soft parameters) 
which results in a large parameter space and a lack of predictivity
in the low energy theory.

However, in general we might expect
that these parameters are determined from
dynamical breaking of SUSY at high energies in a ``hidden sector''.
A supersymmetrized version of QCD - super QCD - with $N_c$ colors and
$N_f$ massive flavors is a natural
candidate for this hidden sector. For $N_c+1 \le N_f < \frac{3}{2}N_c$ it is thought that super QCD has
long lived metastable
SUSY breaking vacua \cite{Intriligator:2006dd}. The lifetimes of these metastable vacua can exceed the age of the Universe and ensure that the
physical vacuum breaks supersymmetry. Within such a vacuum state non-perturbative phenomena such
as confinement and chiral symmetry breaking precipitate a breaking of supersymmetry. 
Furthermore, if the quark masses are small compared to the confinement scale
these vacua have extremely long lifetimes. 
If the standard model fields are coupled to the hidden sector fields in
an appropriate fashion then such 
non-perturbative dynamics arising in the broken phase of this theory can feed down to yield
soft supersymmetry breaking terms in the low energy effective theory.

Thus a detailed understanding of the vacuum structure
and strong coupling dynamics of super QCD can strongly constrain possible supersymmetric models
of BSM physics; in some cases leading detailed predictions of the soft parameters in low energy
BSM theories in terms of
a handful of non-perturbative quantities
 obtained in the hidden sector super QCD theory. Lattice simulations of 
supersymmetric lattice QCD thus have the potential to play an important role in constraining
the parameter space of supersymmetric models and in building realistic supersymmetric
theories of BSM physics.

\begin{figure}[hbt]
\begin{center}
\includegraphics[width=0.7\textwidth]{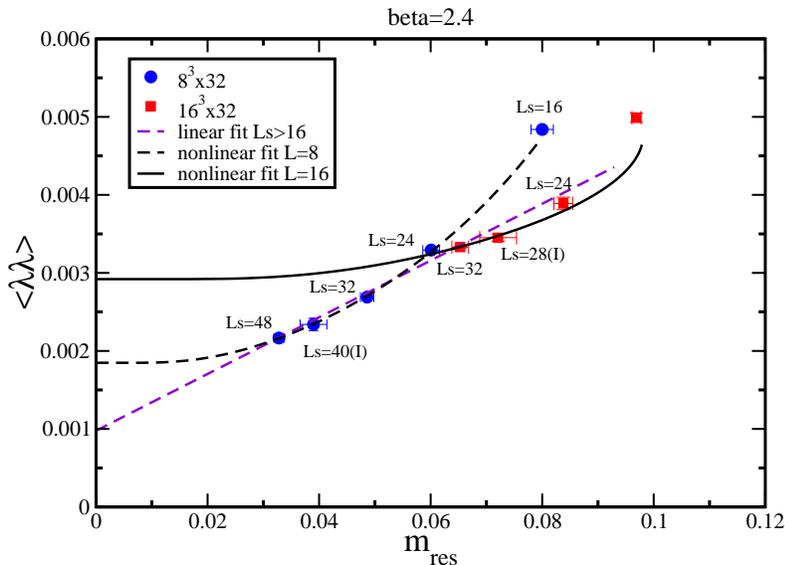}
\caption{\label{N=1cond}Gaugino condensate vs residual mass for $SU(2)$ ${\cal N}=1$ super Yang-Mills regulated using a domain wall fermion action. The gauge coupling $\beta=2.4$ while
the bare fermion mass is set to zero. The data points correspond to different lattice volumes and $L_s$}
\end{center}
\end{figure}

In contrast to the ${\cal N}=4$ super Yang-Mills model discussed
above the technical problem that must be immediately faced in studying generic supersymmetric lattice
theories is that
supersymmetry is broken by discretization and it is non-trivial to regain SUSY
as the lattice spacing is sent to zero with conventional lattice fermion formulations.
Luckily in the case of ${\cal N}=1$ super Yang-Mills theory,
which contains both gluons
and their fermionic superpartners gluinos, this problem can be avoided
by the use of domain wall fermions (DWF).  In this case
the exact lattice chiral symmetry of the fermion action ensures that SUSY is {\it automatically} recovered in 
the chiral limit.

Preliminary work by USQCD has already revealed a non-zero
gluino condensate in the SU(2) theory in agreement with theoretical expectations \cite{Giedt:2008xm, Endres:2009yp}. 
A plot of the gaugino condensate vs residual
mass is shown in Figure~\ref{N=1cond}.
However, to build a realistic theory capable of yielding 
the soft parameters of the MSSM one needs to add $N_f$ quarks and
their scalar superpartners (squarks), and additionally extend the gauge group
to a larger number of colors $N_c$. 
To restore SUSY in the continuum limit now requires {\it both} use of DWF {\it and} tuning of
parameters in the squark sector \cite{Elliott:2008jp}. In principle this can be done by doing
a series of runs over 
a grid in squark parameter space and using (offline) reweighting
techniques in the scalar sector to tune to the supersymmetric point. Clearly such calculations are ambitious
but constitute an important long term goal of USQCD's work on supersymmetric theories of
BSM physics.

\section{Methods and Phenomenological Applications}
\label{sec:methods}

% Anna begin-----------------------------

%\section{Toolset for conformal or near conformal systems}

%\subsection{}

The field of BSM lattice studies is still young, but it has already become clear that methods developed 
for lattice QCD are not always optimal and frequently not sufficient to investigate conformal or near-conformal systems. 
We continuously re-evaluate our methods and develop new techniques that are better suited to  
explore walking or  IR-conformal dynamics that are expected to emerge near the conformal window.
Conformal systems are particularly difficult to investigate as both the finite fermion mass and finite lattice volume break conformal invariance. 
The most promising lattice techniques attempt to use rather then fight this issue.
A repeating theme of BSM model investigations is that it is essential to consider the systems from many different angles, use complementary 
and contrasting methods to fully understand infrared dynamics. 
Exciting phenomenological applications are beginning to emerge from these investigations.

\subsection{Running Coupling and ${\bf \beta}$-function}
\label{subsec:coupling}

Running coupling studies connect the well understood perturbative regime to strong coupling conformal dynamics, or,
near conformality, to chiral symmetry breaking. These are  particularly important  tasks close to the conformal edge.
It is challenging  to  distinguish a  chirally broken system with slowly running coupling from a strongly coupled conformal one. 
In both cases the weak coupling ultraviolet behavior is described by the asymptotically free gaussian fixed point. 
Below the conformal window the system is in the chirally broken phase and in the simplest 
theoretical scenario other phases in the continuum
are not expected to emerge at strong coupling.
Inside the conformal window,  
the coupling of the model runs from the UV phase of perturbation theory to conformal behavior in the infrared limit. 
This evolution along the renormalized trajectory of the critical surface connects the UV fixed point to the
conformal infrared fixed point (IRFP).
Lattice artifacts, which can only create other phases at finite cutoff, have to be identified and separated from
the continuum analysis.
The existence of the conformal phase and the critical exponents of the  IRFP are universal, although the location of the IRFP in 
the multi-parameter action space depends not only on the lattice action but the specific renormalization scheme as well. 

Most lattice running coupling calculations  are done in the chiral limit with vanishing fermion masses as a function of the 
scale set by the finite lattice volume, thus avoiding the two most difficult aspects of lattice simulations. 
These studies aim to identify an IRFP from the flow of the gauge coupling.  
%It is important to note that non-observation of an IRFP in a running coupling lattice study is not sufficient to draw a firm conclusion.
It is  necessary to show in these studies that sufficiently strong continuum couplings were probed in large infrared volumes
to capture the existence, or non-existence of the IRFP. 
Sometimes the emergence of  spurious phase transitions due to discretization artifacts makes 
it challenging to reach continuum strong couplings.

\begin{figure}[h!]
\begin{center}
\begin{tabular}{cc}
\includegraphics[width=0.48\textwidth]{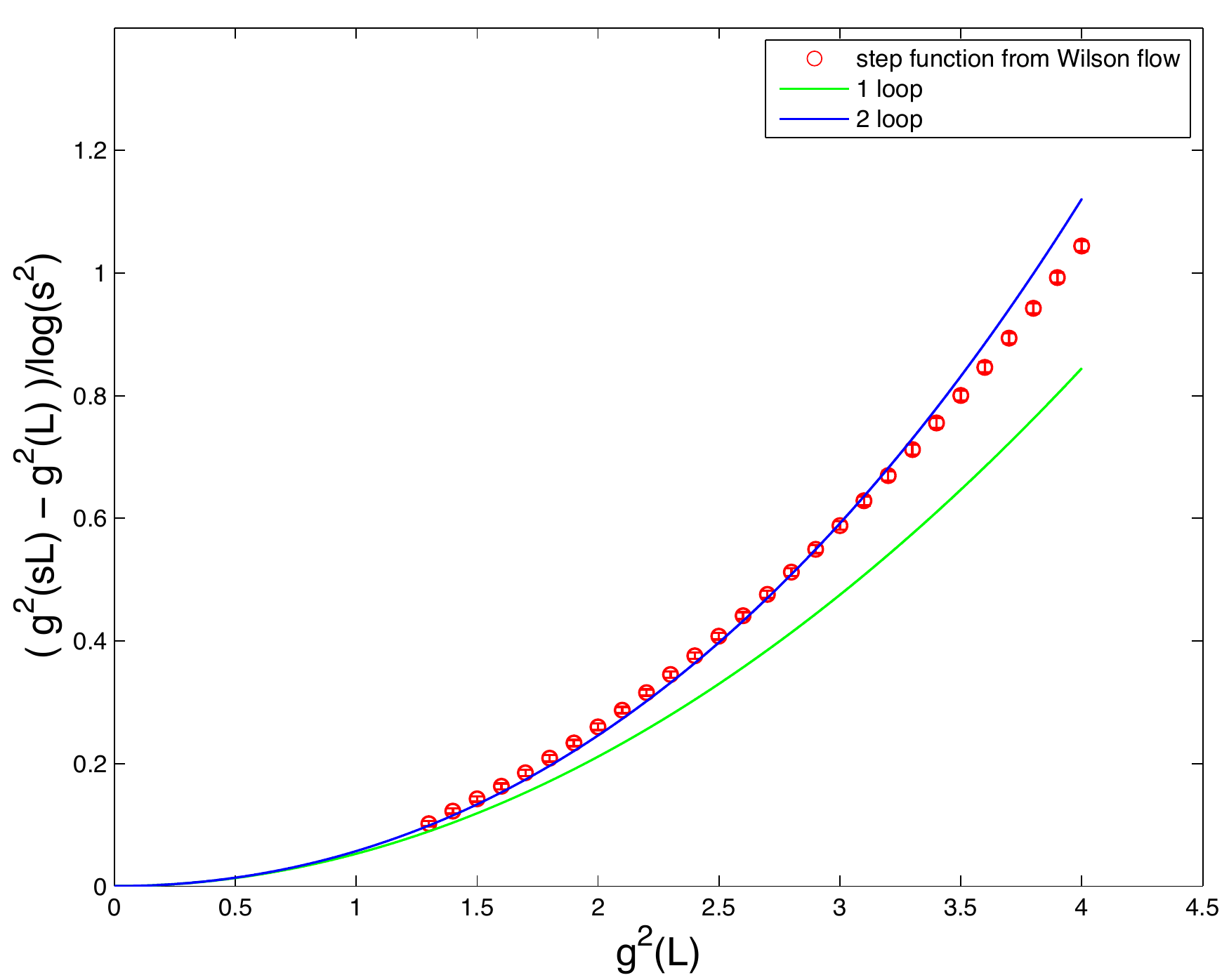} 
\includegraphics[width=0.5\textwidth]{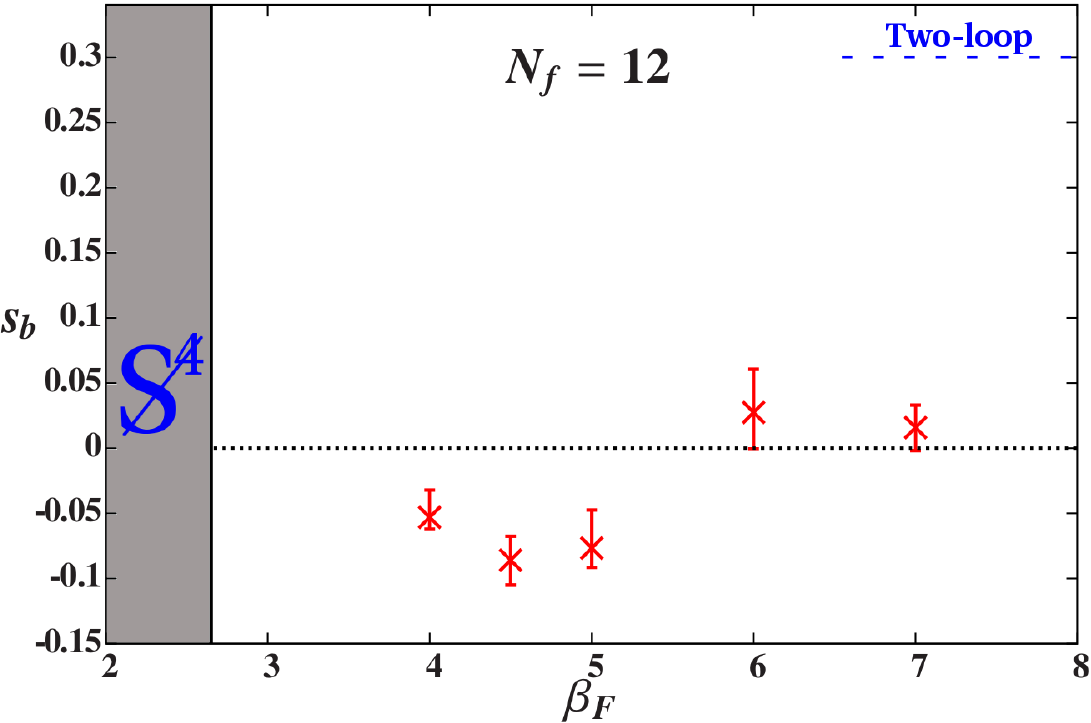} 
\end{tabular}
\end{center}
\vskip -0.25in
\caption{\label{fig:coupling} Two complementary approaches are illustrated for the running coupling and its
$\beta$-function. On the left plot,
the discrete step $\beta$-function of SU(3) continuum gauge theory coupled to $N_f = 4$ flavors 
of massless fundamental fermions is shown for a scale change of s = 3/2. The coupling $g(L)$
runs with the linear size $L$ of the finite continuum volume~\cite{Fodor:2012td,Fodor:2012qh}.
The continuum extrapolated result  is shown together with the 1-loop    
and 2-loop results of perturbation theory.
On the right plot, preliminary results for the bare step-scaling function $s_b$ is shown from Wilson-flowed MCRG two-lattice 
matching for  $N_f = 12$ flavors~\cite{Petropoulos:2012mg}. 
$s_b=0$ around $\beta_F=5.5$ signals an infrared fixed point.  The blue dashed line is the  
perturbative two-loop prediction for asymptotically weak coupling.
}
\end{figure}

There are two complementary directions to follow which are illustrated in Figure~\ref{fig:coupling}. 
Methods that rely on the calculation of a renormalized coupling build 
on the properties of the gaussian fixed point to guide cutoff removal in the model study. 
These methods include Schrodinger functional methods, the twisted Polyakov loop, 
Creutz ratios, and static potential  approaches~\cite{Appelquist:2009ty,DeGrand:2010na,Fodor:2012uw,Lin:2012iw}, 
and the promising new method based on the gradient flow of the Yang-Mills field~\cite{Luscher:2010iy,Fodor:2012td,Fodor:2012qh}.  
The advantage of these methods is that they predict a well defined renormalized coupling in the continuum that can be connected to any 
renormalization scheme. 
It is difficult to control lattice artifacts as one takes  the continuum limit by tuning the bare coupling towards 
zero even when investigating the properties of the system at strong renormalized coupling. 
The problem is particularly perplexing when  a ``backward flow" emerges in the running coupling as a potential signal 
for an infrared fixed point without complete control of the continuum limit.
Theoretical and numerical studies continue  work toward a resolution of these issues.

A complementary approach is to use bare quantities to determine a bare step scaling function, like the Monte Carlo renormalization group (MCRG) methods do~\cite{Hasenfratz:2011xn,Catterall:2011zf,Petropoulos:2012mg}. MCRG methods are very effective in identifying phases and fixed points; 
their  drawback  is the lack of continuum coupling. Optimization of the block transformation of 
earlier MCRG calculations encountered  difficulty in the determination of  a well defined step scaling function. 
 Progress to circumvent this problem has been made with the new Wilson-flow--MCRG approach~\cite{Petropoulos:2012mg}. 
 In MCRG methods the renormalization flow has to reach the
 renormalized trajectory and  simulations  have to be repeated on increasingly larger volumes to ensure this.  
 MCRG matching assumes that the flow is driven to and along a renormalized trajectory, which is problematic
on the strong coupling side of the IRFP.  Theoretical and numerical studies of the MCRG approach continue, in order  to 
provide full control on the remaining important problems.

Running coupling investigations are rather unique in the study of conformal or near-conformal systems. 
There are many new approaches available that were designed for lattice BSM models. 
For the next five years plans include better control of systematic errors and an attempt to combine 
complementary methods. 

\subsection{Chiral Symmetry Breaking, Dirac Spectrum, and Anomalous Mass Dimension}
\label{subsec:dirac}

The absence of spontaneous chiral symmetry breaking at small lattice spacing $a$,  
close to the continuum, 
is a clear indication for the position of the BSM lattice model inside the conformal window. 
The corresponding order parameter, the fermion condensate $\langle \bar{\psi}\psi\rangle$, 
can be measured directly in lattice simulations. 
The condensate is quadratically ultraviolet divergent at finite fermion mass $m$ and  vanishes in finite volumes for $m=0$. 
Lattice calculations have to balance finite volume and finite fermion mass effects making
the  $m\rightarrow 0$ chiral extrapolation of the fermion condensate difficult. 
This makes it challenging to establish  chiral  symmetry breaking. An alternative and promising 
method is to extract the chiral condensate from the spectral density through the 
Banks-Casher formula $\Sigma = - \pi \rho(0)$~\cite{Banks:1979yr}. 
When the Dirac eigenvalue spectrum is calculated with vanishing valence fermion masses,
 $\Sigma$  depends only on the sea fermion masses through the eigenvalue density 
 $\rho(\lambda)$; the ultraviolet divergent parts of  $\langle \bar{\psi}\psi\rangle$ are
 absent, making the chiral extrapolations more reliable. In the chiral limit both the direct  
 and indirect measurements must give the same prediction for the chiral condensate, 
 and we expect that in the next generation of lattice studies the two approaches will be 
 used in a complementary manner.

The eigenvalue spectrum provides an excellent tool to study the anomalous mass dimension as well. 
For small mass deformations of the IRFP, the scheme
independent critical exponent, associated with the anomalous mass dimension, 
can be determined directly from the mode density of the eigenvalue spectrum.  In QCD-like
theories with $\rm{\chi SB}$ the energy dependent anomalous mass dimension is calculated with a particular choice
of renormalization scheme.
In an infrared conformal theory the eigenvalue density scales as $\rho(\lambda) \propto \lambda^\alpha$ for small $\lambda$. 
The mode number  $\nu(\lambda) = V  \int \rho(\lambda') d \lambda'$    
thus scales as $\propto V  \lambda^{\alpha+1}$ and since $\nu(\lambda)$ is a 
renormalization group invariant quantity,
 the scaling exponent $\alpha$ is related to  the mass anomalous dimension 
 as $\gamma_m +1  =  4/(1+\alpha) $~\cite{Giusti:2008vb,DelDebbio:2010ze,Patella:2012da}.
The anomalous dimension depends on the energy scale, or when derived from the mode number, on $\lambda$. 
This energy dependence can be extracted by  fitting the mode number  
in finite intervals with a $\lambda$-dependent exponent not only in conformal but also in 
chirally broken systems as well~\cite{Cheng:2013eu}. 

\begin{figure*}[tb]
  \includegraphics[width=0.45\linewidth]{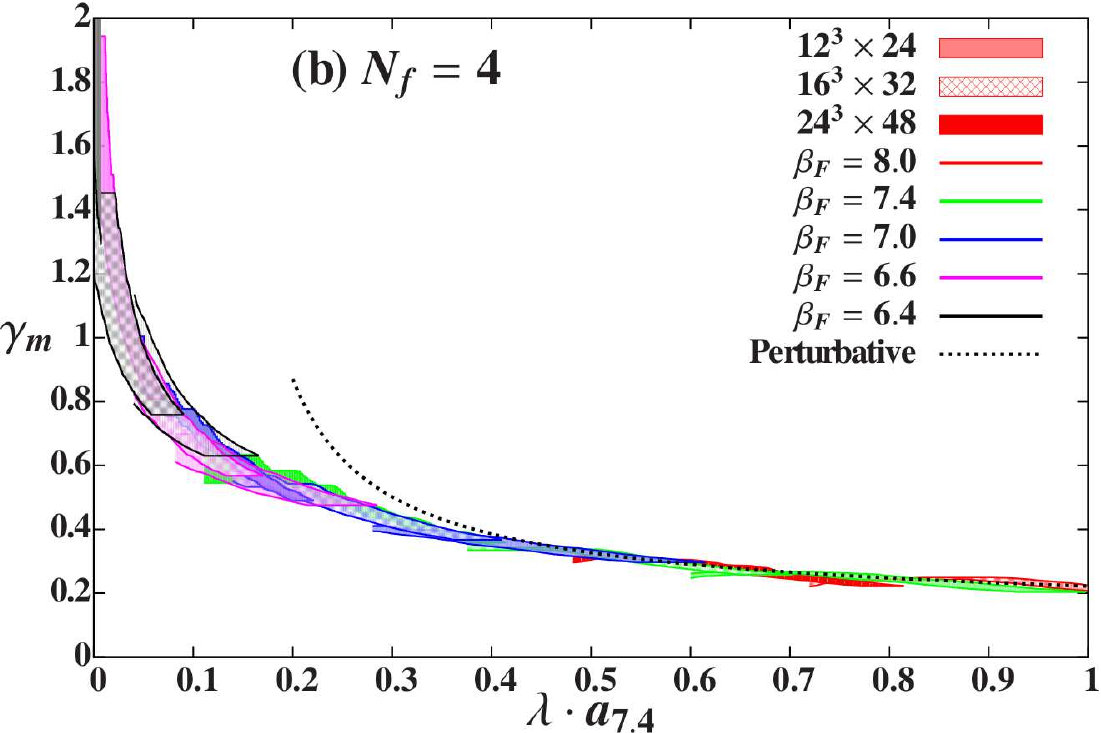}\hfill
  \includegraphics[width=0.45\linewidth]{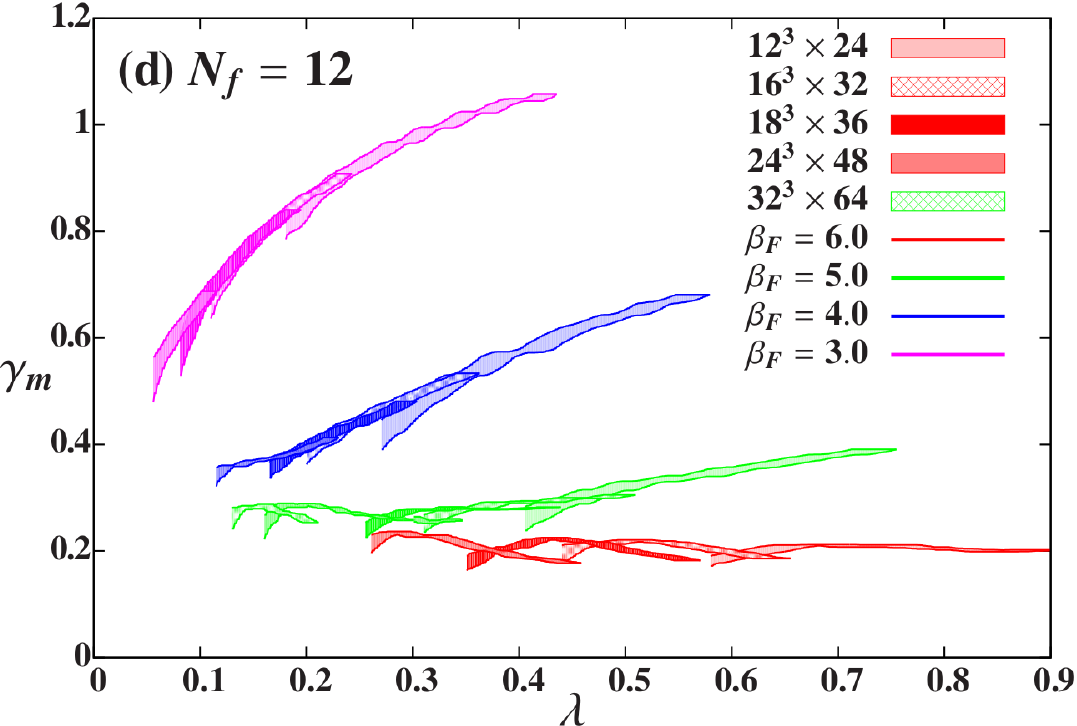}
  \caption{\label{fig:gamma_m} The left panel shows the anomalous dimension as the function of the 
  Dirac operator eigenvalues $\lambda$ for the 4-flavor SU(3) system. The combination of different volumes 
  allows an infinite volume extrapolation and rescaling the eigenvalues to a common lattice spacing predicts a 
  universal curve connecting the perturbative and chirally broken regimes. The right panel is the anomalous dimension at several 
bare couplings for the 12-flavor system. The data are not consistent with ultraviolet 
dynamics governed by the gaussian fixed point 
nor with chiral symmetry breaking in the infrared limit.}
\end{figure*}

The scaling form $\rho(\lambda) \propto \lambda^\alpha$  is valid in the infinite volume chiral limit. 
In conformal systems or even in QCD-like systems in small volumes  simulations are possible in the chiral limit. 
Recent finite volume studies indicate that small volumes can be combined to approximate infinite volume 
results making it possible to map the scale dependent anomalous dimension in QCD-like system from 
the perturbative to the chirally broken regime. This is  illustrated on the left panel of Figure~\ref{fig:gamma_m} 
that shows the anomalous dimension as the function of $\lambda$ for the 4-flavor SU(3) system. 
Over two orders of magnitude in energy scale is covered  by combining different volumes and 
rescaling different gauge couplings to a common lattice spacing. At the large $\lambda$ ultraviolet 
range the data is consistent with the 1-loop perturbative prediction (dashed curve) while in the 
infrared limit chiral symmetry breaking emerges. The right panel of Figure~\ref{fig:gamma_m}
shows the anomalous dimension for the 12 flavor system at several gauge coupling values. 
This data is very different form the $N_f=4$ case. It is not possible to 
rescale the data to form a universal curve.  At stronger couplings the ultraviolet 
scaling is the opposite of what is expected around an asymptotically free fixed point, 
and one can observe a qualitative change around $\beta=5.0$. In this work, the data are consistent 
with an IRFP and an extrapolation to the $\lambda=0$  limit  predicts the conformal anomalous 
mass dimension as $\gamma_m=0.32(3)$~\cite{Cheng:2013eu}.

The scaling of the mode number is a very promising approach to study the infrared dynamics of systems 
near the conformal window. Simulations on larger volumes and the calculation of more eigenmodes, 
possibly with the stochastic estimator method, will allow greater control of finite volume effects and a 
more reliable approximation to the infrared limit. The method described above can be generalized to 
Wilson fermions though it will require additional numerical tuning to the chiral limit.

\subsection{Chiral Perturbation Theory and Condensate Enhancement}
\label{subsec:condensate}

It has been known since the mid-1960's that the low energy behavior of
pseudo-Nambu-Goldstone bosons (PNGBs) was  governed by the decay constant $F$
and the chiral condensate $\sigma(\mu)$, up to corrections of $\mathcal{O}(m)$
and $\mathcal{O}(m \log m)$ in the fermion mass $m(\mu)$, as encoded in the
GMOR relation \cite{GellMann:1968rz}
\begin{equation}
M_\mathrm{PNGB}^2 = \frac{2 m(\mu) \Sigma(\mu)}{F^2} .
\end{equation}
The notation $m(\mu)$ and $\Sigma(\mu)$ indicate that the numerical values for
these quantities depend on the calculational scheme in which they are computed
and all such schemes require the use of an arbitrary scale $\mu$.  Other
quantities $M_\mathrm{PNGB}$ and $F$ do not depend on $\mu$ so the combination
$m(\mu) \Sigma(\mu)$ must be scale-independent.  In the calculation,
we typically define $\mu \equiv a^{-1}$, the inverse lattice spacing, and
$m(a^{-1})$ is the bare lattice mass parameter.

What at first glance may seem to be a rather technical detail about
scale-dependence in calculational schemes can be turned into a powerful
probe of the relative sensitivity of low energy dynamics to high energy vacuum
fluctuations in two different quantum field theories.  We can study the rate of
change of the dimensionless ratio $R(\mu) \equiv \Sigma(\mu) / F^3$ while varying 
$\mu$ as one measure of sensitivity to fluctuations of
$\mathcal{O}(\mu)$.  But, the numerical value of $\Sigma(\mu)$ at any fixed
$\mu$ is somewhat arbitrary because it depends on the specific calculational
scheme, so we compare the ratio computed in two different theories using the same
scale $\mu$ and the same scheme.

One clear hallmark of a walking gauge theory is that the long-distance physics
of the confining regime is dramatically modified by the presence of the
approximately scale-invariant regime at higher energies.  In a QCD-like theory
$R(\mu) \sim \log(\mu)$ with increasing $\mu$ whereas in a walking gauge
theory $R(\mu) \sim \mu^{\gamma_m}$, where the mass anomalous dimension
$\gamma_m \approx 1$. This conjectured phenomenon is called \textit{condensate
enhancement}.

\begin{figure}[ht]
\centering
\includegraphics[width=0.6\textwidth]{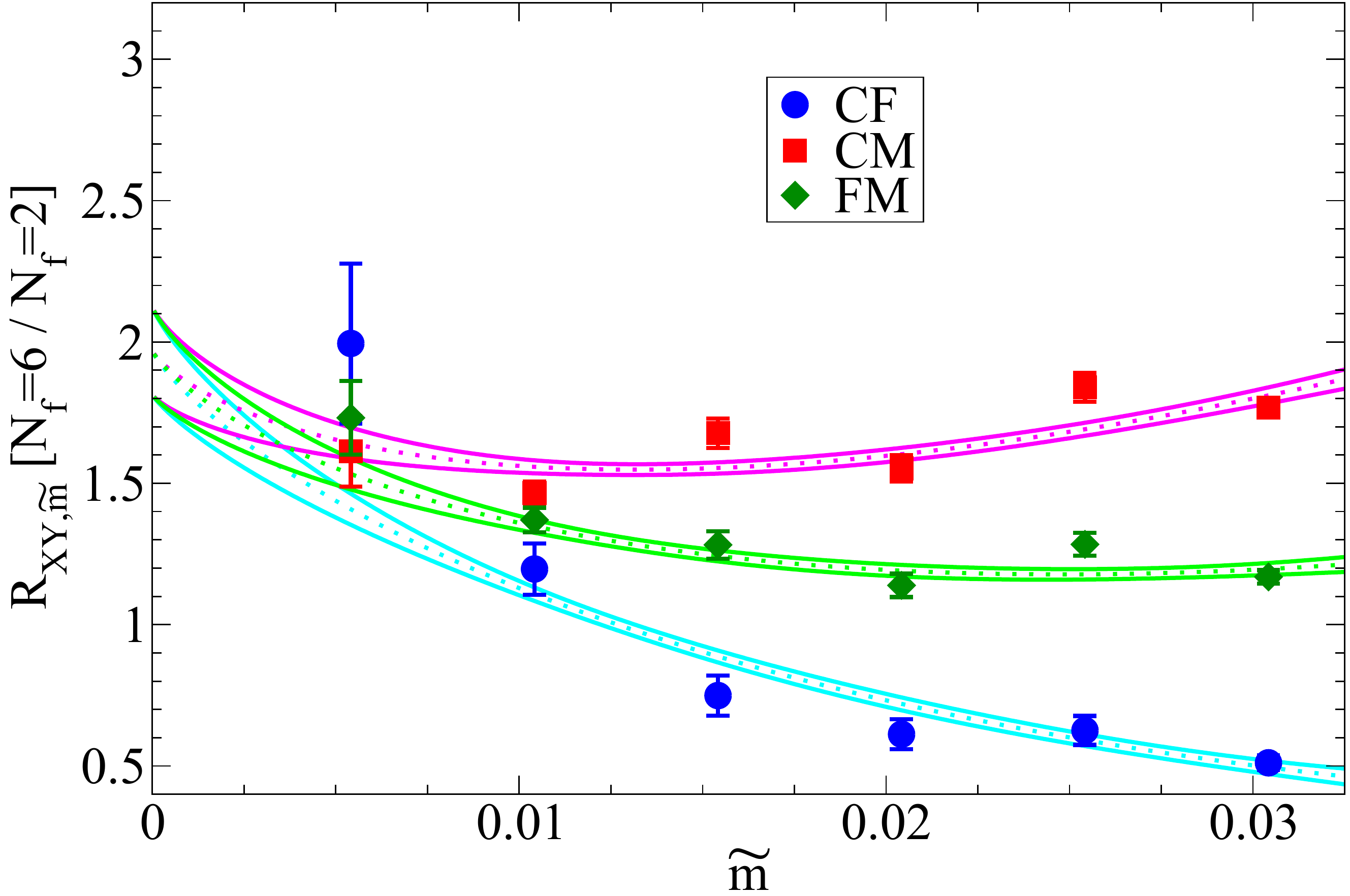}
\caption{\label{fig:R_6_2_combined}The ratio of ratios $\mathcal{R}_{XY}$
described in the text, computed three different ways labeled by $XY$, for $N_f
= 2, 6$ \textit{vs}. geometric mean of the fermion masses $\widetilde{m}$.
The chiral limit suggests a surprisingly large amount of condensate
enhancement, but not enough to claim evidence for walking behavior.}
\end{figure}

Ideally, we can then calculate $R(\mu)$ in two different theories, say SU(3)
gauge theory with different $N_f$ flavors of light fermions, at two different
$\mu = a^{-1}$ inverse lattice spacings.  If the ratio of ratios $\mathcal{R}
\equiv R(N_f) / R(N_f^\prime)$ grows like $(\mu_1 / \mu_2)^{\gamma_m}$ and
$\gamma_m \approx 1$, then this is clear evidence that the theory with $N_f$
fermions exhibits walking behavior between the scales $\mu_1 > \mu > \mu_2$.
In preliminary calculations performed at a single inverse lattice spacing, a
reasonable conjecture can be employed that $\mathcal{R} \sim 1$ at $\mu = 4
\pi F$ in which case $\mathcal{R}$ can be compared to $(\mu / 4 \pi
F)^{\gamma_m}$ \cite{Appelquist:2009ka}.  Figure~\ref{fig:R_6_2_combined} shows
an example of this kind of calculation where the intercept is $\mathcal{R}$
and should be compared to $1 / a 4 \pi F \sim 4$.  Since $\mathcal{R}
\approx 2$ it gives a clear indication the SU(3) $N_f=6$ theory is not a
walking theory. 

Inside the conformal window we have to change our approach. The  finite size scaling of bound state masses 
at a conformal fixed point can be used to predict the mass anomalous dimension, 
but there are systematic effects that have to be investigated carefully. 
An important  issue that has not been considered in depth yet  is the control of the region in 
gauge coupling where scaling according  to the infrared fixed point  (as opposed to the gaussian fixed point) can be expected. 
This could be a serious problem if the gauge coupling changes slowly with the energy as indicated by some of the other methods.
Once the dependence on the irrelevant gauge coupling is understood one still has to find the region in the fermion mass where  
the IRFP is close enough to guarantee scaling.
It is likely that lack of scaling, both due to the gauge coupling or the fermion mass, will show up as scaling violations in any FSS analysis. 
One has to wonder if that is the main reason for controversial results in SU(3) gauge theory with twelve flavors. 

%George end-----------------------------------

\subsection{S-parameter}
\label{subsec:s-parameter}

% ------------------------------------------------------------------
The $S$ parameter was introduced~\cite{Peskin:1991sw} in the early 1990s in analyses of oblique corrections, that is, the effects of new physics on the vacuum polarizations of electroweak gauge bosons.
Along with the Higgs-like particle recently discovered at the LHC, $S$ remains an important experimental constraint on models of electroweak symmetry breaking through new strong dynamics.

Experimentally, $S$ is determined from a global fit to a variety of electroweak measurements, including Z boson partial decay widths and asymmetries, neutrino scattering cross sections, and atomic parity violation.
Since we define $S = 0$ for the standard model with Higgs boson mass $m_h$, the experimental value depends on $m_h$; with $m_h = 126$ GeV, $S = 0.03 \pm 0.10$, completely consistent with the standard model~\cite{Baak:2012kk}.

However, QCD-like new strong dynamics would predict $S \gtrsim 0.3$; this was originally obtained by scaling up experimental QCD data for vector ($V$) and axial-vector ($A$) spectral functions to the electroweak scale~\cite{Peskin:1991sw}, and has been confirmed by lattice QCD calculations~\cite{Shintani:2008qe, Boyle:2009xi, Appelquist:2010xv}.
The discrepancy between experiment and QCD-like technicolor is worsened by the absence of light states in QCD: replacing $m_h = 126$ GeV with a typical technihadronic scale $M_H^{(ref)} \sim 1$ TeV shifts the experimental value to $S \approx -0.15 \pm 0.10$ (the shift is logarithmic in $M_H^{(ref)}$).

To see how non-QCD-like dynamics may change the situation, consider
\begin{equation*}
  S = 4\pi N_D \lim_{Q^2 \to 0}\frac{d}{dQ^2}\Pi_{V - A}(Q^2) - \Delta S_{SM}.
\end{equation*}
There are three important ingredients in this expression:
\begin{enumerate}
  \item $N_D$ is the number of doublets with chiral electroweak couplings; its presence corresponds to the intuition that $S$ measures the ``size'' of the sector that hides electroweak symmetry.
  \item $\Delta S_{SM}$ accounts for the three Nambu--Goldstone bosons (NGBs) eaten by the W and Z bosons, and also sets $S = 0$ for the standard model.
  \item $\Pi_{V - A}(Q^2)$ is the transverse component of the difference between vector and axial-vector vacuum polarization functions, and can also be related to a dispersive integral of spectral functions, $4\pi \Pi_{V - A}'(0) = \frac{1}{3\pi} \int_0^{\infty} \frac{ds}{s}\left[R_V(s) - R_A(s)\right]$.
\end{enumerate}
The near  restoration of chiral symmetry or parity doubling with strong cancellations between $R_V(s)$ and $R_A(s)$ will therefore reduce $S$.
Such dynamics may be expected near to the conformal window.

\begin{figure}[ht]
  \centering
  \includegraphics[width=0.6\linewidth]{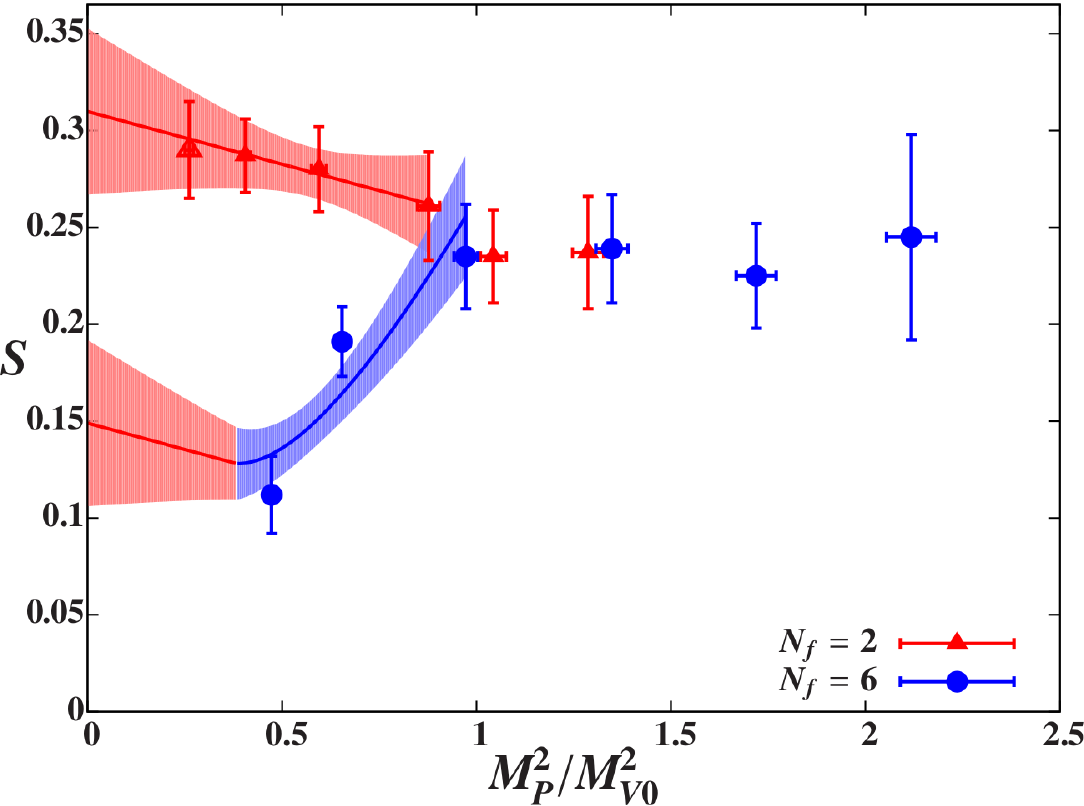}
  \caption{\label{fig:S}The $S$ parameter for SU(3) gauge theories with $N_f = 2$ and 6 fundamental fermions, from Ref.~\cite{Schaich:2011qz}.  $M_P$ is the pseudoscalar mass, while $M_{V0}$ is the vector meson mass in the chiral limit.  The $N_f = 6$ theory has 35 pseudo-NGBs, and here we imagine that 32 of them have mass $\approx 0.6M_{V0}$.}  
\end{figure}

On the lattice, we measure $\Pi_{V - A}(Q^2)$ from $V$ and $A$ two-point current correlation functions.
Existing calculations use overlap or domain wall fermions, for which good chiral and flavor symmetries lead to large cancellations of lattice artifacts in the $V$--$A$ difference, and to equal renormalization factors $Z_V = Z_A$~\cite{Shintani:2008qe, Boyle:2009xi, Schaich:2011qz}.
These formulations also allow direct connection to continuum chiral perturbation theory, where the $S$ parameter corresponds to the low-energy constant $L_{10}$.

So far few lattice studies of $S$ have been completed: Refs.~\cite{Shintani:2008qe, Boyle:2009xi, Appelquist:2010xv} investigated QCD-like technicolor, and relative to those results Ref.~\cite{Appelquist:2010xv} observed a significant reduction in $S$ with $N_f = 6$ light fermions (Figure~\ref{fig:S}).
Ref.~\cite{DeGrand:2010tm} reported some qualitative investigations of $\Pi_{V - A}(Q^2)$ for fermions in the sextet representation of SU(3), and the USQCD BSM Collaboration is currently working to compare mixed-action and staggered measurements.
Some important targets for future work include better understanding and controlling finite-volume effects (which may lead to spurious parity doubling), as well as improving the $Q^2 \to 0$ extrapolation of $\Pi_{V - A}(Q^2)$.
The latter issue has recently attracted a great deal of interest in the context of hadronic contributions to $(g - 2)_{\mu}$, which are sensitive to $\Pi_V(Q^2)$ at very small $Q^2$.

%------------------------------------------------------------------
% WW-scattering

\subsection{$WW$ Scattering on the Lattice}
\label{subsec:ww}

The recent discovery of the Higgs boson, or of some other new physics which would replace it, was widely expected by particle physicists.  A theory including all other known particles and interactions, but no Higgs boson, is well-known to be mathematically inconsistent; the scattering rate of $W$ and $Z$ bosons grows as the energy, eventually occurring with probability greater than 1.  The addition of scattering processes involving an intermediate Higgs boson cuts off the growth in this rate at high energy, solving the problem.  Even so, the scattering of $W$ and $Z$ gauge bosons can be directly sensitive to the presence of deviations from the Standard Model in the electroweak symmetry-breaking sector.  Experimentally, high-energy $W/Z$ scattering is quite clean but occurs with very low rate, so that a detailed study will require many years of data at the upgraded LHC.

\begin{figure}[hbt]
\begin{center}
\includegraphics[width=110mm]{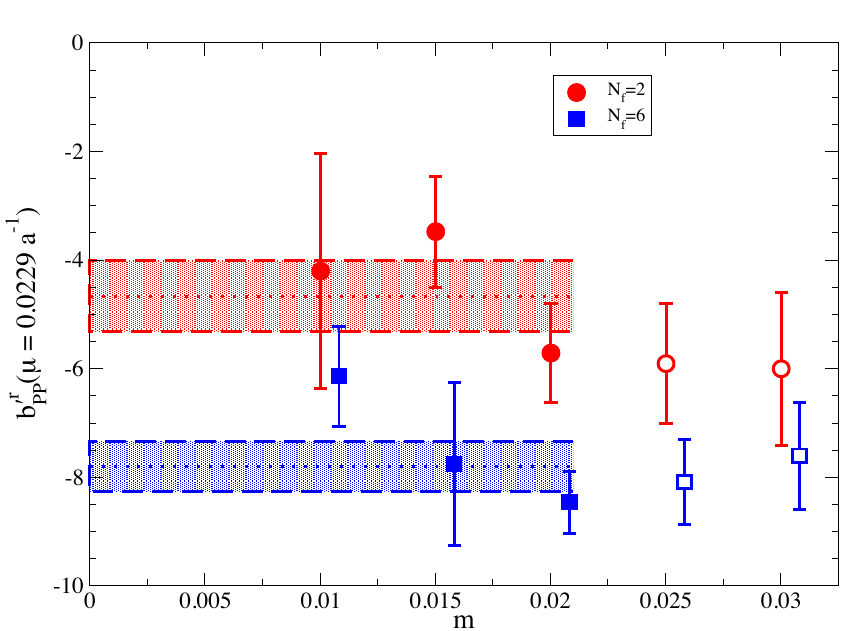}
\caption{\label{fig:wscatt}From \cite{Appelquist:2012sm}, comparison of chiral parameter $b_{PP}^{'r}$ obtained from maximal-isospin $\pi-\pi$ scattering in SU$(3)$ gauge theories with $N_f = 2$ (red) and $N_f = 6$ (blue) light fermion species.  The parameter $b_{PP}^{'r}$ is sensitive to a combination of low-energy constants which would be relevant for $WW$ scattering, hinting that the scattering rate may be enhanced in theories of electroweak symmetry breaking constructed with larger $N_f$.}
\end{center}
\end{figure}

Even with the Higgs boson added, there remain some theoretical questions in the Standard Model, in particular the issue of ``naturalness"; the strong sensitivity of the Higgs boson mass to quantum corrections makes it puzzling that its observed mass should be so close to the electroweak scale.  Composite Higgs theories provide one possible solution, with the quantum corrections effectively cut off at the scale of compositeness.  Within such a theory, there will be many observable composite states aside from the Higgs, but they can be quite heavy and thus difficult to produce directly.  The first hints of new physics may therefore come from indirect observation of effects at low energies.  

Considering $W/Z$ scattering at relatively low energies (compared to the scale at which new physics becomes strongly coupled) allows us to ``integrate out" the heavier states and work in an effective field theory resembling the Standard Model \cite{Appelquist:1980vg, Appelquist:1993ka,Bagger:1993zf}.  The first deviations would appear in two-, three-, and four-point functions of the electroweak gauge bosons, corresponding to oblique corrections \cite{Peskin:1991sw,Appelquist:2010xv}, anomalous triple-boson vertices, and $W/Z$ boson scattering.  In principle, all such corrections (known as ``low-energy constants") are predicted by the dynamics of the underlying strongly-coupled gauge theory under which the composite Higgs is bound together. 
It can be noted that LHC experiments using 7 and 8 TeV data
\cite{CMS:2012zz,ATLAS:2012zz,CMS:2012wv} are already becoming sensitive
to the possibility of anomalous couplings consistent with LEP and Tevatron
limits.
Lattice gauge theory provides an ideal non-perturbative approach for studying the values of these low-energy constants and their relations to one another as the underlying strong dynamics is changed.

A recent lattice study has focused on the low-energy constants relevant for $WW$ scattering for three-color gauge theories with $N_f = 2$ and $6$ light fermion species \cite{Appelquist:2012sm}.  In the strongly-coupled gauge theory, the relevant physical process to determine these low-energy constants is $\pi$-$\pi$ scattering (these ``pions" will become the longitudinal modes of the $W$ bosons in a composite Higgs theory.)  In this initial work, only the extraction of a combination of the relevant low-energy constants was possible; calculation of scattering lengths in other channels and the use of partially-quenched fermion masses may be necessary to separate the individual constants.  Still, a promising trend was shown in the initial study, with the measured combination of low-energy constants showing a significant enhancement from $N_f = 2$ to $N_f = 6$ (see Figure~\ref{fig:wscatt}).  If this trend continues, theories with greater fermion content may give rise to greatly enhanced $WW$ scattering rates, which could be observed at the LHC with a relatively modest amount of data.

% ------------------------------------------------------------------

%------------------------------------------------------------------
% dark matter
\subsection{ Composite Dark Matter}
\label{subsec:dark}

Although only the gravitational effects of dark matter have been observed to date, the existence of other interactions between the dark matter and standard model (SM) particles is well-motivated in many models.  In the standard thermal relic scenario, such interactions keep the dark matter in equilibrium with the thermal bath of SM particles.  Asymmetric production is an alternative mechanism \cite{Kaplan:1991ah,Kaplan:2009ag}, in which the dark matter and ordinary baryons possess particle-antiparticle asymmetries which are connected e.g. by non-perturbative electroweak interactions.  %Asymmetric generation connects the relic densities of dark matter and ordinary matter, explaining why they should differ by only a factor of 5 in our Universe.
In either case, the presence of interactions connecting the dark sector to the SM is a central ingredient.  However, constraints on such interactions from dark matter direct-detection experiments have strengthened by many orders of magnitude in recent years. A fine balance between the presence and absence of such DM-SM interactions when constructing models is therefore needed.

Composite dark matter models provide an attractive mechanism for realizing this balance, and can have unique phenomenological signatures.  In these models, an electroweak-neutral composite dark matter candidate can be formed as a bound state of electroweak-charged fundamental particles, bound together by a new strong gauge force.  The charged constituents give rise to the necessary interactions in the early universe, but the interactions of the neutral state with ordinary matter in the present Universe will be greatly suppressed.  In scenarios of dynamical electroweak symmetry breaking, in which the Higgs boson itself is a composite state, the presence of additional bound states which can play the role of dark matter is natural and the couplings to the electroweak sector are particularly well-motivated~ \cite{Chivukula:1989qb,Kaplan:1991ah,Gudnason:2006yj}.

%The charged fundamental particles can be free in the thermal bath of the early universe for generation of a thermal relic density, and allow the coupling of dark composite states to baryons through e.g. $SU(2)_L$ sphalerons for asymmetric generation.  The presence of a strong force in the dark sector also generally leads to large self-annihilation cross-sections for composite dark states, which is an important feature in the asymmetric generation scenario, suppressing the possible creation of an additional thermal relic component.  In scenarios of dynamical electroweak symmetry breaking, in which the Higgs boson itself is a composite state, the presence of additional bound states which can play the role of dark matter is natural and the couplings to the electroweak sector are particularly well-motivated \cite{techniDM}.

\begin{figure}
\begin{center}
  \includegraphics[width=.6\textwidth]{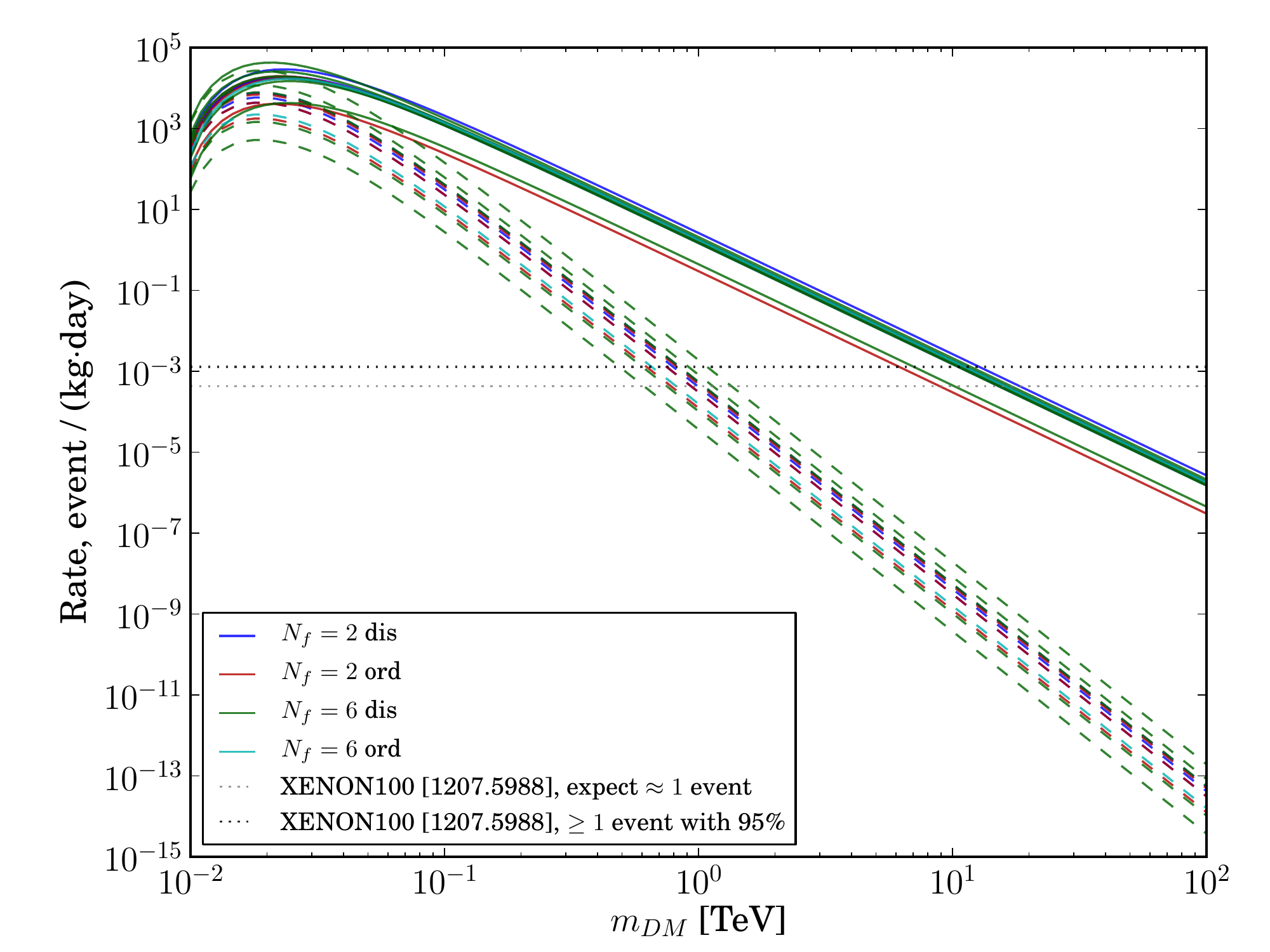}
\caption{\label{fig:X}From \cite{Appelquist:2013ms}, calculated Xenon100 event rates based on simulation results for the charge radius (dashed) and magnetic moment (solid) of baryonic dark matter in three-color gauge theories with $N_f = 2$ and $N_f = 6$ light fermion species.  Bounds arising from the magnetic moment interaction are quite strong, of order 10 TeV based on the latest Xenon100 data \cite{Aprile:2012nq}.}
\end{center}
\end{figure}

The properties of a strongly-coupled dark sector can be roughly estimated from our knowledge of QCD, 
but for theories which differ significantly from QCD, such estimates are unlikely to be reliable.  
Lattice field theory is thus the only systematically controlled approach to calculation in such new strongly-coupled theories.  
In the context of composite dark matter, there are many observables of interest, but one set of observables which are directly 
relevant to experiment are the electromagnetic form factors, which determine the electroweak interaction strength of the 
dark matter with ordinary matter.  Recently, a lattice calculation of the magnetic moment and charge radius has been carried 
out in three-color gauge theories ($N_c = 3$) with $N_f = 2$ and $6$ light fermion species.  The former theory corresponds 
to QCD, while the presence of extra fermions in the $N_f = 6$ theory modifies some of its dynamical properties \cite{Appelquist:2009ka}.  
Results for the charge radius and magnetic moment, converted to direct-detection constraints for the recent Xenon100 experimental 
release, allow the exclusion of composite dark matter based on these models for masses up to 10 TeV \cite{Appelquist:2013ms} - see Figure~\ref{fig:X}.  

This bound is dominated by the magnetic-moment operator, which exists
naturally in any theory with odd $N_c$.  Motivated by this initial
study, a natural next step is the exploration of theories with even
$N_c$, where the bounds will be much weaker.  Work is currently
underway in this direction, requiring code development, theoretical
study, and exploration of the parameter space through simulation.  In
particular, the present focus is on 4-color simulations, where the
baryon is bosonic but not a light pseudo-Goldstone state as in the
$N_c = 2$ theory.  Generation of the necessary lattices to address the
open questions of the $N_c = 4$ theory, including study of the
baryonic charge radius and polarizabilities, will require large-scale
machine resources.  The Blue Gene/Q architecture is ideal for dealing
with the large matrices required for this problem, and simulations are
estimated to require on the order of 24 rack-years on the BG/Q for
completion.

\section{New Methods for BSM Lattice Field Theory}
\label{sec:NewMethods}

Lattice field theory for Beyond the Standard Model physics presents
new challenges to formulating  optimal algorithms and rapidly developing
high performance software.

First, unlike QCD, there is a wide variety of strongly
interacting theories of potential interest. The traditional software
and algorithms must be generalized to accommodate new gauge groups,
fermion representations and large scale separations.  Extensive and on
going research is needed to discover and implement new methods.  In
some cases, just formulating the lattice theory requires considerable
new mathematics and physical insight. Putting supersymmetric field
theories on the lattice discussed above is one dramatic example
requiring new lattice actions based on orbifolding and topological
twists. Even for non-SUSY theories new gauge groups and fermion
representations require substantial code development.  SciDAC-3 is
developing FUEL (Framework for Unified Evolution of Lattices) written
with a LUA front-end to accelerate the adaptation of code for new
theories and to control and autotune them for high performance.

However the most serious new challenge is a
larger range of scales compared to those encountered to date in QCD.
As described above, BSM models must not only have zero mass fermions to
implement the Higgs mechanism, but also are very likely to exhibit near
conformal
behavior in the infrared to satisfy precision electro-weak
experimental constraints.  Nonetheless the problems of scale
separation are not entirely new to lattice field theory so there are
some traditional methods that are available to address them, such as  Wilson
real-space renormalization group, step scaling, finite size analysis,
multigrid solvers etc.
Their role in BSM lattice research will be briefly described. However it is
also important to explore entirely new algorithmic methods. One such idea consists  of
replacing the standard Euclidean hypercubic lattice by one suited to 
Radial Quantization.

%Of course it is likely that
%many of these advances will be shared and leverage new method for QCD
%itself. After all none of these problems are entirely new to lattice
%field theory.  critical phenomena are common place in lattice
%investigation for both quantum field theory and statistical mechanics.
%This research will refine both old methods and search for new
%approaches that should have wider applicability to the rest of the
%High Energy and Nuclear Lattice program.

\subsection{Traditional Multi-scale Tools}
\label{subsec:multiscale}

Several methods to confront scale separation are well known but are
just now being refined for BSM lattice field theory.  The first class
is the use of Wilson real-space renormalization group, step scaling
and finite size effects.  These methods start by acknowledging that
with large scale separations, no single lattice is large enough to
accommodate all the important scales.  So one compares lattices with different scales set
by the bare coupling to interpolate gradually from small to large
scales.  Some of these have been described above but it is clear that
BSM lattice investigation will inevitably refine and develop them
further.

\begin{figure}[h!]
\begin{center}
\begin{tabular}{cc}
\includegraphics[width=0.5\textwidth]{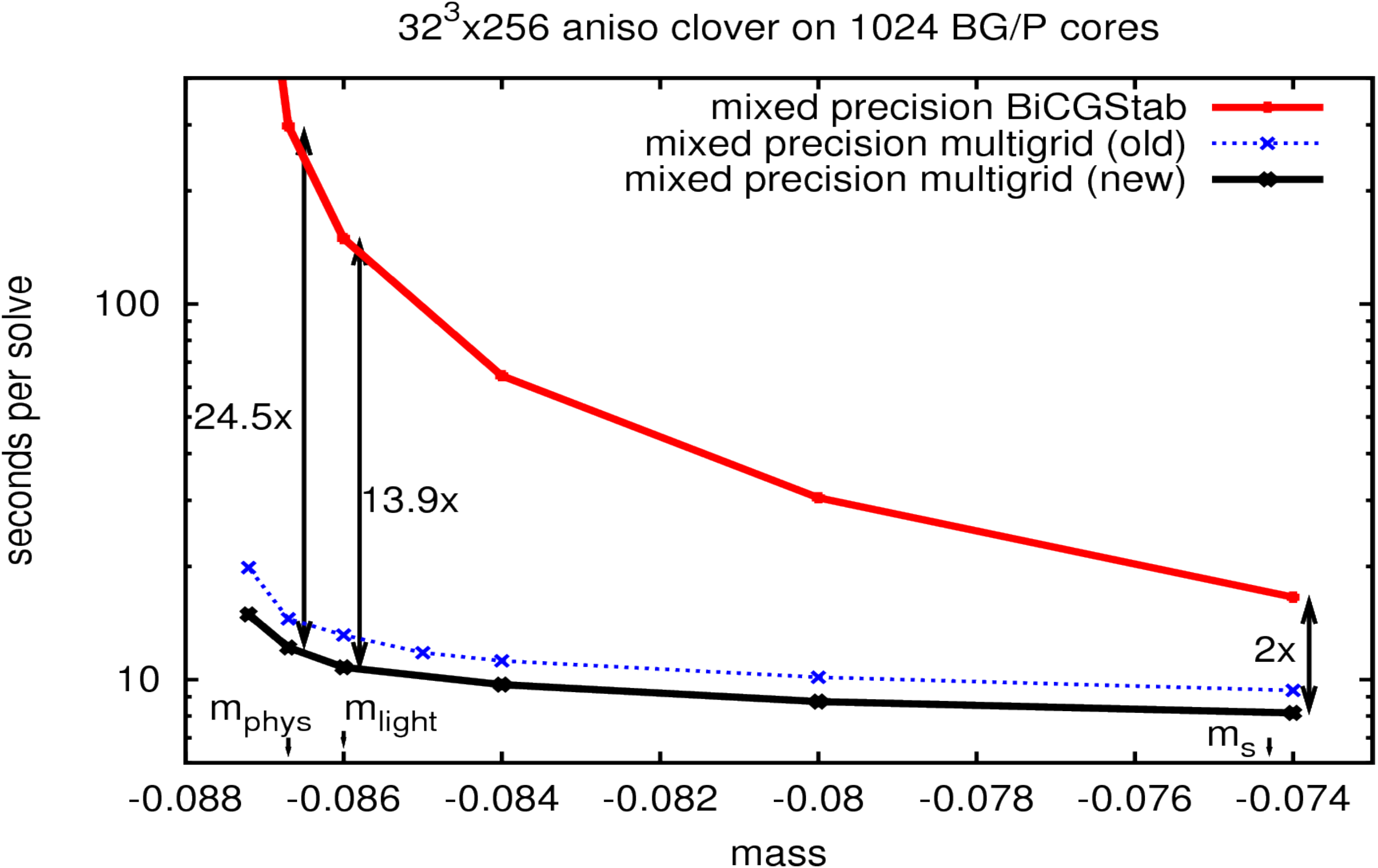} \hskip 1 cm
\includegraphics[width=0.4\textwidth]{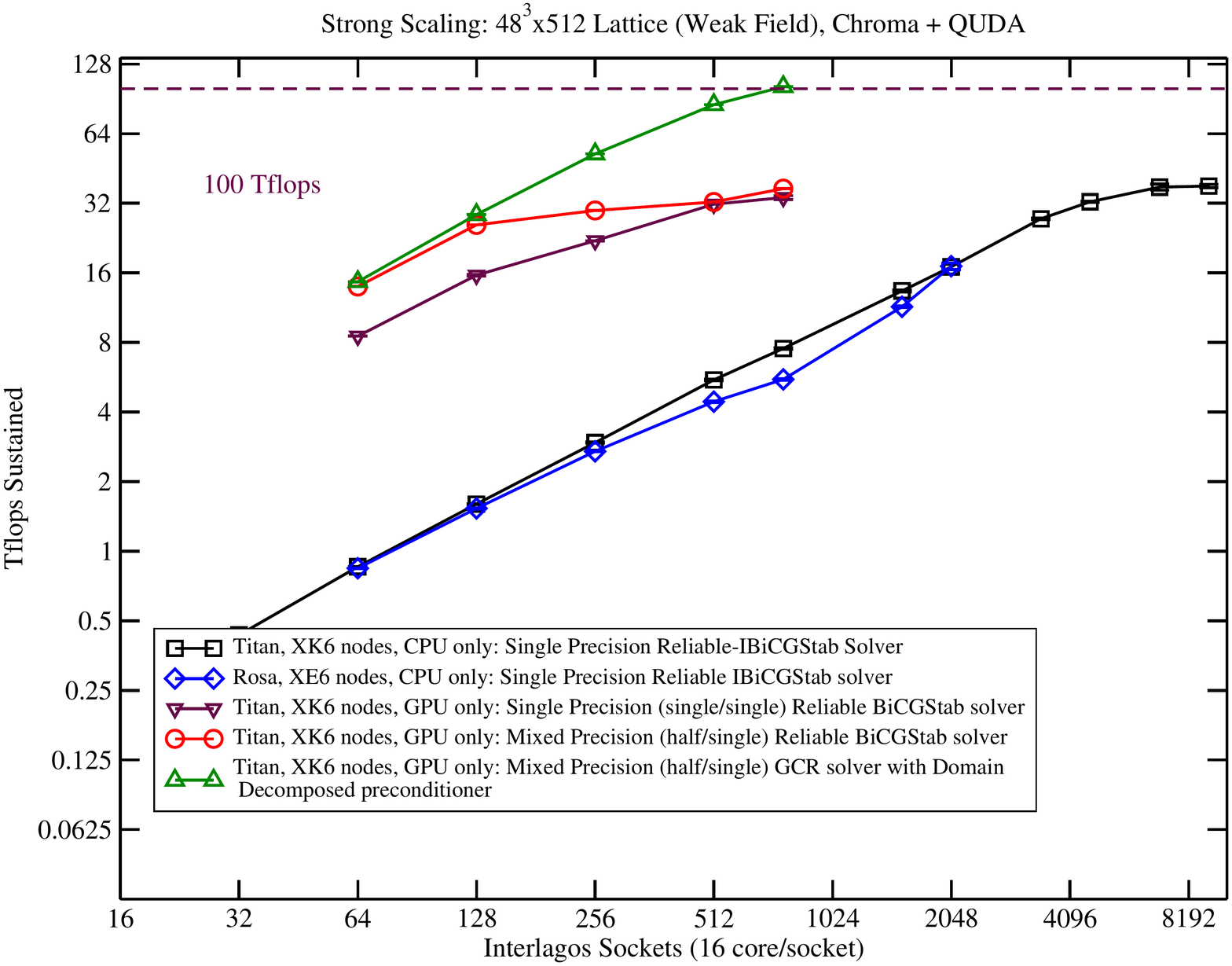}
\end{tabular}
\end{center}
%\vskip -0.25in
\caption{
On the left is a comparison of multigrid performance relative to
the best Krylov solver on the BlueGene/P for the Wilson Dirac solver
as the quark mass is decreased  to its physical value. On the right is
the benefit of using a block Jacobi domain decomposition preconditioner to
reduce inter GPU communication on the  Titan Cray/NVIDIA system.
}
\label{fig:MGdd}
\end{figure}

A second class directly exploits multiple scales on each lattice to
optimize convergence of solvers or to accommodate communication
constraints on hierarchal parallel architectures.  For example new
multi-grid methods have recently been discovered that remove the
divergent computational cost for the Wilson Dirac solvers in the
chiral limit~\cite{Babich:2010qb}. These multigrid algorithms are
beginning to demonstrate more than an order of magnitude improvement
in production codes, when applied to the Wilson Dirac solver near to
physical mass for the light up/down quarks as illustrated in
Figure~\ref{fig:MGdd}. Soon codes for multigrid algorithms will be
available on GPUs decreasing the cost of the Dirac solver by two orders of
magnitude or more.
Future work to be done on these codes includes generalization to domain
wall solvers~\cite{Cohen:2012sh}, integration into standard evolution codes, and tuning to
the chiral limit within specific BSM lattice theories.

A second role of separating short distances from long distance is to
decrease the need for internode communication relative to local
computation. This approach will investigate Schwarz domain decomposition
methods
for solvers. As illustrated in Fig~\ref{fig:MGdd}, even the
simplest example of Jacobi block
decomposition~\cite{Babich:2011np,Babich:2011zz} is enabling improved
performance on the Titan GPU machines and good scaling up to O(1000)
GPUs. This will be  increasingly important on future hierarchical memories
anticipated for future Exascale computing platforms.

\subsection{Radial Quantization}
\label{subsec:radial}

One new approach at the very early stages of research is worth
mentioning as an illustration of a more radical reformulation of
lattice simulation~\cite{Brower:2012vg}. Here
one abandons the traditional Euclidean lattice in favor of one suited
to {\em Radial Quantization}. Indeed this idea has an ancient history
starting with the observation that the early covariant quantization of
the 2-d conformal string action indeed was given as a radial quantized
system with the $L_0$ Virasoro operator replacing the Hamiltonian.  In
1979 Fubini, Hanson and Jackiw~\cite{Fubini:1972mf} suggested radial
quantization of field theory in higher dimensions and later in 1985
Cardy suggested lattice implementations for a ``strip'' in general
dimensions~\cite{Cardy:1985xx}.

For an exactly conformal field theory, the  idea is straight forward.  
The flat metric  for
any Euclidean field theory on $\mathbb R^D$ can obviously be expressed in
radial co-ordinates,
\begin{equation}
d^2s = dx^\mu dx ^\mu =  r^2_0 e^{ 2t} ( dt^2 + d\Omega^2_{D-1} ) \; ,
\end{equation}
where $t = \log(r/r_0)$, introducing an arbitrary reference scale
$r_0$, and where $d\Omega^2_{D-1}$ is the metric on the $\mathbb
S^{D-1}$ sphere of unit radius. However in the case of an exactly
conformal field theory, a local Weyl transformation will also remove
the conformal factor, $\exp[ 2t]$, from the Lagrangian of the quantum
theory. The resultant theory is mapped from the Euclidean space
$\mathbb R^D$ to a D dimensional cylinder, $\mathbb R\times \mathbb
S^{D-1}$. Now one introduces a ``uniform'' radial lattice in $\tau =
\log (r)$ and lattice in angles on concentric spheres $\mathbb{S}^d$
with fixed radius $R$ with the potential advantage that
a finite lattice in $log(r)$ grows  exponential long relative to
the conventional Euclidean lattice.  This lattice is a discrete
approximation to
dilatation $\tau \rightarrow \tau + a_t$, inversions $\tau \rightarrow
- \tau$ and discrete rotations in $O(d\!- \!1)$. The full conformal
group, including Poincare invariance, should be realized in the
continuum limit. A preliminary test of this idea  has been made for the
Wilson-Fisher fixed point in the 3D Ising model~\cite{Brower:2012vg} on a lattice
discretization
of $\mathbb R \times S^2$.

The study of lattice field theories for BSM models does call for new
theoretical and algorithmic tools many of which are almost certain to
significantly strengthen methods even for lattice QCD.
For lattice field theory, like many areas of computational physics,  the
advances
in computer hardware at the rate of Moore's law has generally been
matched by equal contributions due to fundamental theoretical and
algorithmic advances.  
%While future progress is never guaranteed neither is it naive to anticipate it based on past experience. 
This should extend the range of practicable BSM projects  beyond those
suggested in the next section.

\section{RESOURCES FOR STUDIES AT THE ENERGY FRONTIER }
\label{sec:resources}

In this section we discuss the computational resources needed to reach 
the scientific goals set out above. At present, members of USQCD 
are making use of dedicated hardware funded by the DOE through the 
LQCD-ext Computing Project, as well as a Cray XE/XK computer, and
IBM Blue Gene/Q and Blue Gene/P computers, made available by the DOE's 
INCITE Program. During 2013, USQCD, as a whole, expects to sustain 
approximately 300 Tflop/s on these machines.  USQCD has a PRAC grant 
for the development of code for the NSF's petascale computing facility, 
Blue Waters, and expects to obtain a significant allocation on this computer
during 2013. Subgroups within USQCD also make use of computing facilities
at the DOE's National Energy Research Scientific Computing Center (NERSC), the 
Lawrence Livermore National Laboratory (LLNL), and centers supported
by the NSF's XSEDE Program. In addition,
RBC Collaboration, has access to dedicated Blue Gene/Q computers at
Brookhaven National Laboratory and the University of Edinburgh. 
For some time, the resources we have obtained have grown with a doubling 
time of approximately 1.5~years, consistent with Moore's law, and this 
growth rate will need to continue if we are to meet our scientific objectives. 

The software developed by USQCD under our SciDAC grant enables us to use a wide
variety of architectures with very high efficiency, and it is critical
that our software efforts continue at their current pace. Over time,
the development of new algorithms has had at least as important an
impact on our field as advances in hardware, and we expect this trend
to continue, although the rate of algorithmic advances is not as smooth 
or easy to predict as that of hardware. Consequently while  algorithmic
improvements are expected  to play an important role in advancing the
 USQCD BSM research program, the sustained Teraflops years in the tables below  are
based solely  on the performance of our current code base. 

%Strong coupling BSM attice gauge calculations  need resources 
%to generate gauge field configurations in Monte Carlo simulations with probabilities proportional 
%to their weight in the Feynman path integrals of the BSM model. Sophisticated and demanding
%physics measurements using the configurations typically require more resources than configuration generation. 
%In BSM models all the relevant results have to be extrapolated to the massless fermion limit which is an added difficulty
%in comparison with lattice QCD calculations. At each gauge coupling, this requires several simulations with a range of fermion masses 
%to enable accurate extrapolation. Each choice of the gauge coupling and fermion mass requires 
%a range of lattice volumes for control on finite size scaling behavior.  Simulations at several lattice spacings are needed to
%perform extrapolations to the continuum (zero lattice spacing) limit. 

In our work at the energy frontier, members of USQCD are currently
using two choices of fermion actions, critically important to
achieve physics goals: highly improved staggered fermions and domain
wall fermions. However for  the pseudo-Nambu-Goldstone boson project, we are
developing two color gauge code using Wilson fermions both because
they offer a first level project less computationally demanding than
domain wall fermions, but also because we are rapidly implementing the Wilson multigrid
algorithm into evolution code for both the BlueGene and GPU clusters, which has the potential to reduce the
cost substantially. Similar multigrid  procedures for domain wall code is under active
study~\cite{Cohen:2012sh}.  For each subfield, we pick a project that is in
our current plans for the next two years, although with the rapid
experimental and theoretical developments in BSM physics 
priorities should clearly be reevaluated  beyond this time frame.

\subsection{Scalar Spectrum and ${\bf\rm \chi SB}$ Project}
\label{subsec:scalarResource} 

Spectroscopy and the determination of the chiral condensate are essential  to
evaluate the experimental viability of many BSM phenomenological models. In view of the
recent discovery at the LHC, the scalar spectrum with Higgs quantum numbers is particularly 
pressing and becomes a significant part of the measurement time.
The main goal is to determine whether a composite dilaton-like 
particle or light Higgs can emerge in the scalar spectrum of the model~\cite{Fodor:2012ty}. 
\begin{table}[h!]
\begin{center}
%\quad
%\parbox{1.0\linewidth}{
%\centering
\begin{tabular}{|c|c|c|c|c|}
\hline\hline
lattice spacing $a$ &fermion mass & lattice volume & config generation & spectrum and ${\rm \chi SB}$\\
(in fermi) &(in $a$ units) & $V\times T$& (TF-Years) &  (TF-Years) \\
\hline
\hline $2.25\times 10^{-5}$  & 0.003& $48^3\times  96$ &  5 &  5\\ 
\hline $2.25\times 10^{-5}$  & 0.004& $48^3\times  96$ &  4.8 &  5\\ 
\hline $2.25\times 10^{-5}$  & 0.005& $48^3\times  96$ &  4.4 &  5\\ 
\hline $1.75\times 10^{-5}$  & 0.0023& $64^3\times 128$ &  9 &  9\\ 
\hline $1.75\times 10^{-5}$  & 0.0030& $64^3\times 128$ &  8.6 &  9\\ 
\hline $1.75\times 10^{-5}$  & 0.0035& $64^3\times 128$ &  8.2 &  9\\ 
\hline
\hline
\end{tabular} 
\caption{\label{tab:Dilaton}  Resources for SU(3) two flavor sextet project. 
The first column is the approximate lattice spacing $a$ in fermi, as set from the measured VEV
in the chiral limit at fixed gauge coupling using the physical value  ${\rm v=246~GeV}$ for ${\rm a\cdot v=0.029}$
on the coarser lattice. 
The second column is the fermion mass of the run and the third is the lattice volume with twice the lattice points in the temporal direction. 
The fourth  column gives the resources needed to generate 1,000 configurations from 10,000 MD time units. 
The fifth column shows the required resources to determine the full spectrum of the model based on the generally
well-tested balance between configuration generation and measurements.}
\end{center}
\end{table}
While the generic mechanism for a light dilaton may well be model independent, the
detailed parameters are not.  We chose
a representative model with a fermion flavor doublet in the two-index symmetric (sextet) representation
of the SU(3) color gauge group which was introduced in Section 3~\cite{Fodor:2012ty}. The resource estimates are
based on theses simulations which provided the first results consistent with near-conformal behavior, and benchmarks for
future studies  of the sextet model. The scaling of resources
with the fermion mass and the lattice volume does not follow closely the pattern of lattice QCD.

The resources are shown in Table~\ref{tab:Dilaton}  in teraflop/s-years (TF-Years) to generate the required gauge configurations using the
staggered fermion action with stout smeared gauge links and with a minimal number of
fermion masses necessary to extrapolate to the chiral limit at two different lattice spacings with appropriate scaling in the lattice volume.  
The measurements include the scalar spectrum with ${\rm 0^{++}}$ quantum numbers, the spectroscopy of composite states in the
TeV region, and the extrapolation of the chiral condensate to vanishing fermion mass 
(one TF-Year is defined to be the number of floating point operations produced in a year by a computer sustaining one teraflop/s). 
The required resource is 82 TF-Years for the project. 

Other model studies with the SU(3) color gauge group would require similar resources near conformality. 
A great deal of effort will be needed to narrow down the model choices to the most promising scenario to capture 
BSM physics close to the conformal window. The required resources presented in Table I are large compared with
lattice QCD. The physical point of BSM models is at vanishing Goldstone pion mass making the chiral extrapolation
more difficult and more costly. The small finite Goldstone masses required for extrapolation
to the physical point is a significant demand on resources.
Not unrelated, increased finite volume effects and strong chiral condensate enhancement near the conformal window 
are new and costly challenges compared with lattice QCD.

%Since domain wall fermions with excellent chiral properties
%require 10-20 times the resources needed in similar physics enviroinment, they are realistic in models built%
%with the SU(2) color gauge group. 
%This has very interesting physics application in search for the PNGB paradigm of the light Higgs particle.

%\newpage
\subsection{PNGB SU(2) Color Higgs Project}
\label{subsec:pngbResource}

The  investigation of strongly coupled theories with a composite Higgs as a pseudo-Goldstone boson has
received considerable attention in the last few years. Lattice field theory is
needed to give a definitive non-perturbative realization  of this mechanism.  We propose to undertake a careful study of the dynamics of the minimal
composite PNGB Higgs model: SU(2) gauge theory with $N_f = 2$ fermions in the
fundamental representation, which lies at the core of many phenomenological extensions.  In much the same manner as described above for the technicolor example, the low energy effective chiral parameters, $F$ and
$\Sigma(\mu)$ can be found, the $S$ parameter determined and the masses for
higher spin resonances (e.g.  $\rho$, $a_1$ mesons, \textit{etc.}) computed.

\begin{table}[h!]
\begin{center}
\begin{tabular}{|c|c|c|c|}
\hline
 min.\ $M_H$ &   lattice volume  & MD trajectory   & config generation  \\
 (GeV) &   $V\times T$     & (time units)        & (TF-Years)  \\
 \hline\hline
 \hline  650       &   $32^3 \times 64$       & 10000   & 1.2            \\
 \hline 520         & $40^3 \times 80$       &  10000   & 8.6            \\
 \hline 433         &  $48^3 \times 96$       &10000   & 44              \\
 \hline 371         & $56^3 \times 112$      &  10000   & 180          \\
 \hline\hline
\end{tabular}
\caption{\label{tab:PNGB}  Resources to generate gauge
  configuration ensembles in SU(2) gauge theory with $N_f = 2$ fermions in the
  fundamental representation. The inverse lattice spacing is held fixed at
  $a^{-1} =$ 5~TeV.   The first 
  column gives the minimum Higgs mass that can fit in the volume assuming $L M_H
  \ge 4$ and the second column gives the corresponding lattice volume.  The  fourth column gives the resources in teraflop/s-years
  (TF-Years) needed to generate 10,000 molecular dynamics time units (1,000
  equilibrated gauge configurations) for each ensemble for the Wilson fermions.  (One TF-Year is defined to be the number
  of floating point operations produced in a year by a computer sustaining one
  teraflop/s).}
\end{center}
\end{table}

%In short the same general strategy and methods developed over the last several
%years for ``technicolor" are applicable.  
Over the past few years, the USQCD SciDAC software libraries (QLA, QDP/C,
QDP++, FUEL, Chroma) has been extended to support SU(2) gauge
theories for both Wilson and domain wall fermions. Several exploratory calculations have started for SU(2) wiht $N_f = 2$ and $6$ fundamental matter fields.  This phase has focused on mapping out the bare parameter space using the finite temperature
phase transition and the Schr{\"o}dinger Functional running coupling.

The next step is to move gradually to  larger volumes at  zero
temperature with Wilson fermions to study the low energy spectrum
nearer to the chiral limit. The resources are give in
Table~\ref{tab:PNGB}. The estimates are based on the benchmarks of the
USQCD software system for SU(2) gauge theories and scaling experience
from SU(3) lattice QCD with 2 flavors.  The first  priority of this project 
 is to use Wilson fermions for investigating
the most important features of the PNGB mechanism. The initial
inverse lattice spacing will be $( a F )^{-1} \approx 20$ or $a^{-1}
\approx$ 5~TeV since $F \approx$ 250~GeV is the Higgs vev.  This will
enable study of Higgs masses in the range $M_H \sim$ 650--1000~GeV,
well above the physical Higgs mass of 126~GeV.  As in the case of lattice QCD for the past two decades, we will have to
perform calculations where the Higgs mass is heavier than the physical Higgs
mass and extrapolate our results using chiral perturbation theory. Also it should
be noted that  the experimental  mass of 
the Higgs maybe  reduced substantially by perturbative contributions when the this
strong dynamics composite Higgs is coupled  to the electro-weak sector of the standard model.

In addition to the proposed Wilson simulations, exploratory study will
be continued with domain wall fermions. While both the Wilson and
domain wall formulations have an exact Sp($2 N_f$) global symmetry at
finite lattice spacing, only domain wall fermions have
the larger SU($2 N_f$) global symmetry in the chiral limit at finite
lattice spacing. The enhanced chiral symmetries of domain wall
fermions maybe important to rigorously establish the
precise nature of the vacuum alignment mechanism of the model and in
chiral extrapolations to the physical point, which for BSM models requires an exactly vanishing techni-pion
mass. However the prices of domain wall fermions, as
for lattice QCD, is at least an order of magnitude increase in
computational cost. Consequently production using domain
wall fermions for large lattice volumes is  out of reach in the next five years unless ongoing
algorithmic optimizations for the M\"{o}bius DW action~\cite{Brower:2004xi,
  Brower:2012vg} and multigrid solvers~\cite{Babich:2010qb,
  Cohen:2012sh}
 are able  to reducing their cost substantial.

%The cost to lower the Higgs mass while increasing $L$ such that $M_H L \ge 4$
%can be determined from the cost model \cite{Christ:2002pb}
%
%\begin{equation}
%\mathrm{cost} \propto ( L F )^{4+y} \left( \frac{F}{M_H} \right)^{4+z}
%\left( \frac{1}{a F} \right)^{6+y+z} .
%\end{equation}
%
%where we will assume the dynamical exponent $z \approx 1$ and $y = 1$ in an
%exact hybrid Monte Carlo (HMC) algorithm \cite{Creutz:1988wv}.  So, doubling
%the box size $L \to 2 L$ to halve the Higgs mass requires a thousand-fold
%increase in computational capability.  Table~\ref{tab:PNGB_Higgs_resources}
%shows the resources required to perform a chiral extrapolation at fixed
%lattice spacing.

\subsection{Supersymmetric Yang-Mills theories}
\label{subsec:susyResource}

Supersymmetry is interesting both for its potential for phenomenological
extensions of the standard model with light scalars, and for fundamental properties of
field theory and their gravity  duals in the AdS/CFT correspondance. We 
discuss resource needs for both of these programs in the context of 1.)  ${\cal N}=1$ Yang-Mills  and
super QCD and 2.) ${\cal N}=4$ Yang-Mills.
Resource estimates are based on  investigations of ${\cal N} = 1$ supersymmetric 
Yang Mills theory (gauge bosons and gauginos) with phyics results of the 
gluino condensate and string tension in these theories.

\subsubsection{${\cal N}=1$ Yang-Mills and super QCD}
%\noindent {\bf ${\cal N}=1$ Yang-Mills and super QCD}: 
Current simulations of ${\cal N}=1$ super Yang-Mills using domain wall
fermions the trajectory takes 30 mins on a $16^3\times 32\times
L_s=24$ lattice on a single rack of the BG/L. Since our initial work
has revealed that the residual mass is larger than typically
encountered in QCD and falls off slowly as a function of $L_s$, we
will also implement the M\"{o}bius DWF
improvement~\cite{Brower:2004xi, Brower:2012vg} which falls of like
$1/L^2_s$.

\begin{table}[h!]
\begin{center}
\begin{tabular}{||c|c|c|c|c||}
\hline
lattice volume & wall separation & bare coupling & trajectory. & config generation \\
$V\times T$ & $L_s$ & $\beta = 4/g^2_0$ & (time units) & (TF-Years) \\
\hline \hline
\hline $16^3\times 32$ & $16$ & $2.3$ & $10000$ & $2.3$\\
\hline $16^3\times 32$ & $24$ & $2.3$ & $10000$ & $3.5$\\
%\hline $16^3\times 32$ & $48$ & $2.3$ & $10000$ & $7.0$\\
\hline $16^3\times 32$ & $16$ & $2.4$ & $10000$ & $3.6$\\
\hline $16^3\times 32$ & $24$ & $2.4$ & $10000$ & $5.5$\\
\hline $16^3\times 32$ & $48$ & $2.4$ & $10000$ & $11.0$\\
\hline $24^3\times 48$ & $24$ & $2.4$ & $10000$ & $41.8$ \\
\hline $24^3\times 48$ & $48$ & $2.4$ & $10000$ & $83.6$ \\
%\hline $24^3\times 48$ & $64$ & $2.4$ & $10000$ & $111.5$\\
\hline $24^3\times 48$ & $48$ & $2.5$ & $10000$ & $200$ \\
%\hline $24^3\times 48$ & $64$ & $2.5$ & $10000$ & $300$\\
\hline
\end{tabular}
\caption{\label{tab:SUSY} Resouces needed for DWF simulation of $SU(2)$ ${\cal N}=1$ Yang-Mills theory. As in
previous studies, we set the bare fermion mass $m_f=0$ for these estimates. Residual masses fall in the range 0.02-0.1 for these values of the parameters using Shamir
(non-M\"{o}bius) domain wall fermions. Using two lattice volumes, three lattice spacings and several values of $L_s$ should allow for careful extrapolation to the chiral continuum limit while maintaining control over finite volume effects}
\end{center}
\end{table}
The spectrum of ${\cal N}=1$ super Yang-Mills requires the computation
of disconnected correlators. Using stochastic estimators and dilution
techniques these can be computed in a time comparable to the time
needed to generate configurations. Table~\ref{tab:SUSY} indicates the
scale of resources that will be needed to compute the condensate and
low lying spectrum of ${\cal N}=1$ super Yang-Mills. These
calculations can be completed in principle within a two year timeframe
on leadership class machines like the BG/Q.

%\paragraph{Super QCD}

While the spectrum and structure of ${\cal N}=1$ Yang-Mills is clearly
very interesting this work also forms the basis for the much more
difficult study of super QCD.  Super QCD will require significantly
more resources; a series of DWF simulations must be performed to tune
the scalar (squark) sector of the theory for a fixed value of the
gauge coupling and gaugino mass. In addition it is necessary to
simulate for larger numbers of colors and flavors to access the
metastable susy breaking vacua. The simplest system which is expected
to exhibit metastable vacua corresponds to four colors and five
flavors.  If we model the dependence on $N_f$ and $N_c$ via the naive
formula $T\sim N_f\times (N_c^2-1)$ we find that a single DWF
simulation on a $16^3\times 32\times 24$ lattice will require
approximately $137$ Teraflop-years. Clearly super QCD will require
Petascale resources and a longer time frame to complete.

\subsubsection{${\cal N}=4$ Yang-Mills}
%\noindent {\bf ${\cal N}=4$ Yang-Mills}:
\vskip -0.1in
Another high priority is to push into large scale simulation of ${\cal
  N}=4$ Yang-Mills. Preliminary work has focused on surveying the
phase diagram of the theory with lattices of size $8^3\times 16$ over
a wide range of gauge coupling and scalar mass. The current parallel
code (based on the FermiQCD communication libraries) is capable of
producing 1 trajectory in 1.25 hrs CPU time on a single node of the
Fermilab Ds cluster (32 cores).  More realistic simulations on a
$16^3\times 32$ lattices correspond to a 5.8 Teraflop-yr
calculation. Long term goals would require simulating lattices as
large as $32^3\times 64$ and would clearly require Petascale
resources.

\section*{Acknowledgements}
We gratefully acknowledge discussions, input, and comments from Ayana Arce,
Tulika Bose, Ben Grinstein, Ken Intriligator, Martin Schmaltz, Jesse Thaler,  Ruth Van de Water, 
and Uwe-Jens Wiese.

\newpage

\bibliography{BSMwhitepaper, revtex-custm}

\begin{thebibliography}{132}%
\makeatletter
\providecommand \@ifxundefined [1]{%
 \@ifx{#1\undefined}
}%
\providecommand \@ifnum [1]{%
 \ifnum #1\expandafter \@firstoftwo
 \else \expandafter \@secondoftwo
 \fi
}%
\providecommand \@ifx [1]{%
 \ifx #1\expandafter \@firstoftwo
 \else \expandafter \@secondoftwo
 \fi
}%
\providecommand \natexlab [1]{#1}%
\providecommand \enquote  [1]{``#1''}%
\providecommand \bibnamefont  [1]{#1}%
\providecommand \bibfnamefont [1]{#1}%
\providecommand \citenamefont [1]{#1}%
\providecommand \href@noop [0]{\@secondoftwo}%
\providecommand \href [0]{\begingroup \@sanitize@url \@href}%
\providecommand \@href[1]{\@@startlink{#1}\@@href}%
\providecommand \@@href[1]{\endgroup#1\@@endlink}%
\providecommand \@sanitize@url [0]{\catcode `\\12\catcode `\$12\catcode
  `\&12\catcode `\#12\catcode `\^12\catcode `\_12\catcode `\%12\relax}%
\providecommand \@@startlink[1]{}%
\providecommand \@@endlink[0]{}%
\providecommand \url  [0]{\begingroup\@sanitize@url \@url }%
\providecommand \@url [1]{\endgroup\@href {#1}{\urlprefix }}%
\providecommand \urlprefix  [0]{URL }%
\providecommand \Eprint [0]{\href }%
\providecommand \doibase [0]{http://dx.doi.org/}%
\providecommand \selectlanguage [0]{\@gobble}%
\providecommand \bibinfo  [0]{\@secondoftwo}%
\providecommand \bibfield  [0]{\@secondoftwo}%
\providecommand \translation [1]{[#1]}%
\providecommand \BibitemOpen [0]{}%
\providecommand \bibitemStop [0]{}%
\providecommand \bibitemNoStop [0]{.\EOS\space}%
\providecommand \EOS [0]{\spacefactor3000\relax}%
\providecommand \BibitemShut  [1]{\csname bibitem#1\endcsname}%
\let\auto@bib@innerbib\@empty
%</preamble>
\bibitem [{\citenamefont {Chatrchyan}\ \emph
  {et~al.}(2012{\natexlab{a}})\citenamefont {Chatrchyan} \emph
  {et~al.}}]{CMS:2012gu}%
  \BibitemOpen
  \bibfield  {author} {\bibinfo {author} {\bibfnamefont {Serguei}\ \bibnamefont
  {Chatrchyan}} \emph {et~al.} (\bibinfo {collaboration} {CMS Collaboration}),\
  }\bibfield  {title} {\enquote {\bibinfo {title} {{Observation of a new boson
  at a mass of 125 GeV with the CMS experiment at the LHC}},}\ }\href {\doibase
  10.1016/j.physletb.2012.08.021} {\bibfield  {journal} {\bibinfo  {journal}
  {Phys.Lett.}\ }\textbf {\bibinfo {volume} {B716}},\ \bibinfo {pages} {30--61}
  (\bibinfo {year} {2012}{\natexlab{a}})},\ \Eprint
  {http://arxiv.org/abs/1207.7235} {arXiv:1207.7235 [hep-ex]} \BibitemShut
  {NoStop}%
%%CITATION = ARXIV:1207.7235;%%
\bibitem [{\citenamefont {Aad}\ \emph {et~al.}(2012{\natexlab{a}})\citenamefont
  {Aad} \emph {et~al.}}]{ATLAS:2012gk}%
  \BibitemOpen
  \bibfield  {author} {\bibinfo {author} {\bibfnamefont {Georges}\ \bibnamefont
  {Aad}} \emph {et~al.} (\bibinfo {collaboration} {ATLAS Collaboration}),\
  }\bibfield  {title} {\enquote {\bibinfo {title} {{Observation of a new
  particle in the search for the Standard Model Higgs boson with the ATLAS
  detector at the LHC}},}\ }\href {\doibase 10.1016/j.physletb.2012.08.020}
  {\bibfield  {journal} {\bibinfo  {journal} {Phys.Lett.}\ }\textbf {\bibinfo
  {volume} {B716}},\ \bibinfo {pages} {1--29} (\bibinfo {year}
  {2012}{\natexlab{a}})},\ \Eprint {http://arxiv.org/abs/1207.7214}
  {arXiv:1207.7214 [hep-ex]} \BibitemShut {NoStop}%
%%CITATION = ARXIV:1207.7214;%%
\bibitem [{\citenamefont {Weinberg}(1979)}]{Weinberg:1979bn}%
  \BibitemOpen
  \bibfield  {author} {\bibinfo {author} {\bibfnamefont {Steven}\ \bibnamefont
  {Weinberg}},\ }\bibfield  {title} {\enquote {\bibinfo {title} {{Implications
  of Dynamical Symmetry Breaking: An Addendum}},}\ }\href {\doibase
  10.1103/PhysRevD.19.1277} {\bibfield  {journal} {\bibinfo  {journal}
  {Phys.Rev.}\ }\textbf {\bibinfo {volume} {D19}},\ \bibinfo {pages}
  {1277--1280} (\bibinfo {year} {1979})}\BibitemShut {NoStop}%
%%CITATION = PHRVA,D19,1277;%%
\bibitem [{\citenamefont {Susskind}(1979)}]{Susskind:1978ms}%
  \BibitemOpen
  \bibfield  {author} {\bibinfo {author} {\bibfnamefont {Leonard}\ \bibnamefont
  {Susskind}},\ }\bibfield  {title} {\enquote {\bibinfo {title} {{Dynamics of
  Spontaneous Symmetry Breaking in the Weinberg-Salam Theory}},}\ }\href
  {\doibase 10.1103/PhysRevD.20.2619} {\bibfield  {journal} {\bibinfo
  {journal} {Phys.Rev.}\ }\textbf {\bibinfo {volume} {D20}},\ \bibinfo {pages}
  {2619--2625} (\bibinfo {year} {1979})}\BibitemShut {NoStop}%
%%CITATION = PHRVA,D20,2619;%%
\bibitem [{\citenamefont {Dimopoulos}\ and\ \citenamefont
  {Susskind}(1979)}]{Dimopoulos:1979es}%
  \BibitemOpen
  \bibfield  {author} {\bibinfo {author} {\bibfnamefont {Savas}\ \bibnamefont
  {Dimopoulos}}\ and\ \bibinfo {author} {\bibfnamefont {Leonard}\ \bibnamefont
  {Susskind}},\ }\bibfield  {title} {\enquote {\bibinfo {title} {{Mass Without
  Scalars}},}\ }\href {\doibase 10.1016/0550-3213(79)90364-X} {\bibfield
  {journal} {\bibinfo  {journal} {Nucl.Phys.}\ }\textbf {\bibinfo {volume}
  {B155}},\ \bibinfo {pages} {237--252} (\bibinfo {year} {1979})}\BibitemShut
  {NoStop}%
%%CITATION = NUPHA,B155,237;%%
\bibitem [{\citenamefont {Eichten}\ and\ \citenamefont
  {Lane}(1980)}]{Eichten:1979ah}%
  \BibitemOpen
  \bibfield  {author} {\bibinfo {author} {\bibfnamefont {Estia}\ \bibnamefont
  {Eichten}}\ and\ \bibinfo {author} {\bibfnamefont {Kenneth~D.}\ \bibnamefont
  {Lane}},\ }\bibfield  {title} {\enquote {\bibinfo {title} {{Dynamical
  Breaking of Weak Interaction Symmetries}},}\ }\href {\doibase
  10.1016/0370-2693(80)90065-9} {\bibfield  {journal} {\bibinfo  {journal}
  {Phys.Lett.}\ }\textbf {\bibinfo {volume} {B90}},\ \bibinfo {pages}
  {125--130} (\bibinfo {year} {1980})}\BibitemShut {NoStop}%
%%CITATION = PHLTA,B90,125;%%
\bibitem [{\citenamefont {Farhi}\ and\ \citenamefont
  {Susskind}(1981)}]{Farhi:1980xs}%
  \BibitemOpen
  \bibfield  {author} {\bibinfo {author} {\bibfnamefont {Edward}\ \bibnamefont
  {Farhi}}\ and\ \bibinfo {author} {\bibfnamefont {Leonard}\ \bibnamefont
  {Susskind}},\ }\bibfield  {title} {\enquote {\bibinfo {title}
  {{Technicolor}},}\ }\href {\doibase 10.1016/0370-1573(81)90173-3} {\bibfield
  {journal} {\bibinfo  {journal} {Phys.Rept.}\ }\textbf {\bibinfo {volume}
  {74}},\ \bibinfo {pages} {277} (\bibinfo {year} {1981})}\BibitemShut
  {NoStop}%
%%CITATION = PRPLC,74,277;%%
\bibitem [{\citenamefont {Holdom}(1985)}]{Holdom:1984sk}%
  \BibitemOpen
  \bibfield  {author} {\bibinfo {author} {\bibfnamefont {Bob}\ \bibnamefont
  {Holdom}},\ }\bibfield  {title} {\enquote {\bibinfo {title} {{Techniodor}},}\
  }\href {\doibase 10.1016/0370-2693(85)91015-9} {\bibfield  {journal}
  {\bibinfo  {journal} {Phys.Lett.}\ }\textbf {\bibinfo {volume} {B150}},\
  \bibinfo {pages} {301} (\bibinfo {year} {1985})}\BibitemShut {NoStop}%
%%CITATION = PHLTA,B150,301;%%
\bibitem [{\citenamefont {Appelquist}\ and\ \citenamefont
  {Wijewardhana}(1987)}]{Appelquist:1987fc}%
  \BibitemOpen
  \bibfield  {author} {\bibinfo {author} {\bibfnamefont {Thomas}\ \bibnamefont
  {Appelquist}}\ and\ \bibinfo {author} {\bibfnamefont {L.C.R.}\ \bibnamefont
  {Wijewardhana}},\ }\bibfield  {title} {\enquote {\bibinfo {title} {{Chiral
  Hierarchies from Slowly Running Couplings in Technicolor Theories}},}\ }\href
  {\doibase 10.1103/PhysRevD.36.568} {\bibfield  {journal} {\bibinfo  {journal}
  {Phys.Rev.}\ }\textbf {\bibinfo {volume} {D36}},\ \bibinfo {pages} {568}
  (\bibinfo {year} {1987})}\BibitemShut {NoStop}%
%%CITATION = PHRVA,D36,568;%%
\bibitem [{\citenamefont {Yamawaki}\ \emph {et~al.}(1986)\citenamefont
  {Yamawaki}, \citenamefont {Bando},\ and\ \citenamefont
  {Matumoto}}]{Yamawaki:1985zg}%
  \BibitemOpen
  \bibfield  {author} {\bibinfo {author} {\bibfnamefont {Koichi}\ \bibnamefont
  {Yamawaki}}, \bibinfo {author} {\bibfnamefont {Masako}\ \bibnamefont
  {Bando}}, \ and\ \bibinfo {author} {\bibfnamefont {Ken-iti}\ \bibnamefont
  {Matumoto}},\ }\bibfield  {title} {\enquote {\bibinfo {title} {{Scale
  Invariant Technicolor Model and a Technidilaton}},}\ }\href {\doibase
  10.1103/PhysRevLett.56.1335} {\bibfield  {journal} {\bibinfo  {journal}
  {Phys.Rev.Lett.}\ }\textbf {\bibinfo {volume} {56}},\ \bibinfo {pages} {1335}
  (\bibinfo {year} {1986})}\BibitemShut {NoStop}%
%%CITATION = PRLTA,56,1335;%%
\bibitem [{\citenamefont {Caswell}(1974)}]{Caswell:1974gg}%
  \BibitemOpen
  \bibfield  {author} {\bibinfo {author} {\bibfnamefont {William~E.}\
  \bibnamefont {Caswell}},\ }\bibfield  {title} {\enquote {\bibinfo {title}
  {{Asymptotic Behavior of Nonabelian Gauge Theories to Two Loop Order}},}\
  }\href {\doibase 10.1103/PhysRevLett.33.244} {\bibfield  {journal} {\bibinfo
  {journal} {Phys.Rev.Lett.}\ }\textbf {\bibinfo {volume} {33}},\ \bibinfo
  {pages} {244} (\bibinfo {year} {1974})}\BibitemShut {NoStop}%
%%CITATION = PRLTA,33,244;%%
\bibitem [{\citenamefont {Banks}\ and\ \citenamefont
  {Zaks}(1982)}]{Banks:1981nn}%
  \BibitemOpen
  \bibfield  {author} {\bibinfo {author} {\bibfnamefont {Tom}\ \bibnamefont
  {Banks}}\ and\ \bibinfo {author} {\bibfnamefont {A.}~\bibnamefont {Zaks}},\
  }\bibfield  {title} {\enquote {\bibinfo {title} {{On the Phase Structure of
  Vector-Like Gauge Theories with Massless Fermions}},}\ }\href {\doibase
  10.1016/0550-3213(82)90035-9} {\bibfield  {journal} {\bibinfo  {journal}
  {Nucl.Phys.}\ }\textbf {\bibinfo {volume} {B196}},\ \bibinfo {pages} {189}
  (\bibinfo {year} {1982})}\BibitemShut {NoStop}%
%%CITATION = NUPHA,B196,189;%%
\bibitem [{\citenamefont {Marciano}(1980)}]{Marciano:1980zf}%
  \BibitemOpen
  \bibfield  {author} {\bibinfo {author} {\bibfnamefont {William~J.}\
  \bibnamefont {Marciano}},\ }\bibfield  {title} {\enquote {\bibinfo {title}
  {{Exotic New Quarks and Dynamical Symmetry Breaking}},}\ }\href {\doibase
  10.1103/PhysRevD.21.2425} {\bibfield  {journal} {\bibinfo  {journal}
  {Phys.Rev.}\ }\textbf {\bibinfo {volume} {D21}},\ \bibinfo {pages} {2425}
  (\bibinfo {year} {1980})}\BibitemShut {NoStop}%
%%CITATION = PHRVA,D21,2425;%%
\bibitem [{\citenamefont {Kogut}\ \emph {et~al.}(1984)\citenamefont {Kogut},
  \citenamefont {Shigemitsu},\ and\ \citenamefont {Sinclair}}]{Kogut:1984sb}%
  \BibitemOpen
  \bibfield  {author} {\bibinfo {author} {\bibfnamefont {J.B.}\ \bibnamefont
  {Kogut}}, \bibinfo {author} {\bibfnamefont {J.}~\bibnamefont {Shigemitsu}}, \
  and\ \bibinfo {author} {\bibfnamefont {D.K.}\ \bibnamefont {Sinclair}},\
  }\bibfield  {title} {\enquote {\bibinfo {title} {{CHIRAL SYMMETRY BREAKING
  WITH OCTET AND SEXTET QUARKS}},}\ }\href {\doibase
  10.1016/0370-2693(84)90346-0} {\bibfield  {journal} {\bibinfo  {journal}
  {Phys.Lett.}\ }\textbf {\bibinfo {volume} {B145}},\ \bibinfo {pages} {239}
  (\bibinfo {year} {1984})}\BibitemShut {NoStop}%
%%CITATION = PHLTA,B145,239;%%
\bibitem [{\citenamefont {Appelquist}\ \emph {et~al.}(2004)\citenamefont
  {Appelquist}, \citenamefont {Piai},\ and\ \citenamefont
  {Shrock}}]{Appelquist:2003hn}%
  \BibitemOpen
  \bibfield  {author} {\bibinfo {author} {\bibfnamefont {Thomas}\ \bibnamefont
  {Appelquist}}, \bibinfo {author} {\bibfnamefont {Maurizio}\ \bibnamefont
  {Piai}}, \ and\ \bibinfo {author} {\bibfnamefont {Robert}\ \bibnamefont
  {Shrock}},\ }\bibfield  {title} {\enquote {\bibinfo {title} {{Fermion masses
  and mixing in extended technicolor models}},}\ }\href {\doibase
  10.1103/PhysRevD.69.015002} {\bibfield  {journal} {\bibinfo  {journal}
  {Phys.Rev.}\ }\textbf {\bibinfo {volume} {D69}},\ \bibinfo {pages} {015002}
  (\bibinfo {year} {2004})},\ \Eprint {http://arxiv.org/abs/hep-ph/0308061}
  {arXiv:hep-ph/0308061 [hep-ph]} \BibitemShut {NoStop}%
%%CITATION = HEP-PH/0308061;%%
\bibitem [{\citenamefont {Sannino}\ and\ \citenamefont
  {Tuominen}(2005)}]{Sannino:2004qp}%
  \BibitemOpen
  \bibfield  {author} {\bibinfo {author} {\bibfnamefont {Francesco}\
  \bibnamefont {Sannino}}\ and\ \bibinfo {author} {\bibfnamefont {Kimmo}\
  \bibnamefont {Tuominen}},\ }\bibfield  {title} {\enquote {\bibinfo {title}
  {{Orientifold theory dynamics and symmetry breaking}},}\ }\href {\doibase
  10.1103/PhysRevD.71.051901} {\bibfield  {journal} {\bibinfo  {journal}
  {Phys.Rev.}\ }\textbf {\bibinfo {volume} {D71}},\ \bibinfo {pages} {051901}
  (\bibinfo {year} {2005})},\ \Eprint {http://arxiv.org/abs/hep-ph/0405209}
  {arXiv:hep-ph/0405209 [hep-ph]} \BibitemShut {NoStop}%
%%CITATION = HEP-PH/0405209;%%
\bibitem [{\citenamefont {Dietrich}\ \emph {et~al.}(2005)\citenamefont
  {Dietrich}, \citenamefont {Sannino},\ and\ \citenamefont
  {Tuominen}}]{Dietrich:2005jn}%
  \BibitemOpen
  \bibfield  {author} {\bibinfo {author} {\bibfnamefont {Dennis~D.}\
  \bibnamefont {Dietrich}}, \bibinfo {author} {\bibfnamefont {Francesco}\
  \bibnamefont {Sannino}}, \ and\ \bibinfo {author} {\bibfnamefont {Kimmo}\
  \bibnamefont {Tuominen}},\ }\bibfield  {title} {\enquote {\bibinfo {title}
  {{Light composite Higgs from higher representations versus electroweak
  precision measurements: Predictions for CERN LHC}},}\ }\href {\doibase
  10.1103/PhysRevD.72.055001} {\bibfield  {journal} {\bibinfo  {journal}
  {Phys.Rev.}\ }\textbf {\bibinfo {volume} {D72}},\ \bibinfo {pages} {055001}
  (\bibinfo {year} {2005})},\ \Eprint {http://arxiv.org/abs/hep-ph/0505059}
  {arXiv:hep-ph/0505059 [hep-ph]} \BibitemShut {NoStop}%
%%CITATION = HEP-PH/0505059;%%
\bibitem [{\citenamefont {Luty}\ and\ \citenamefont
  {Okui}(2006)}]{Luty:2004ye}%
  \BibitemOpen
  \bibfield  {author} {\bibinfo {author} {\bibfnamefont {Markus~A.}\
  \bibnamefont {Luty}}\ and\ \bibinfo {author} {\bibfnamefont {Takemichi}\
  \bibnamefont {Okui}},\ }\bibfield  {title} {\enquote {\bibinfo {title}
  {{Conformal technicolor}},}\ }\href {\doibase 10.1088/1126-6708/2006/09/070}
  {\bibfield  {journal} {\bibinfo  {journal} {JHEP}\ }\textbf {\bibinfo
  {volume} {0609}},\ \bibinfo {pages} {070} (\bibinfo {year} {2006})},\ \Eprint
  {http://arxiv.org/abs/hep-ph/0409274} {arXiv:hep-ph/0409274 [hep-ph]}
  \BibitemShut {NoStop}%
%%CITATION = HEP-PH/0409274;%%
\bibitem [{\citenamefont {Dietrich}\ and\ \citenamefont
  {Sannino}(2007)}]{Dietrich:2006cm}%
  \BibitemOpen
  \bibfield  {author} {\bibinfo {author} {\bibfnamefont {Dennis~D.}\
  \bibnamefont {Dietrich}}\ and\ \bibinfo {author} {\bibfnamefont {Francesco}\
  \bibnamefont {Sannino}},\ }\bibfield  {title} {\enquote {\bibinfo {title}
  {{Conformal window of SU(N) gauge theories with fermions in higher
  dimensional representations}},}\ }\href {\doibase 10.1103/PhysRevD.75.085018}
  {\bibfield  {journal} {\bibinfo  {journal} {Phys.Rev.}\ }\textbf {\bibinfo
  {volume} {D75}},\ \bibinfo {pages} {085018} (\bibinfo {year} {2007})},\
  \Eprint {http://arxiv.org/abs/hep-ph/0611341} {arXiv:hep-ph/0611341 [hep-ph]}
  \BibitemShut {NoStop}%
%%CITATION = HEP-PH/0611341;%%
\bibitem [{\citenamefont {Kurachi}\ and\ \citenamefont
  {Shrock}(2006)}]{Kurachi:2006ej}%
  \BibitemOpen
  \bibfield  {author} {\bibinfo {author} {\bibfnamefont {Masafumi}\
  \bibnamefont {Kurachi}}\ and\ \bibinfo {author} {\bibfnamefont {Robert}\
  \bibnamefont {Shrock}},\ }\bibfield  {title} {\enquote {\bibinfo {title}
  {{Study of the Change from Walking to Non-Walking Behavior in a Vectorial
  Gauge Theory as a Function of N(f)}},}\ }\href {\doibase
  10.1088/1126-6708/2006/12/034} {\bibfield  {journal} {\bibinfo  {journal}
  {JHEP}\ }\textbf {\bibinfo {volume} {0612}},\ \bibinfo {pages} {034}
  (\bibinfo {year} {2006})},\ \Eprint {http://arxiv.org/abs/hep-ph/0605290}
  {arXiv:hep-ph/0605290 [hep-ph]} \BibitemShut {NoStop}%
%%CITATION = HEP-PH/0605290;%%
\bibitem [{\citenamefont {Ellis}\ and\ \citenamefont
  {You}(2012)}]{Ellis:2012hz}%
  \BibitemOpen
  \bibfield  {author} {\bibinfo {author} {\bibfnamefont {John}\ \bibnamefont
  {Ellis}}\ and\ \bibinfo {author} {\bibfnamefont {Tevong}\ \bibnamefont
  {You}},\ }\bibfield  {title} {\enquote {\bibinfo {title} {{Global Analysis of
  the Higgs Candidate with Mass ~ 125 GeV}},}\ }\href {\doibase
  10.1007/JHEP09(2012)123} {\bibfield  {journal} {\bibinfo  {journal} {JHEP}\
  }\textbf {\bibinfo {volume} {1209}},\ \bibinfo {pages} {123} (\bibinfo {year}
  {2012})},\ \Eprint {http://arxiv.org/abs/1207.1693} {arXiv:1207.1693
  [hep-ph]} \BibitemShut {NoStop}%
%%CITATION = ARXIV:1207.1693;%%
\bibitem [{\citenamefont {Low}\ \emph {et~al.}(2012)\citenamefont {Low},
  \citenamefont {Lykken},\ and\ \citenamefont {Shaughnessy}}]{Low:2012rj}%
  \BibitemOpen
  \bibfield  {author} {\bibinfo {author} {\bibfnamefont {Ian}\ \bibnamefont
  {Low}}, \bibinfo {author} {\bibfnamefont {Joseph}\ \bibnamefont {Lykken}}, \
  and\ \bibinfo {author} {\bibfnamefont {Gabe}\ \bibnamefont {Shaughnessy}},\
  }\bibfield  {title} {\enquote {\bibinfo {title} {{Have We Observed the Higgs
  (Imposter)?}}}\ }\href {\doibase 10.1103/PhysRevD.86.093012} {\bibfield
  {journal} {\bibinfo  {journal} {Phys.Rev.}\ }\textbf {\bibinfo {volume}
  {D86}},\ \bibinfo {pages} {093012} (\bibinfo {year} {2012})},\ \Eprint
  {http://arxiv.org/abs/1207.1093} {arXiv:1207.1093 [hep-ph]} \BibitemShut
  {NoStop}%
%%CITATION = ARXIV:1207.1093;%%
\bibitem [{\citenamefont {Elander}\ and\ \citenamefont
  {Piai}(2012)}]{Elander:2012fk}%
  \BibitemOpen
  \bibfield  {author} {\bibinfo {author} {\bibfnamefont {Daniel}\ \bibnamefont
  {Elander}}\ and\ \bibinfo {author} {\bibfnamefont {Maurizio}\ \bibnamefont
  {Piai}},\ }\bibfield  {title} {\enquote {\bibinfo {title} {{The decay
  constant of the holographic techni-dilaton and the 125 GeV boson}},}\
  }\href@noop {} {\  (\bibinfo {year} {2012})},\ \Eprint
  {http://arxiv.org/abs/1208.0546} {arXiv:1208.0546 [hep-ph]} \BibitemShut
  {NoStop}%
%%CITATION = ARXIV:1208.0546;%%
\bibitem [{\citenamefont {Bardeen}\ \emph {et~al.}(1986)\citenamefont
  {Bardeen}, \citenamefont {Leung},\ and\ \citenamefont
  {Love}}]{Bardeen:1985sm}%
  \BibitemOpen
  \bibfield  {author} {\bibinfo {author} {\bibfnamefont {William~A.}\
  \bibnamefont {Bardeen}}, \bibinfo {author} {\bibfnamefont {Chung~Ngoc}\
  \bibnamefont {Leung}}, \ and\ \bibinfo {author} {\bibfnamefont {S.T.}\
  \bibnamefont {Love}},\ }\bibfield  {title} {\enquote {\bibinfo {title} {{The
  Dilaton and Chiral Symmetry Breaking}},}\ }\href {\doibase
  10.1103/PhysRevLett.56.1230} {\bibfield  {journal} {\bibinfo  {journal}
  {Phys.Rev.Lett.}\ }\textbf {\bibinfo {volume} {56}},\ \bibinfo {pages} {1230}
  (\bibinfo {year} {1986})}\BibitemShut {NoStop}%
%%CITATION = PRLTA,56,1230;%%
\bibitem [{\citenamefont {Holdom}\ and\ \citenamefont
  {Terning}(1987)}]{Holdom:1986ub}%
  \BibitemOpen
  \bibfield  {author} {\bibinfo {author} {\bibfnamefont {Bob}\ \bibnamefont
  {Holdom}}\ and\ \bibinfo {author} {\bibfnamefont {John}\ \bibnamefont
  {Terning}},\ }\bibfield  {title} {\enquote {\bibinfo {title} {{A Light
  Dilaton in Gauge Theories?}}}\ }\href {\doibase 10.1016/0370-2693(87)91109-9}
  {\bibfield  {journal} {\bibinfo  {journal} {Phys.Lett.}\ }\textbf {\bibinfo
  {volume} {B187}},\ \bibinfo {pages} {357} (\bibinfo {year}
  {1987})}\BibitemShut {NoStop}%
%%CITATION = PHLTA,B187,357;%%
\bibitem [{\citenamefont {Miransky}\ and\ \citenamefont
  {Yamawaki}(1997)}]{Miransky:1996pd}%
  \BibitemOpen
  \bibfield  {author} {\bibinfo {author} {\bibfnamefont {V.A.}\ \bibnamefont
  {Miransky}}\ and\ \bibinfo {author} {\bibfnamefont {Koichi}\ \bibnamefont
  {Yamawaki}},\ }\bibfield  {title} {\enquote {\bibinfo {title} {{Conformal
  phase transition in gauge theories}},}\ }\href {\doibase
  10.1103/PhysRevD.56.3768, 10.1103/PhysRevD.55.5051} {\bibfield  {journal}
  {\bibinfo  {journal} {Phys.Rev.}\ }\textbf {\bibinfo {volume} {D55}},\
  \bibinfo {pages} {5051--5066} (\bibinfo {year} {1997})},\ \Eprint
  {http://arxiv.org/abs/hep-th/9611142} {arXiv:hep-th/9611142 [hep-th]}
  \BibitemShut {NoStop}%
%%CITATION = HEP-TH/9611142;%%
\bibitem [{\citenamefont {Goldberger}\ \emph {et~al.}(2008)\citenamefont
  {Goldberger}, \citenamefont {Grinstein},\ and\ \citenamefont
  {Skiba}}]{Goldberger:2007zk}%
  \BibitemOpen
  \bibfield  {author} {\bibinfo {author} {\bibfnamefont {Walter~D.}\
  \bibnamefont {Goldberger}}, \bibinfo {author} {\bibfnamefont {Benjamin}\
  \bibnamefont {Grinstein}}, \ and\ \bibinfo {author} {\bibfnamefont {Witold}\
  \bibnamefont {Skiba}},\ }\bibfield  {title} {\enquote {\bibinfo {title}
  {{Distinguishing the Higgs boson from the dilaton at the Large Hadron
  Collider}},}\ }\href {\doibase 10.1103/PhysRevLett.100.111802} {\bibfield
  {journal} {\bibinfo  {journal} {Phys.Rev.Lett.}\ }\textbf {\bibinfo {volume}
  {100}},\ \bibinfo {pages} {111802} (\bibinfo {year} {2008})},\ \Eprint
  {http://arxiv.org/abs/0708.1463} {arXiv:0708.1463 [hep-ph]} \BibitemShut
  {NoStop}%
%%CITATION = ARXIV:0708.1463;%%
\bibitem [{\citenamefont {Appelquist}\ and\ \citenamefont
  {Bai}(2010)}]{Appelquist:2010gy}%
  \BibitemOpen
  \bibfield  {author} {\bibinfo {author} {\bibfnamefont {Thomas}\ \bibnamefont
  {Appelquist}}\ and\ \bibinfo {author} {\bibfnamefont {Yang}\ \bibnamefont
  {Bai}},\ }\bibfield  {title} {\enquote {\bibinfo {title} {{A Light Dilaton in
  Walking Gauge Theories}},}\ }\href {\doibase 10.1103/PhysRevD.82.071701}
  {\bibfield  {journal} {\bibinfo  {journal} {Phys.Rev.}\ }\textbf {\bibinfo
  {volume} {D82}},\ \bibinfo {pages} {071701} (\bibinfo {year} {2010})},\
  \Eprint {http://arxiv.org/abs/1006.4375} {arXiv:1006.4375 [hep-ph]}
  \BibitemShut {NoStop}%
%%CITATION = ARXIV:1006.4375;%%
\bibitem [{\citenamefont {Grinstein}\ and\ \citenamefont
  {Uttayarat}(2011)}]{Grinstein:2011dq}%
  \BibitemOpen
  \bibfield  {author} {\bibinfo {author} {\bibfnamefont {Benjamin}\
  \bibnamefont {Grinstein}}\ and\ \bibinfo {author} {\bibfnamefont {Patipan}\
  \bibnamefont {Uttayarat}},\ }\bibfield  {title} {\enquote {\bibinfo {title}
  {{A Very Light Dilaton}},}\ }\href {\doibase 10.1007/JHEP07(2011)038}
  {\bibfield  {journal} {\bibinfo  {journal} {JHEP}\ }\textbf {\bibinfo
  {volume} {1107}},\ \bibinfo {pages} {038} (\bibinfo {year} {2011})},\ \Eprint
  {http://arxiv.org/abs/1105.2370} {arXiv:1105.2370 [hep-ph]} \BibitemShut
  {NoStop}%
%%CITATION = ARXIV:1105.2370;%%
\bibitem [{\citenamefont {Antipin}\ \emph {et~al.}(2012)\citenamefont
  {Antipin}, \citenamefont {Mojaza},\ and\ \citenamefont
  {Sannino}}]{Antipin:2011aa}%
  \BibitemOpen
  \bibfield  {author} {\bibinfo {author} {\bibfnamefont {Oleg}\ \bibnamefont
  {Antipin}}, \bibinfo {author} {\bibfnamefont {Matin}\ \bibnamefont {Mojaza}},
  \ and\ \bibinfo {author} {\bibfnamefont {Francesco}\ \bibnamefont
  {Sannino}},\ }\bibfield  {title} {\enquote {\bibinfo {title} {{Light Dilaton
  at Fixed Points and Ultra Light Scale Super Yang Mills}},}\ }\href {\doibase
  10.1016/j.physletb.2012.04.050} {\bibfield  {journal} {\bibinfo  {journal}
  {Phys.Lett.}\ }\textbf {\bibinfo {volume} {B712}},\ \bibinfo {pages}
  {119--125} (\bibinfo {year} {2012})},\ \Eprint
  {http://arxiv.org/abs/1107.2932} {arXiv:1107.2932 [hep-ph]} \BibitemShut
  {NoStop}%
%%CITATION = ARXIV:1107.2932;%%
\bibitem [{\citenamefont {Hashimoto}\ and\ \citenamefont
  {Yamawaki}(2011)}]{Hashimoto:2010nw}%
  \BibitemOpen
  \bibfield  {author} {\bibinfo {author} {\bibfnamefont {Michio}\ \bibnamefont
  {Hashimoto}}\ and\ \bibinfo {author} {\bibfnamefont {Koichi}\ \bibnamefont
  {Yamawaki}},\ }\bibfield  {title} {\enquote {\bibinfo {title}
  {{Techni-dilaton at Conformal Edge}},}\ }\href {\doibase
  10.1103/PhysRevD.83.015008} {\bibfield  {journal} {\bibinfo  {journal}
  {Phys.Rev.}\ }\textbf {\bibinfo {volume} {D83}},\ \bibinfo {pages} {015008}
  (\bibinfo {year} {2011})},\ \Eprint {http://arxiv.org/abs/1009.5482}
  {arXiv:1009.5482 [hep-ph]} \BibitemShut {NoStop}%
%%CITATION = ARXIV:1009.5482;%%
\bibitem [{\citenamefont {Matsuzaki}\ and\ \citenamefont
  {Yamawaki}(2012{\natexlab{a}})}]{Matsuzaki:2012xx}%
  \BibitemOpen
  \bibfield  {author} {\bibinfo {author} {\bibfnamefont {Shinya}\ \bibnamefont
  {Matsuzaki}}\ and\ \bibinfo {author} {\bibfnamefont {Koichi}\ \bibnamefont
  {Yamawaki}},\ }\bibfield  {title} {\enquote {\bibinfo {title} {{Holographic
  techni-dilaton at 125 GeV}},}\ }\href {\doibase 10.1103/PhysRevD.86.115004}
  {\bibfield  {journal} {\bibinfo  {journal} {Phys.Rev.}\ }\textbf {\bibinfo
  {volume} {D86}},\ \bibinfo {pages} {115004} (\bibinfo {year}
  {2012}{\natexlab{a}})},\ \Eprint {http://arxiv.org/abs/1209.2017}
  {arXiv:1209.2017 [hep-ph]} \BibitemShut {NoStop}%
%%CITATION = ARXIV:1209.2017;%%
\bibitem [{\citenamefont {Matsuzaki}\ and\ \citenamefont
  {Yamawaki}(2012{\natexlab{b}})}]{Matsuzaki:2012vc}%
  \BibitemOpen
  \bibfield  {author} {\bibinfo {author} {\bibfnamefont {Shinya}\ \bibnamefont
  {Matsuzaki}}\ and\ \bibinfo {author} {\bibfnamefont {Koichi}\ \bibnamefont
  {Yamawaki}},\ }\bibfield  {title} {\enquote {\bibinfo {title} {{Discovering
  125 GeV techni-dilaton at LHC}},}\ }\href {\doibase
  10.1103/PhysRevD.86.035025} {\bibfield  {journal} {\bibinfo  {journal}
  {Phys.Rev.}\ }\textbf {\bibinfo {volume} {D86}},\ \bibinfo {pages} {035025}
  (\bibinfo {year} {2012}{\natexlab{b}})},\ \Eprint
  {http://arxiv.org/abs/1206.6703} {arXiv:1206.6703 [hep-ph]} \BibitemShut
  {NoStop}%
%%CITATION = ARXIV:1206.6703;%%
\bibitem [{\citenamefont {Arkani-Hamed}\ \emph {et~al.}(2001)\citenamefont
  {Arkani-Hamed}, \citenamefont {Cohen},\ and\ \citenamefont
  {Georgi}}]{ArkaniHamed:2001nc}%
  \BibitemOpen
  \bibfield  {author} {\bibinfo {author} {\bibfnamefont {Nima}\ \bibnamefont
  {Arkani-Hamed}}, \bibinfo {author} {\bibfnamefont {Andrew~G.}\ \bibnamefont
  {Cohen}}, \ and\ \bibinfo {author} {\bibfnamefont {Howard}\ \bibnamefont
  {Georgi}},\ }\bibfield  {title} {\enquote {\bibinfo {title} {{Electroweak
  symmetry breaking from dimensional deconstruction}},}\ }\href {\doibase
  10.1016/S0370-2693(01)00741-9} {\bibfield  {journal} {\bibinfo  {journal}
  {Phys.Lett.}\ }\textbf {\bibinfo {volume} {B513}},\ \bibinfo {pages}
  {232--240} (\bibinfo {year} {2001})},\ \Eprint
  {http://arxiv.org/abs/hep-ph/0105239} {arXiv:hep-ph/0105239 [hep-ph]}
  \BibitemShut {NoStop}%
%%CITATION = HEP-PH/0105239;%%
\bibitem [{\citenamefont {Arkani-Hamed}\ \emph
  {et~al.}(2002{\natexlab{a}})\citenamefont {Arkani-Hamed}, \citenamefont
  {Cohen}, \citenamefont {Katz}, \citenamefont {Nelson}, \citenamefont
  {Gregoire} \emph {et~al.}}]{ArkaniHamed:2002qx}%
  \BibitemOpen
  \bibfield  {author} {\bibinfo {author} {\bibfnamefont {N.}~\bibnamefont
  {Arkani-Hamed}}, \bibinfo {author} {\bibfnamefont {A.G.}\ \bibnamefont
  {Cohen}}, \bibinfo {author} {\bibfnamefont {E.}~\bibnamefont {Katz}},
  \bibinfo {author} {\bibfnamefont {A.E.}\ \bibnamefont {Nelson}}, \bibinfo
  {author} {\bibfnamefont {T.}~\bibnamefont {Gregoire}},  \emph {et~al.},\
  }\bibfield  {title} {\enquote {\bibinfo {title} {{The Minimal moose for a
  little Higgs}},}\ }\href@noop {} {\bibfield  {journal} {\bibinfo  {journal}
  {JHEP}\ }\textbf {\bibinfo {volume} {0208}},\ \bibinfo {pages} {021}
  (\bibinfo {year} {2002}{\natexlab{a}})},\ \Eprint
  {http://arxiv.org/abs/hep-ph/0206020} {arXiv:hep-ph/0206020 [hep-ph]}
  \BibitemShut {NoStop}%
%%CITATION = HEP-PH/0206020;%%
\bibitem [{\citenamefont {Arkani-Hamed}\ \emph
  {et~al.}(2002{\natexlab{b}})\citenamefont {Arkani-Hamed}, \citenamefont
  {Cohen}, \citenamefont {Katz},\ and\ \citenamefont
  {Nelson}}]{ArkaniHamed:2002qy}%
  \BibitemOpen
  \bibfield  {author} {\bibinfo {author} {\bibfnamefont {N.}~\bibnamefont
  {Arkani-Hamed}}, \bibinfo {author} {\bibfnamefont {A.G.}\ \bibnamefont
  {Cohen}}, \bibinfo {author} {\bibfnamefont {E.}~\bibnamefont {Katz}}, \ and\
  \bibinfo {author} {\bibfnamefont {A.E.}\ \bibnamefont {Nelson}},\ }\bibfield
  {title} {\enquote {\bibinfo {title} {{The Littlest Higgs}},}\ }\href@noop {}
  {\bibfield  {journal} {\bibinfo  {journal} {JHEP}\ }\textbf {\bibinfo
  {volume} {0207}},\ \bibinfo {pages} {034} (\bibinfo {year}
  {2002}{\natexlab{b}})},\ \Eprint {http://arxiv.org/abs/hep-ph/0206021}
  {arXiv:hep-ph/0206021 [hep-ph]} \BibitemShut {NoStop}%
%%CITATION = HEP-PH/0206021;%%
\bibitem [{\citenamefont {Kaplan}\ and\ \citenamefont
  {Georgi}(1984)}]{Kaplan:1983fs}%
  \BibitemOpen
  \bibfield  {author} {\bibinfo {author} {\bibfnamefont {David~B.}\
  \bibnamefont {Kaplan}}\ and\ \bibinfo {author} {\bibfnamefont {Howard}\
  \bibnamefont {Georgi}},\ }\bibfield  {title} {\enquote {\bibinfo {title}
  {{SU(2) x U(1) Breaking by Vacuum Misalignment}},}\ }\href {\doibase
  10.1016/0370-2693(84)91177-8} {\bibfield  {journal} {\bibinfo  {journal}
  {Phys.Lett.}\ }\textbf {\bibinfo {volume} {B136}},\ \bibinfo {pages} {183}
  (\bibinfo {year} {1984})}\BibitemShut {NoStop}%
%%CITATION = PHLTA,B136,183;%%
\bibitem [{\citenamefont {Kaplan}\ \emph {et~al.}(1984)\citenamefont {Kaplan},
  \citenamefont {Georgi},\ and\ \citenamefont {Dimopoulos}}]{Kaplan:1983sm}%
  \BibitemOpen
  \bibfield  {author} {\bibinfo {author} {\bibfnamefont {David~B.}\
  \bibnamefont {Kaplan}}, \bibinfo {author} {\bibfnamefont {Howard}\
  \bibnamefont {Georgi}}, \ and\ \bibinfo {author} {\bibfnamefont {Savas}\
  \bibnamefont {Dimopoulos}},\ }\bibfield  {title} {\enquote {\bibinfo {title}
  {{Composite Higgs Scalars}},}\ }\href {\doibase 10.1016/0370-2693(84)91178-X}
  {\bibfield  {journal} {\bibinfo  {journal} {Phys.Lett.}\ }\textbf {\bibinfo
  {volume} {B136}},\ \bibinfo {pages} {187} (\bibinfo {year}
  {1984})}\BibitemShut {NoStop}%
%%CITATION = PHLTA,B136,187;%%
\bibitem [{\citenamefont {Georgi}\ and\ \citenamefont
  {Kaplan}(1984)}]{Georgi:1984af}%
  \BibitemOpen
  \bibfield  {author} {\bibinfo {author} {\bibfnamefont {Howard}\ \bibnamefont
  {Georgi}}\ and\ \bibinfo {author} {\bibfnamefont {David~B.}\ \bibnamefont
  {Kaplan}},\ }\bibfield  {title} {\enquote {\bibinfo {title} {{Composite Higgs
  and Custodial SU(2)}},}\ }\href {\doibase 10.1016/0370-2693(84)90341-1}
  {\bibfield  {journal} {\bibinfo  {journal} {Phys.Lett.}\ }\textbf {\bibinfo
  {volume} {B145}},\ \bibinfo {pages} {216} (\bibinfo {year}
  {1984})}\BibitemShut {NoStop}%
%%CITATION = PHLTA,B145,216;%%
\bibitem [{\citenamefont {Thaler}(2005)}]{Thaler:2005kr}%
  \BibitemOpen
  \bibfield  {author} {\bibinfo {author} {\bibfnamefont {Jesse}\ \bibnamefont
  {Thaler}},\ }\bibfield  {title} {\enquote {\bibinfo {title} {{Little
  technicolor}},}\ }\href {\doibase 10.1088/1126-6708/2005/07/024} {\bibfield
  {journal} {\bibinfo  {journal} {JHEP}\ }\textbf {\bibinfo {volume} {0507}},\
  \bibinfo {pages} {024} (\bibinfo {year} {2005})},\ \Eprint
  {http://arxiv.org/abs/hep-ph/0502175} {arXiv:hep-ph/0502175 [hep-ph]}
  \BibitemShut {NoStop}%
%%CITATION = HEP-PH/0502175;%%
\bibitem [{\citenamefont {Schmaltz}\ and\ \citenamefont
  {Thaler}(2009)}]{Schmaltz:2008vd}%
  \BibitemOpen
  \bibfield  {author} {\bibinfo {author} {\bibfnamefont {Martin}\ \bibnamefont
  {Schmaltz}}\ and\ \bibinfo {author} {\bibfnamefont {Jesse}\ \bibnamefont
  {Thaler}},\ }\bibfield  {title} {\enquote {\bibinfo {title} {{Collective
  Quartics and Dangerous Singlets in Little Higgs}},}\ }\href {\doibase
  10.1088/1126-6708/2009/03/137} {\bibfield  {journal} {\bibinfo  {journal}
  {JHEP}\ }\textbf {\bibinfo {volume} {0903}},\ \bibinfo {pages} {137}
  (\bibinfo {year} {2009})},\ \Eprint {http://arxiv.org/abs/0812.2477}
  {arXiv:0812.2477 [hep-ph]} \BibitemShut {NoStop}%
%%CITATION = ARXIV:0812.2477;%%
\bibitem [{\citenamefont {Foadi}\ \emph {et~al.}(2012)\citenamefont {Foadi},
  \citenamefont {Frandsen},\ and\ \citenamefont {Francesco}}]{Foadi:2012bb}%
  \BibitemOpen
  \bibfield  {author} {\bibinfo {author} {\bibfnamefont {Roshan}\ \bibnamefont
  {Foadi}}, \bibinfo {author} {\bibfnamefont {Mads~T.}\ \bibnamefont
  {Frandsen}}, \ and\ \bibinfo {author} {\bibfnamefont {Sannino}\ \bibnamefont
  {Francesco}},\ }\bibfield  {title} {\enquote {\bibinfo {title} {{125 GeV
  Higgs from a not so light Technicolor Scalar}},}\ }\href@noop {} {\
  (\bibinfo {year} {2012})},\ \Eprint {http://arxiv.org/abs/1211.1083}
  {arXiv:1211.1083 [hep-ph]} \BibitemShut {NoStop}%
%%CITATION = ARXIV:1211.1083;%%
\bibitem [{\citenamefont {Appelquist}\ \emph {et~al.}(2008)\citenamefont
  {Appelquist}, \citenamefont {Fleming},\ and\ \citenamefont
  {Neil}}]{Appelquist:2007hu}%
  \BibitemOpen
  \bibfield  {author} {\bibinfo {author} {\bibfnamefont {Thomas}\ \bibnamefont
  {Appelquist}}, \bibinfo {author} {\bibfnamefont {George~T.}\ \bibnamefont
  {Fleming}}, \ and\ \bibinfo {author} {\bibfnamefont {Ethan~T.}\ \bibnamefont
  {Neil}},\ }\bibfield  {title} {\enquote {\bibinfo {title} {{Lattice study of
  the conformal window in QCD-like theories}},}\ }\href {\doibase
  10.1103/PhysRevLett.100.171607} {\bibfield  {journal} {\bibinfo  {journal}
  {Phys.Rev.Lett.}\ }\textbf {\bibinfo {volume} {100}},\ \bibinfo {pages}
  {171607} (\bibinfo {year} {2008})},\ \Eprint {http://arxiv.org/abs/0712.0609}
  {arXiv:0712.0609 [hep-ph]} \BibitemShut {NoStop}%
%%CITATION = ARXIV:0712.0609;%%
\bibitem [{\citenamefont {Fodor}\ \emph {et~al.}(2009)\citenamefont {Fodor},
  \citenamefont {Holland}, \citenamefont {Kuti}, \citenamefont {Nogradi},\ and\
  \citenamefont {Schroeder}}]{Fodor:2009wk}%
  \BibitemOpen
  \bibfield  {author} {\bibinfo {author} {\bibfnamefont {Zoltan}\ \bibnamefont
  {Fodor}}, \bibinfo {author} {\bibfnamefont {Kieran}\ \bibnamefont {Holland}},
  \bibinfo {author} {\bibfnamefont {Julius}\ \bibnamefont {Kuti}}, \bibinfo
  {author} {\bibfnamefont {Daniel}\ \bibnamefont {Nogradi}}, \ and\ \bibinfo
  {author} {\bibfnamefont {Chris}\ \bibnamefont {Schroeder}},\ }\bibfield
  {title} {\enquote {\bibinfo {title} {{Nearly conformal gauge theories in
  finite volume}},}\ }\href {\doibase 10.1016/j.physletb.2009.10.040}
  {\bibfield  {journal} {\bibinfo  {journal} {Phys.Lett.}\ }\textbf {\bibinfo
  {volume} {B681}},\ \bibinfo {pages} {353--361} (\bibinfo {year} {2009})},\
  \Eprint {http://arxiv.org/abs/0907.4562} {arXiv:0907.4562 [hep-lat]}
  \BibitemShut {NoStop}%
%%CITATION = ARXIV:0907.4562;%%
\bibitem [{\citenamefont {Fodor}\ \emph
  {et~al.}(2011{\natexlab{a}})\citenamefont {Fodor}, \citenamefont {Holland},
  \citenamefont {Kuti}, \citenamefont {Nogradi},\ and\ \citenamefont
  {Schroeder}}]{Fodor:2011tw}%
  \BibitemOpen
  \bibfield  {author} {\bibinfo {author} {\bibfnamefont {Zoltan}\ \bibnamefont
  {Fodor}}, \bibinfo {author} {\bibfnamefont {Kieran}\ \bibnamefont {Holland}},
  \bibinfo {author} {\bibfnamefont {Julius}\ \bibnamefont {Kuti}}, \bibinfo
  {author} {\bibfnamefont {Daniel}\ \bibnamefont {Nogradi}}, \ and\ \bibinfo
  {author} {\bibfnamefont {Chris}\ \bibnamefont {Schroeder}},\ }\bibfield
  {title} {\enquote {\bibinfo {title} {{Chiral symmetry breaking in fundamental
  and sextet fermion representations of SU(3) color}},}\ }\href@noop {} {\
  (\bibinfo {year} {2011}{\natexlab{a}})},\ \Eprint
  {http://arxiv.org/abs/1103.5998} {arXiv:1103.5998 [hep-lat]} \BibitemShut
  {NoStop}%
%%CITATION = ARXIV:1103.5998;%%
\bibitem [{\citenamefont {Fodor}\ \emph
  {et~al.}(2011{\natexlab{b}})\citenamefont {Fodor}, \citenamefont {Holland},
  \citenamefont {Kuti}, \citenamefont {Nogradi},\ and\ \citenamefont
  {Schroeder}}]{Fodor:2011tu}%
  \BibitemOpen
  \bibfield  {author} {\bibinfo {author} {\bibfnamefont {Zoltan}\ \bibnamefont
  {Fodor}}, \bibinfo {author} {\bibfnamefont {Kieran}\ \bibnamefont {Holland}},
  \bibinfo {author} {\bibfnamefont {Julius}\ \bibnamefont {Kuti}}, \bibinfo
  {author} {\bibfnamefont {Daniel}\ \bibnamefont {Nogradi}}, \ and\ \bibinfo
  {author} {\bibfnamefont {Chris}\ \bibnamefont {Schroeder}},\ }\bibfield
  {title} {\enquote {\bibinfo {title} {{Twelve massless flavors and three
  colors below the conformal window}},}\ }\href {\doibase
  10.1016/j.physletb.2011.07.037} {\bibfield  {journal} {\bibinfo  {journal}
  {Phys. Lett.}\ }\textbf {\bibinfo {volume} {B703}},\ \bibinfo {pages}
  {348--358} (\bibinfo {year} {2011}{\natexlab{b}})},\ \Eprint
  {http://arxiv.org/abs/1104.3124} {arXiv:1104.3124} \BibitemShut {NoStop}%
\bibitem [{\citenamefont {Fodor}\ \emph
  {et~al.}(2012{\natexlab{a}})\citenamefont {Fodor}, \citenamefont {Holland},
  \citenamefont {Kuti}, \citenamefont {Nogradi}, \citenamefont {Schroeder},\
  and\ \citenamefont {Wong}}]{Fodor:2012et}%
  \BibitemOpen
  \bibfield  {author} {\bibinfo {author} {\bibfnamefont {Zoltan}\ \bibnamefont
  {Fodor}}, \bibinfo {author} {\bibfnamefont {Kieran}\ \bibnamefont {Holland}},
  \bibinfo {author} {\bibfnamefont {Julius}\ \bibnamefont {Kuti}}, \bibinfo
  {author} {\bibfnamefont {Daniel}\ \bibnamefont {Nogradi}}, \bibinfo {author}
  {\bibfnamefont {Chris}\ \bibnamefont {Schroeder}}, \ and\ \bibinfo {author}
  {\bibfnamefont {Chik~Him}\ \bibnamefont {Wong}},\ }\bibfield  {title}
  {\enquote {\bibinfo {title} {{Conformal finite size scaling of twelve fermion
  flavors}},}\ }\href@noop {} {\  (\bibinfo {year} {2012}{\natexlab{a}})},\
  \Eprint {http://arxiv.org/abs/1211.4238} {arXiv:1211.4238} \BibitemShut
  {NoStop}%
\bibitem [{\citenamefont {Lin}\ \emph {et~al.}(2012)\citenamefont {Lin},
  \citenamefont {Ogawa}, \citenamefont {Ohki},\ and\ \citenamefont
  {Shintani}}]{Lin:2012iw}%
  \BibitemOpen
  \bibfield  {author} {\bibinfo {author} {\bibfnamefont {C.-J.~David}\
  \bibnamefont {Lin}}, \bibinfo {author} {\bibfnamefont {Kenji}\ \bibnamefont
  {Ogawa}}, \bibinfo {author} {\bibfnamefont {Hiroshi}\ \bibnamefont {Ohki}}, \
  and\ \bibinfo {author} {\bibfnamefont {Eigo}\ \bibnamefont {Shintani}},\
  }\bibfield  {title} {\enquote {\bibinfo {title} {{Lattice study of infrared
  behaviour in SU(3) gauge theory with twelve massless flavours}},}\ }\href
  {\doibase 10.1007/JHEP08(2012)096} {\bibfield  {journal} {\bibinfo  {journal}
  {JHEP}\ }\textbf {\bibinfo {volume} {1208}},\ \bibinfo {pages} {096}
  (\bibinfo {year} {2012})},\ \Eprint {http://arxiv.org/abs/1205.6076}
  {arXiv:1205.6076} \BibitemShut {NoStop}%
\bibitem [{\citenamefont {Deuzeman}\ \emph {et~al.}(2010)\citenamefont
  {Deuzeman}, \citenamefont {Lombardo},\ and\ \citenamefont
  {Pallante}}]{Deuzeman:2009mh}%
  \BibitemOpen
  \bibfield  {author} {\bibinfo {author} {\bibfnamefont {A.}~\bibnamefont
  {Deuzeman}}, \bibinfo {author} {\bibfnamefont {M.P.}\ \bibnamefont
  {Lombardo}}, \ and\ \bibinfo {author} {\bibfnamefont {E.}~\bibnamefont
  {Pallante}},\ }\bibfield  {title} {\enquote {\bibinfo {title} {{Evidence for
  a conformal phase in SU(N) gauge theories}},}\ }\href {\doibase
  10.1103/PhysRevD.82.074503} {\bibfield  {journal} {\bibinfo  {journal}
  {Phys.Rev.}\ }\textbf {\bibinfo {volume} {D82}},\ \bibinfo {pages} {074503}
  (\bibinfo {year} {2010})},\ \Eprint {http://arxiv.org/abs/0904.4662}
  {arXiv:0904.4662 [hep-ph]} \BibitemShut {NoStop}%
%%CITATION = ARXIV:0904.4662;%%
\bibitem [{\citenamefont {Deuzeman}\ \emph {et~al.}(2011)\citenamefont
  {Deuzeman}, \citenamefont {Lombardo}, \citenamefont {da~Silva},\ and\
  \citenamefont {Pallante}}]{Deuzeman:2011pa}%
  \BibitemOpen
  \bibfield  {author} {\bibinfo {author} {\bibfnamefont {Albert}\ \bibnamefont
  {Deuzeman}}, \bibinfo {author} {\bibfnamefont {Maria~Paola}\ \bibnamefont
  {Lombardo}}, \bibinfo {author} {\bibfnamefont {Tiago~Nunes}\ \bibnamefont
  {da~Silva}}, \ and\ \bibinfo {author} {\bibfnamefont {Elisabetta}\
  \bibnamefont {Pallante}},\ }\bibfield  {title} {\enquote {\bibinfo {title}
  {{Bulk transitions of twelve flavor QCD and $U_A(1)$ symmetry}},}\
  }\href@noop {} {\bibfield  {journal} {\bibinfo  {journal} {PoS}\ }\textbf
  {\bibinfo {volume} {LATTICE2011}},\ \bibinfo {pages} {321} (\bibinfo {year}
  {2011})},\ \Eprint {http://arxiv.org/abs/1111.2590} {arXiv:1111.2590
  [hep-lat]} \BibitemShut {NoStop}%
%%CITATION = ARXIV:1111.2590;%%
\bibitem [{\citenamefont {Hasenfratz}(2009)}]{Hasenfratz:2009ea}%
  \BibitemOpen
  \bibfield  {author} {\bibinfo {author} {\bibfnamefont {Anna}\ \bibnamefont
  {Hasenfratz}},\ }\bibfield  {title} {\enquote {\bibinfo {title}
  {{Investigating the critical properties of beyond-QCD theories using Monte
  Carlo Renormalization Group matching}},}\ }\href {\doibase
  10.1103/PhysRevD.80.034505} {\bibfield  {journal} {\bibinfo  {journal}
  {Phys.Rev.}\ }\textbf {\bibinfo {volume} {D80}},\ \bibinfo {pages} {034505}
  (\bibinfo {year} {2009})},\ \Eprint {http://arxiv.org/abs/0907.0919}
  {arXiv:0907.0919 [hep-lat]} \BibitemShut {NoStop}%
%%CITATION = ARXIV:0907.0919;%%
\bibitem [{\citenamefont {Hasenfratz}(2010)}]{Hasenfratz:2010fi}%
  \BibitemOpen
  \bibfield  {author} {\bibinfo {author} {\bibfnamefont {Anna}\ \bibnamefont
  {Hasenfratz}},\ }\bibfield  {title} {\enquote {\bibinfo {title} {{Conformal
  or Walking? Monte Carlo renormalization group studies of SU(3) gauge models
  with fundamental fermions}},}\ }\href {\doibase 10.1103/PhysRevD.82.014506}
  {\bibfield  {journal} {\bibinfo  {journal} {Phys.Rev.}\ }\textbf {\bibinfo
  {volume} {D82}},\ \bibinfo {pages} {014506} (\bibinfo {year} {2010})},\
  \Eprint {http://arxiv.org/abs/1004.1004} {arXiv:1004.1004 [hep-lat]}
  \BibitemShut {NoStop}%
%%CITATION = ARXIV:1004.1004;%%
\bibitem [{\citenamefont {Cheng}\ \emph {et~al.}(2012)\citenamefont {Cheng},
  \citenamefont {Hasenfratz},\ and\ \citenamefont {Schaich}}]{Cheng:2011ic}%
  \BibitemOpen
  \bibfield  {author} {\bibinfo {author} {\bibfnamefont {Anqi}\ \bibnamefont
  {Cheng}}, \bibinfo {author} {\bibfnamefont {Anna}\ \bibnamefont
  {Hasenfratz}}, \ and\ \bibinfo {author} {\bibfnamefont {David}\ \bibnamefont
  {Schaich}},\ }\bibfield  {title} {\enquote {\bibinfo {title} {{Novel phase in
  SU(3) lattice gauge theory with 12 light fermions}},}\ }\href {\doibase
  10.1103/PhysRevD.85.094509} {\bibfield  {journal} {\bibinfo  {journal} {Phys.
  Rev.}\ }\textbf {\bibinfo {volume} {D85}},\ \bibinfo {pages} {094509}
  (\bibinfo {year} {2012})},\ \Eprint {http://arxiv.org/abs/1111.2317}
  {arXiv:1111.2317} \BibitemShut {NoStop}%
\bibitem [{\citenamefont {Jin}\ and\ \citenamefont
  {Mawhinney}(2009)}]{Jin:2009mc}%
  \BibitemOpen
  \bibfield  {author} {\bibinfo {author} {\bibfnamefont {Xiao-Yong}\
  \bibnamefont {Jin}}\ and\ \bibinfo {author} {\bibfnamefont {Robert~D.}\
  \bibnamefont {Mawhinney}},\ }\bibfield  {title} {\enquote {\bibinfo {title}
  {{Lattice QCD with 8 and 12 degenerate quark flavors}},}\ }\href@noop {}
  {\bibfield  {journal} {\bibinfo  {journal} {PoS}\ }\textbf {\bibinfo {volume}
  {LAT2009}},\ \bibinfo {pages} {049} (\bibinfo {year} {2009})},\ \Eprint
  {http://arxiv.org/abs/0910.3216} {arXiv:0910.3216 [hep-lat]} \BibitemShut
  {NoStop}%
%%CITATION = ARXIV:0910.3216;%%
\bibitem [{\citenamefont {Jin}\ and\ \citenamefont
  {Mawhinney}(2011)}]{Jin:2012dw}%
  \BibitemOpen
  \bibfield  {author} {\bibinfo {author} {\bibfnamefont {Xiao-Yong}\
  \bibnamefont {Jin}}\ and\ \bibinfo {author} {\bibfnamefont {Robert~D.}\
  \bibnamefont {Mawhinney}},\ }\bibfield  {title} {\enquote {\bibinfo {title}
  {{Lattice QCD with 12 Degenerate Quark Flavors}},}\ }\href@noop {} {\bibfield
   {journal} {\bibinfo  {journal} {PoS}\ }\textbf {\bibinfo {volume}
  {LATTICE2011}},\ \bibinfo {pages} {066} (\bibinfo {year} {2011})},\ \Eprint
  {http://arxiv.org/abs/1203.5855} {arXiv:1203.5855 [hep-lat]} \BibitemShut
  {NoStop}%
%%CITATION = ARXIV:1203.5855;%%
\bibitem [{\citenamefont {Aoki}\ \emph {et~al.}(2012)\citenamefont {Aoki},
  \citenamefont {Aoyama}, \citenamefont {Kurachi}, \citenamefont {Maskawa},
  \citenamefont {Nagai} \emph {et~al.}}]{Aoki:2012eq}%
  \BibitemOpen
  \bibfield  {author} {\bibinfo {author} {\bibfnamefont {Yasumichi}\
  \bibnamefont {Aoki}}, \bibinfo {author} {\bibfnamefont {Tatsumi}\
  \bibnamefont {Aoyama}}, \bibinfo {author} {\bibfnamefont {Masafumi}\
  \bibnamefont {Kurachi}}, \bibinfo {author} {\bibfnamefont {Toshihide}\
  \bibnamefont {Maskawa}}, \bibinfo {author} {\bibfnamefont {Kei-ichi}\
  \bibnamefont {Nagai}},  \emph {et~al.},\ }\bibfield  {title} {\enquote
  {\bibinfo {title} {{Lattice study of conformality in twelve-flavor QCD}},}\
  }\href {\doibase 10.1103/PhysRevD.86.059903, 10.1103/PhysRevD.86.054506}
  {\bibfield  {journal} {\bibinfo  {journal} {Phys.Rev.}\ }\textbf {\bibinfo
  {volume} {D86}},\ \bibinfo {pages} {054506} (\bibinfo {year} {2012})},\
  \Eprint {http://arxiv.org/abs/1207.3060} {arXiv:1207.3060 [hep-lat]}
  \BibitemShut {NoStop}%
%%CITATION = ARXIV:1207.3060;%%
\bibitem [{\citenamefont {Fodor}\ \emph
  {et~al.}(2012{\natexlab{b}})\citenamefont {Fodor}, \citenamefont {Holland},
  \citenamefont {Kuti}, \citenamefont {Nogradi}, \citenamefont {Schroeder}
  \emph {et~al.}}]{Fodor:2012ty}%
  \BibitemOpen
  \bibfield  {author} {\bibinfo {author} {\bibfnamefont {Zoltan}\ \bibnamefont
  {Fodor}}, \bibinfo {author} {\bibfnamefont {Kieran}\ \bibnamefont {Holland}},
  \bibinfo {author} {\bibfnamefont {Julius}\ \bibnamefont {Kuti}}, \bibinfo
  {author} {\bibfnamefont {Daniel}\ \bibnamefont {Nogradi}}, \bibinfo {author}
  {\bibfnamefont {Chris}\ \bibnamefont {Schroeder}},  \emph {et~al.},\
  }\bibfield  {title} {\enquote {\bibinfo {title} {{Can the nearly conformal
  sextet gauge model hide the Higgs impostor?}}}\ }\href {\doibase
  10.1016/j.physletb.2012.10.079} {\bibfield  {journal} {\bibinfo  {journal}
  {Phys.Lett.}\ }\textbf {\bibinfo {volume} {B718}},\ \bibinfo {pages}
  {657--666} (\bibinfo {year} {2012}{\natexlab{b}})},\ \Eprint
  {http://arxiv.org/abs/1209.0391} {arXiv:1209.0391 [hep-lat]} \BibitemShut
  {NoStop}%
%%CITATION = ARXIV:1209.0391;%%
\bibitem [{\citenamefont {Kogut}\ and\ \citenamefont
  {Sinclair}(2010)}]{Kogut:2010cz}%
  \BibitemOpen
  \bibfield  {author} {\bibinfo {author} {\bibfnamefont {J.B.}\ \bibnamefont
  {Kogut}}\ and\ \bibinfo {author} {\bibfnamefont {D.K.}\ \bibnamefont
  {Sinclair}},\ }\bibfield  {title} {\enquote {\bibinfo {title}
  {{Thermodynamics of lattice QCD with 2 flavours of colour-sextet quarks: A
  model of walking/conformal Technicolor}},}\ }\href {\doibase
  10.1103/PhysRevD.81.114507} {\bibfield  {journal} {\bibinfo  {journal}
  {Phys.Rev.}\ }\textbf {\bibinfo {volume} {D81}},\ \bibinfo {pages} {114507}
  (\bibinfo {year} {2010})},\ \Eprint {http://arxiv.org/abs/1002.2988}
  {arXiv:1002.2988 [hep-lat]} \BibitemShut {NoStop}%
%%CITATION = ARXIV:1002.2988;%%
\bibitem [{\citenamefont {Kogut}\ and\ \citenamefont
  {Sinclair}(2011)}]{Kogut:2011ty}%
  \BibitemOpen
  \bibfield  {author} {\bibinfo {author} {\bibfnamefont {J.B.}\ \bibnamefont
  {Kogut}}\ and\ \bibinfo {author} {\bibfnamefont {D.K.}\ \bibnamefont
  {Sinclair}},\ }\bibfield  {title} {\enquote {\bibinfo {title}
  {{Thermodynamics of lattice QCD with 2 sextet quarks on $N_t$=8 lattices}},}\
  }\href {\doibase 10.1103/PhysRevD.84.074504} {\bibfield  {journal} {\bibinfo
  {journal} {Phys.Rev.}\ }\textbf {\bibinfo {volume} {D84}},\ \bibinfo {pages}
  {074504} (\bibinfo {year} {2011})},\ \Eprint {http://arxiv.org/abs/1105.3749}
  {arXiv:1105.3749 [hep-lat]} \BibitemShut {NoStop}%
%%CITATION = ARXIV:1105.3749;%%
\bibitem [{\citenamefont {Fodor}\ \emph {et~al.}(2013)\citenamefont {Fodor},
  \citenamefont {Holland}, \citenamefont {Kuti}, \citenamefont {Nogradi},\ and\
  \citenamefont {Wong}}]{Fodor:2013xxx}%
  \BibitemOpen
  \bibfield  {author} {\bibinfo {author} {\bibfnamefont {Zoltan}\ \bibnamefont
  {Fodor}}, \bibinfo {author} {\bibfnamefont {Kieran}\ \bibnamefont {Holland}},
  \bibinfo {author} {\bibfnamefont {Julius}\ \bibnamefont {Kuti}}, \bibinfo
  {author} {\bibfnamefont {Daniel}\ \bibnamefont {Nogradi}}, \ and\ \bibinfo
  {author} {\bibfnamefont {Chik~Him}\ \bibnamefont {Wong}},\ }\bibfield
  {title} {\enquote {\bibinfo {title} {{In preparation for publication}},}\
  }\href@noop {} {\  (\bibinfo {year} {2013})}\BibitemShut {NoStop}%
\bibitem [{\citenamefont {Rinaldi}(2012)}]{Rinaldi:2012xxx}%
  \BibitemOpen
  \bibfield  {author} {\bibinfo {author} {\bibfnamefont {Enrico}\ \bibnamefont
  {Rinaldi}},\ }\bibfield  {title} {\enquote {\bibinfo {title} {{Talk,
  presented at SCGT 2012 on behalf of the LatKMI collaboration}},}\ }\href@noop
  {} {\  (\bibinfo {year} {2012})}\BibitemShut {NoStop}%
\bibitem [{\citenamefont {DeGrand}\ \emph {et~al.}(2012)\citenamefont
  {DeGrand}, \citenamefont {Shamir},\ and\ \citenamefont
  {Svetitsky}}]{DeGrand:2012yq}%
  \BibitemOpen
  \bibfield  {author} {\bibinfo {author} {\bibfnamefont {Thomas}\ \bibnamefont
  {DeGrand}}, \bibinfo {author} {\bibfnamefont {Yigal}\ \bibnamefont {Shamir}},
  \ and\ \bibinfo {author} {\bibfnamefont {Benjamin}\ \bibnamefont
  {Svetitsky}},\ }\bibfield  {title} {\enquote {\bibinfo {title} {{Mass
  anomalous dimension in sextet QCD}},}\ }\href@noop {} {\  (\bibinfo {year}
  {2012})},\ \Eprint {http://arxiv.org/abs/1201.0935} {arXiv:1201.0935
  [hep-lat]} \BibitemShut {NoStop}%
%%CITATION = ARXIV:1201.0935;%%
\bibitem [{\citenamefont {Luscher}(2010)}]{Luscher:2010iy}%
  \BibitemOpen
  \bibfield  {author} {\bibinfo {author} {\bibfnamefont {Martin}\ \bibnamefont
  {Luscher}},\ }\bibfield  {title} {\enquote {\bibinfo {title} {{Properties and
  uses of the Wilson flow in lattice QCD}},}\ }\href {\doibase
  10.1007/JHEP08(2010)071} {\bibfield  {journal} {\bibinfo  {journal} {JHEP}\
  }\textbf {\bibinfo {volume} {1008}},\ \bibinfo {pages} {071} (\bibinfo {year}
  {2010})},\ \Eprint {http://arxiv.org/abs/1006.4518} {arXiv:1006.4518
  [hep-lat]} \BibitemShut {NoStop}%
%%CITATION = ARXIV:1006.4518;%%
\bibitem [{\citenamefont {Luscher}\ and\ \citenamefont
  {Weisz}(2011)}]{Luscher:2011bx}%
  \BibitemOpen
  \bibfield  {author} {\bibinfo {author} {\bibfnamefont {Martin}\ \bibnamefont
  {Luscher}}\ and\ \bibinfo {author} {\bibfnamefont {Peter}\ \bibnamefont
  {Weisz}},\ }\bibfield  {title} {\enquote {\bibinfo {title} {{Perturbative
  analysis of the gradient flow in non-abelian gauge theories}},}\ }\href
  {\doibase 10.1007/JHEP02(2011)051} {\bibfield  {journal} {\bibinfo  {journal}
  {JHEP}\ }\textbf {\bibinfo {volume} {1102}},\ \bibinfo {pages} {051}
  (\bibinfo {year} {2011})},\ \Eprint {http://arxiv.org/abs/1101.0963}
  {arXiv:1101.0963 [hep-th]} \BibitemShut {NoStop}%
%%CITATION = ARXIV:1101.0963;%%
\bibitem [{\citenamefont {Peskin}(1980)}]{Peskin:1980gc}%
  \BibitemOpen
  \bibfield  {author} {\bibinfo {author} {\bibfnamefont {Michael~E.}\
  \bibnamefont {Peskin}},\ }\bibfield  {title} {\enquote {\bibinfo {title}
  {{The Alignment of the Vacuum in Theories of Technicolor}},}\ }\href
  {\doibase 10.1016/0550-3213(80)90051-6} {\bibfield  {journal} {\bibinfo
  {journal} {Nucl.Phys.}\ }\textbf {\bibinfo {volume} {B175}},\ \bibinfo
  {pages} {197--233} (\bibinfo {year} {1980})}\BibitemShut {NoStop}%
%%CITATION = NUPHA,B175,197;%%
\bibitem [{\citenamefont {Schmaltz}\ and\ \citenamefont
  {Tucker-Smith}(2005)}]{Schmaltz:2005ky}%
  \BibitemOpen
  \bibfield  {author} {\bibinfo {author} {\bibfnamefont {Martin}\ \bibnamefont
  {Schmaltz}}\ and\ \bibinfo {author} {\bibfnamefont {David}\ \bibnamefont
  {Tucker-Smith}},\ }\bibfield  {title} {\enquote {\bibinfo {title} {{Little
  Higgs review}},}\ }\href {\doibase 10.1146/annurev.nucl.55.090704.151502}
  {\bibfield  {journal} {\bibinfo  {journal} {Ann.Rev.Nucl.Part.Sci.}\ }\textbf
  {\bibinfo {volume} {55}},\ \bibinfo {pages} {229--270} (\bibinfo {year}
  {2005})},\ \Eprint {http://arxiv.org/abs/hep-ph/0502182}
  {arXiv:hep-ph/0502182 [hep-ph]} \BibitemShut {NoStop}%
%%CITATION = HEP-PH/0502182;%%
\bibitem [{\citenamefont {Katz}\ \emph {et~al.}(2005)\citenamefont {Katz},
  \citenamefont {Lee}, \citenamefont {Nelson},\ and\ \citenamefont
  {Walker}}]{Katz:2003sn}%
  \BibitemOpen
  \bibfield  {author} {\bibinfo {author} {\bibfnamefont {Emanuel}\ \bibnamefont
  {Katz}}, \bibinfo {author} {\bibfnamefont {Jae-yong}\ \bibnamefont {Lee}},
  \bibinfo {author} {\bibfnamefont {Ann~E.}\ \bibnamefont {Nelson}}, \ and\
  \bibinfo {author} {\bibfnamefont {Devin~G.E.}\ \bibnamefont {Walker}},\
  }\bibfield  {title} {\enquote {\bibinfo {title} {{A Composite little Higgs
  model}},}\ }\href {\doibase 10.1088/1126-6708/2005/10/088} {\bibfield
  {journal} {\bibinfo  {journal} {JHEP}\ }\textbf {\bibinfo {volume} {0510}},\
  \bibinfo {pages} {088} (\bibinfo {year} {2005})},\ \Eprint
  {http://arxiv.org/abs/hep-ph/0312287} {arXiv:hep-ph/0312287 [hep-ph]}
  \BibitemShut {NoStop}%
%%CITATION = HEP-PH/0312287;%%
\bibitem [{\citenamefont {Galloway}\ \emph {et~al.}(2010)\citenamefont
  {Galloway}, \citenamefont {Evans}, \citenamefont {Luty},\ and\ \citenamefont
  {Tacchi}}]{Galloway:2010bp}%
  \BibitemOpen
  \bibfield  {author} {\bibinfo {author} {\bibfnamefont {Jamison}\ \bibnamefont
  {Galloway}}, \bibinfo {author} {\bibfnamefont {Jared~A.}\ \bibnamefont
  {Evans}}, \bibinfo {author} {\bibfnamefont {Markus~A.}\ \bibnamefont {Luty}},
  \ and\ \bibinfo {author} {\bibfnamefont {Ruggero~Altair}\ \bibnamefont
  {Tacchi}},\ }\bibfield  {title} {\enquote {\bibinfo {title} {{Minimal
  Conformal Technicolor and Precision Electroweak Tests}},}\ }\href {\doibase
  10.1007/JHEP10(2010)086} {\bibfield  {journal} {\bibinfo  {journal} {JHEP}\
  }\textbf {\bibinfo {volume} {1010}},\ \bibinfo {pages} {086} (\bibinfo {year}
  {2010})},\ \Eprint {http://arxiv.org/abs/1001.1361} {arXiv:1001.1361
  [hep-ph]} \BibitemShut {NoStop}%
%%CITATION = ARXIV:1001.1361;%%
\bibitem [{\citenamefont {Kogut}\ \emph {et~al.}(2000)\citenamefont {Kogut},
  \citenamefont {Stephanov}, \citenamefont {Toublan}, \citenamefont
  {Verbaarschot},\ and\ \citenamefont {Zhitnitsky}}]{Kogut:2000ek}%
  \BibitemOpen
  \bibfield  {author} {\bibinfo {author} {\bibfnamefont {J.B.}\ \bibnamefont
  {Kogut}}, \bibinfo {author} {\bibfnamefont {Misha~A.}\ \bibnamefont
  {Stephanov}}, \bibinfo {author} {\bibfnamefont {D.}~\bibnamefont {Toublan}},
  \bibinfo {author} {\bibfnamefont {J.J.M.}\ \bibnamefont {Verbaarschot}}, \
  and\ \bibinfo {author} {\bibfnamefont {A.}~\bibnamefont {Zhitnitsky}},\
  }\bibfield  {title} {\enquote {\bibinfo {title} {{QCD - like theories at
  finite baryon density}},}\ }\href {\doibase 10.1016/S0550-3213(00)00242-X}
  {\bibfield  {journal} {\bibinfo  {journal} {Nucl.Phys.}\ }\textbf {\bibinfo
  {volume} {B582}},\ \bibinfo {pages} {477--513} (\bibinfo {year} {2000})},\
  \Eprint {http://arxiv.org/abs/hep-ph/0001171} {arXiv:hep-ph/0001171 [hep-ph]}
  \BibitemShut {NoStop}%
%%CITATION = HEP-PH/0001171;%%
\bibitem [{\citenamefont {Lewis}\ \emph {et~al.}(2012)\citenamefont {Lewis},
  \citenamefont {Pica},\ and\ \citenamefont {Sannino}}]{Lewis:2011zb}%
  \BibitemOpen
  \bibfield  {author} {\bibinfo {author} {\bibfnamefont {Randy}\ \bibnamefont
  {Lewis}}, \bibinfo {author} {\bibfnamefont {Claudio}\ \bibnamefont {Pica}}, \
  and\ \bibinfo {author} {\bibfnamefont {Francesco}\ \bibnamefont {Sannino}},\
  }\bibfield  {title} {\enquote {\bibinfo {title} {{Light Asymmetric Dark
  Matter on the Lattice: SU(2) Technicolor with Two Fundamental Flavors}},}\
  }\href {\doibase 10.1103/PhysRevD.85.014504} {\bibfield  {journal} {\bibinfo
  {journal} {Phys.Rev.}\ }\textbf {\bibinfo {volume} {D85}},\ \bibinfo {pages}
  {014504} (\bibinfo {year} {2012})},\ \Eprint {http://arxiv.org/abs/1109.3513}
  {arXiv:1109.3513 [hep-ph]} \BibitemShut {NoStop}%
%%CITATION = ARXIV:1109.3513;%%
\bibitem [{\citenamefont {Kogut}\ \emph {et~al.}(1999)\citenamefont {Kogut},
  \citenamefont {Stephanov},\ and\ \citenamefont {Toublan}}]{Kogut:1999iv}%
  \BibitemOpen
  \bibfield  {author} {\bibinfo {author} {\bibfnamefont {J.B.}\ \bibnamefont
  {Kogut}}, \bibinfo {author} {\bibfnamefont {Misha~A.}\ \bibnamefont
  {Stephanov}}, \ and\ \bibinfo {author} {\bibfnamefont {D.}~\bibnamefont
  {Toublan}},\ }\bibfield  {title} {\enquote {\bibinfo {title} {{On two color
  QCD with baryon chemical potential}},}\ }\href {\doibase
  10.1016/S0370-2693(99)00971-5} {\bibfield  {journal} {\bibinfo  {journal}
  {Phys.Lett.}\ }\textbf {\bibinfo {volume} {B464}},\ \bibinfo {pages}
  {183--191} (\bibinfo {year} {1999})},\ \Eprint
  {http://arxiv.org/abs/hep-ph/9906346} {arXiv:hep-ph/9906346 [hep-ph]}
  \BibitemShut {NoStop}%
%%CITATION = HEP-PH/9906346;%%
\bibitem [{\citenamefont {Hands}\ \emph {et~al.}(2000)\citenamefont {Hands},
  \citenamefont {Montvay}, \citenamefont {Morrison}, \citenamefont {Oevers},
  \citenamefont {Scorzato} \emph {et~al.}}]{Hands:2000ei}%
  \BibitemOpen
  \bibfield  {author} {\bibinfo {author} {\bibfnamefont {Simon}\ \bibnamefont
  {Hands}}, \bibinfo {author} {\bibfnamefont {Istvan}\ \bibnamefont {Montvay}},
  \bibinfo {author} {\bibfnamefont {Susan}\ \bibnamefont {Morrison}}, \bibinfo
  {author} {\bibfnamefont {Manfred}\ \bibnamefont {Oevers}}, \bibinfo {author}
  {\bibfnamefont {Luigi}\ \bibnamefont {Scorzato}},  \emph {et~al.},\
  }\bibfield  {title} {\enquote {\bibinfo {title} {{Numerical study of dense
  adjoint matter in two color QCD}},}\ }\href {\doibase 10.1007/s100520000477}
  {\bibfield  {journal} {\bibinfo  {journal} {Eur.Phys.J.}\ }\textbf {\bibinfo
  {volume} {C17}},\ \bibinfo {pages} {285--302} (\bibinfo {year} {2000})},\
  \Eprint {http://arxiv.org/abs/hep-lat/0006018} {arXiv:hep-lat/0006018
  [hep-lat]} \BibitemShut {NoStop}%
%%CITATION = HEP-LAT/0006018;%%
\bibitem [{\citenamefont {Hands}\ \emph {et~al.}(2008)\citenamefont {Hands},
  \citenamefont {Sitch},\ and\ \citenamefont {Skullerud}}]{Hands:2007uc}%
  \BibitemOpen
  \bibfield  {author} {\bibinfo {author} {\bibfnamefont {Simon}\ \bibnamefont
  {Hands}}, \bibinfo {author} {\bibfnamefont {Peter}\ \bibnamefont {Sitch}}, \
  and\ \bibinfo {author} {\bibfnamefont {Jon-Ivar}\ \bibnamefont {Skullerud}},\
  }\bibfield  {title} {\enquote {\bibinfo {title} {{Hadron Spectrum in a
  Two-Colour Baryon-Rich Medium}},}\ }\href {\doibase
  10.1016/j.physletb.2008.01.078} {\bibfield  {journal} {\bibinfo  {journal}
  {Phys.Lett.}\ }\textbf {\bibinfo {volume} {B662}},\ \bibinfo {pages}
  {405--412} (\bibinfo {year} {2008})},\ \Eprint
  {http://arxiv.org/abs/0710.1966} {arXiv:0710.1966 [hep-lat]} \BibitemShut
  {NoStop}%
%%CITATION = ARXIV:0710.1966;%%
\bibitem [{\citenamefont {Brower}\ \emph {et~al.}(1995)\citenamefont {Brower},
  \citenamefont {Orginos},\ and\ \citenamefont {Tan}}]{Brower:1995vf}%
  \BibitemOpen
  \bibfield  {author} {\bibinfo {author} {\bibfnamefont {R.C.}\ \bibnamefont
  {Brower}}, \bibinfo {author} {\bibfnamefont {K.}~\bibnamefont {Orginos}}, \
  and\ \bibinfo {author} {\bibfnamefont {C.I.}\ \bibnamefont {Tan}},\
  }\bibfield  {title} {\enquote {\bibinfo {title} {{The Chiral extension of
  lattice QCD}},}\ }\href {\doibase 10.1016/0920-5632(95)00185-C} {\bibfield
  {journal} {\bibinfo  {journal} {Nucl.Phys.Proc.Suppl.}\ }\textbf {\bibinfo
  {volume} {42}},\ \bibinfo {pages} {42--48} (\bibinfo {year} {1995})},\
  \Eprint {http://arxiv.org/abs/hep-lat/9501026} {arXiv:hep-lat/9501026
  [hep-lat]} \BibitemShut {NoStop}%
%%CITATION = HEP-LAT/9501026;%%
\bibitem [{\citenamefont {Catterall}\ and\ \citenamefont
  {Veernala}(2012)}]{Catterall:2012vu}%
  \BibitemOpen
  \bibfield  {author} {\bibinfo {author} {\bibfnamefont {Simon}\ \bibnamefont
  {Catterall}}\ and\ \bibinfo {author} {\bibfnamefont {Aarti}\ \bibnamefont
  {Veernala}},\ }\bibfield  {title} {\enquote {\bibinfo {title} {{Four fermion
  operators and the search for BSM Physics}},}\ }\href@noop {} {\  (\bibinfo
  {year} {2012})},\ \Eprint {http://arxiv.org/abs/1211.3622} {arXiv:1211.3622
  [hep-lat]} \BibitemShut {NoStop}%
%%CITATION = ARXIV:1211.3622;%%
\bibitem [{\citenamefont {Endres}(2009)}]{Endres:2009yp}%
  \BibitemOpen
  \bibfield  {author} {\bibinfo {author} {\bibfnamefont {Michael~G.}\
  \bibnamefont {Endres}},\ }\bibfield  {title} {\enquote {\bibinfo {title}
  {{Dynamical simulation of N=1 supersymmetric Yang-Mills theory with domain
  wall fermions}},}\ }\href {\doibase 10.1103/PhysRevD.79.094503} {\bibfield
  {journal} {\bibinfo  {journal} {Phys. Rev.}\ }\textbf {\bibinfo {volume}
  {D79}},\ \bibinfo {pages} {094503} (\bibinfo {year} {2009})},\ \Eprint
  {http://arxiv.org/abs/0902.4267} {arXiv:0902.4267 [hep-lat]} \BibitemShut
  {NoStop}%
%%CITATION = 0902.4267;%%
\bibitem [{\citenamefont {Catterall}\ \emph
  {et~al.}(2012{\natexlab{a}})\citenamefont {Catterall}, \citenamefont
  {Damgaard}, \citenamefont {Degrand}, \citenamefont {Galvez},\ and\
  \citenamefont {Mehta}}]{Catterall:2012yq}%
  \BibitemOpen
  \bibfield  {author} {\bibinfo {author} {\bibfnamefont {Simon}\ \bibnamefont
  {Catterall}}, \bibinfo {author} {\bibfnamefont {Poul~H.}\ \bibnamefont
  {Damgaard}}, \bibinfo {author} {\bibfnamefont {Thomas}\ \bibnamefont
  {Degrand}}, \bibinfo {author} {\bibfnamefont {Richard}\ \bibnamefont
  {Galvez}}, \ and\ \bibinfo {author} {\bibfnamefont {Dhagash}\ \bibnamefont
  {Mehta}},\ }\bibfield  {title} {\enquote {\bibinfo {title} {{Phase Structure
  of Lattice N=4 Super Yang-Mills}},}\ }\href {\doibase
  10.1007/JHEP11(2012)072} {\bibfield  {journal} {\bibinfo  {journal} {JHEP}\
  }\textbf {\bibinfo {volume} {11}},\ \bibinfo {pages} {072} (\bibinfo {year}
  {2012}{\natexlab{a}})},\ \Eprint {http://arxiv.org/abs/1209.5285}
  {arXiv:1209.5285 [hep-lat]} \BibitemShut {NoStop}%
%%CITATION = 1209.5285;%%
\bibitem [{\citenamefont {Intriligator}\ \emph {et~al.}(2006)\citenamefont
  {Intriligator}, \citenamefont {Seiberg},\ and\ \citenamefont
  {Shih}}]{Intriligator:2006dd}%
  \BibitemOpen
  \bibfield  {author} {\bibinfo {author} {\bibfnamefont {Kenneth~A.}\
  \bibnamefont {Intriligator}}, \bibinfo {author} {\bibfnamefont {Nathan}\
  \bibnamefont {Seiberg}}, \ and\ \bibinfo {author} {\bibfnamefont {David}\
  \bibnamefont {Shih}},\ }\bibfield  {title} {\enquote {\bibinfo {title}
  {{Dynamical SUSY breaking in meta-stable vacua}},}\ }\href {\doibase
  10.1088/1126-6708/2006/04/021} {\bibfield  {journal} {\bibinfo  {journal}
  {JHEP}\ }\textbf {\bibinfo {volume} {04}},\ \bibinfo {pages} {021} (\bibinfo
  {year} {2006})},\ \Eprint {http://arxiv.org/abs/hep-th/0602239}
  {arXiv:hep-th/0602239} \BibitemShut {NoStop}%
%%CITATION = HEP-TH/0602239;%%
\bibitem [{\citenamefont {Catterall}\ \emph {et~al.}(2009)\citenamefont
  {Catterall}, \citenamefont {Kaplan},\ and\ \citenamefont
  {Unsal}}]{Catterall:2009it}%
  \BibitemOpen
  \bibfield  {author} {\bibinfo {author} {\bibfnamefont {Simon}\ \bibnamefont
  {Catterall}}, \bibinfo {author} {\bibfnamefont {David~B.}\ \bibnamefont
  {Kaplan}}, \ and\ \bibinfo {author} {\bibfnamefont {Mithat}\ \bibnamefont
  {Unsal}},\ }\bibfield  {title} {\enquote {\bibinfo {title} {{Exact lattice
  supersymmetry}},}\ }\href {\doibase 10.1016/j.physrep.2009.09.001} {\bibfield
   {journal} {\bibinfo  {journal} {Phys. Rept.}\ }\textbf {\bibinfo {volume}
  {484}},\ \bibinfo {pages} {71--130} (\bibinfo {year} {2009})},\ \Eprint
  {http://arxiv.org/abs/0903.4881} {arXiv:0903.4881 [hep-lat]} \BibitemShut
  {NoStop}%
%%CITATION = 0903.4881;%%
\bibitem [{\citenamefont {Catterall}(2010{\natexlab{a}})}]{Catterall:2010jh}%
  \BibitemOpen
  \bibfield  {author} {\bibinfo {author} {\bibfnamefont {Simon}\ \bibnamefont
  {Catterall}},\ }\bibfield  {title} {\enquote {\bibinfo {title} {{Twisted
  lattice supersymmetry and applications to AdS/CFT}},}\ }\href@noop {}
  {\bibfield  {journal} {\bibinfo  {journal} {PoS}\ }\textbf {\bibinfo {volume}
  {LATTICE2010}},\ \bibinfo {pages} {002} (\bibinfo {year}
  {2010}{\natexlab{a}})},\ \Eprint {http://arxiv.org/abs/1010.6224}
  {arXiv:1010.6224 [hep-lat]} \BibitemShut {NoStop}%
%%CITATION = 1010.6224;%%
\bibitem [{\citenamefont {Catterall}\ \emph {et~al.}(2010)\citenamefont
  {Catterall}, \citenamefont {Joseph},\ and\ \citenamefont
  {Wiseman}}]{Catterall:2010fx}%
  \BibitemOpen
  \bibfield  {author} {\bibinfo {author} {\bibfnamefont {Simon}\ \bibnamefont
  {Catterall}}, \bibinfo {author} {\bibfnamefont {Anosh}\ \bibnamefont
  {Joseph}}, \ and\ \bibinfo {author} {\bibfnamefont {Toby}\ \bibnamefont
  {Wiseman}},\ }\bibfield  {title} {\enquote {\bibinfo {title} {{Thermal phases
  of D1-branes on a circle from lattice super Yang-Mills}},}\ }\href {\doibase
  10.1007/JHEP12(2010)022} {\bibfield  {journal} {\bibinfo  {journal} {JHEP}\
  }\textbf {\bibinfo {volume} {12}},\ \bibinfo {pages} {022} (\bibinfo {year}
  {2010})},\ \Eprint {http://arxiv.org/abs/1008.4964} {arXiv:1008.4964
  [hep-th]} \BibitemShut {NoStop}%
%%CITATION = 1008.4964;%%
\bibitem [{\citenamefont {Catterall}(2010{\natexlab{b}})}]{Catterall:2010ng}%
  \BibitemOpen
  \bibfield  {author} {\bibinfo {author} {\bibfnamefont {Simon}\ \bibnamefont
  {Catterall}},\ }\bibfield  {title} {\enquote {\bibinfo {title} {{Topological
  gravity on the lattice}},}\ }\href {\doibase 10.1007/JHEP07(2010)066}
  {\bibfield  {journal} {\bibinfo  {journal} {JHEP}\ }\textbf {\bibinfo
  {volume} {07}},\ \bibinfo {pages} {066} (\bibinfo {year}
  {2010}{\natexlab{b}})},\ \Eprint {http://arxiv.org/abs/1003.5202}
  {arXiv:1003.5202 [hep-lat]} \BibitemShut {NoStop}%
%%CITATION = 1003.5202;%%
\bibitem [{\citenamefont {Catterall}\ and\ \citenamefont {van
  Anders}(2010)}]{Catterall:2010gf}%
  \BibitemOpen
  \bibfield  {author} {\bibinfo {author} {\bibfnamefont {Simon}\ \bibnamefont
  {Catterall}}\ and\ \bibinfo {author} {\bibfnamefont {Greg}\ \bibnamefont {van
  Anders}},\ }\bibfield  {title} {\enquote {\bibinfo {title} {{First Results
  from Lattice Simulation of the PWMM}},}\ }\href {\doibase
  10.1007/JHEP09(2010)088} {\bibfield  {journal} {\bibinfo  {journal} {JHEP}\
  }\textbf {\bibinfo {volume} {09}},\ \bibinfo {pages} {088} (\bibinfo {year}
  {2010})},\ \Eprint {http://arxiv.org/abs/1003.4952} {arXiv:1003.4952
  [hep-th]} \BibitemShut {NoStop}%
%%CITATION = 1003.4952;%%
\bibitem [{\citenamefont {Catterall}\ and\ \citenamefont
  {Wiseman}(2010)}]{Catterall:2009xn}%
  \BibitemOpen
  \bibfield  {author} {\bibinfo {author} {\bibfnamefont {Simon}\ \bibnamefont
  {Catterall}}\ and\ \bibinfo {author} {\bibfnamefont {Toby}\ \bibnamefont
  {Wiseman}},\ }\bibfield  {title} {\enquote {\bibinfo {title} {{Extracting
  black hole physics from the lattice}},}\ }\href {\doibase
  10.1007/JHEP04(2010)077} {\bibfield  {journal} {\bibinfo  {journal} {JHEP}\
  }\textbf {\bibinfo {volume} {04}},\ \bibinfo {pages} {077} (\bibinfo {year}
  {2010})},\ \Eprint {http://arxiv.org/abs/0909.4947} {arXiv:0909.4947
  [hep-th]} \BibitemShut {NoStop}%
%%CITATION = 0909.4947;%%
\bibitem [{\citenamefont {Catterall}\ and\ \citenamefont
  {Wiseman}(2008)}]{Catterall:2008yz}%
  \BibitemOpen
  \bibfield  {author} {\bibinfo {author} {\bibfnamefont {Simon}\ \bibnamefont
  {Catterall}}\ and\ \bibinfo {author} {\bibfnamefont {Toby}\ \bibnamefont
  {Wiseman}},\ }\bibfield  {title} {\enquote {\bibinfo {title} {{Black hole
  thermodynamics from simulations of lattice Yang-Mills theory}},}\ }\href
  {\doibase 10.1103/PhysRevD.78.041502} {\bibfield  {journal} {\bibinfo
  {journal} {Phys. Rev.}\ }\textbf {\bibinfo {volume} {D78}},\ \bibinfo {pages}
  {041502} (\bibinfo {year} {2008})},\ \Eprint {http://arxiv.org/abs/0803.4273}
  {arXiv:0803.4273 [hep-th]} \BibitemShut {NoStop}%
%%CITATION = 0803.4273;%%
\bibitem [{\citenamefont {Catterall}\ \emph {et~al.}(2011)\citenamefont
  {Catterall}, \citenamefont {Dzienkowski}, \citenamefont {Giedt},
  \citenamefont {Joseph},\ and\ \citenamefont {Wells}}]{Catterall:2011pd}%
  \BibitemOpen
  \bibfield  {author} {\bibinfo {author} {\bibfnamefont {Simon}\ \bibnamefont
  {Catterall}}, \bibinfo {author} {\bibfnamefont {Eric}\ \bibnamefont
  {Dzienkowski}}, \bibinfo {author} {\bibfnamefont {Joel}\ \bibnamefont
  {Giedt}}, \bibinfo {author} {\bibfnamefont {Anosh}\ \bibnamefont {Joseph}}, \
  and\ \bibinfo {author} {\bibfnamefont {Robert}\ \bibnamefont {Wells}},\
  }\bibfield  {title} {\enquote {\bibinfo {title} {{Perturbative
  renormalization of lattice N=4 super Yang- Mills theory}},}\ }\href {\doibase
  10.1007/JHEP04(2011)074} {\bibfield  {journal} {\bibinfo  {journal} {JHEP}\
  }\textbf {\bibinfo {volume} {04}},\ \bibinfo {pages} {074} (\bibinfo {year}
  {2011})},\ \Eprint {http://arxiv.org/abs/1102.1725} {arXiv:1102.1725
  [hep-th]} \BibitemShut {NoStop}%
%%CITATION = 1102.1725;%%
\bibitem [{\citenamefont {Catterall}\ \emph
  {et~al.}(2012{\natexlab{b}})\citenamefont {Catterall}, \citenamefont
  {Galvez}, \citenamefont {Joseph},\ and\ \citenamefont
  {Mehta}}]{Catterall:2011aa}%
  \BibitemOpen
  \bibfield  {author} {\bibinfo {author} {\bibfnamefont {Simon}\ \bibnamefont
  {Catterall}}, \bibinfo {author} {\bibfnamefont {Richard}\ \bibnamefont
  {Galvez}}, \bibinfo {author} {\bibfnamefont {Anosh}\ \bibnamefont {Joseph}},
  \ and\ \bibinfo {author} {\bibfnamefont {Dhagash}\ \bibnamefont {Mehta}},\
  }\bibfield  {title} {\enquote {\bibinfo {title} {{On the sign problem in 2D
  lattice super Yang--Mills}},}\ }\href {\doibase 10.1007/JHEP01(2012)108}
  {\bibfield  {journal} {\bibinfo  {journal} {JHEP}\ }\textbf {\bibinfo
  {volume} {01}},\ \bibinfo {pages} {108} (\bibinfo {year}
  {2012}{\natexlab{b}})},\ \Eprint {http://arxiv.org/abs/1112.3588}
  {arXiv:1112.3588 [hep-lat]} \BibitemShut {NoStop}%
%%CITATION = 1112.3588;%%
\bibitem [{\citenamefont {Giedt}\ \emph {et~al.}(2009)\citenamefont {Giedt},
  \citenamefont {Brower}, \citenamefont {Catterall}, \citenamefont {Fleming},\
  and\ \citenamefont {Vranas}}]{Giedt:2008xm}%
  \BibitemOpen
  \bibfield  {author} {\bibinfo {author} {\bibfnamefont {Joel}\ \bibnamefont
  {Giedt}}, \bibinfo {author} {\bibfnamefont {Richard}\ \bibnamefont {Brower}},
  \bibinfo {author} {\bibfnamefont {Simon}\ \bibnamefont {Catterall}}, \bibinfo
  {author} {\bibfnamefont {George~T.}\ \bibnamefont {Fleming}}, \ and\ \bibinfo
  {author} {\bibfnamefont {Pavlos}\ \bibnamefont {Vranas}},\ }\bibfield
  {title} {\enquote {\bibinfo {title} {{Lattice super-Yang-Mills using domain
  wall fermions in the chiral limit}},}\ }\href {\doibase
  10.1103/PhysRevD.79.025015} {\bibfield  {journal} {\bibinfo  {journal} {Phys.
  Rev.}\ }\textbf {\bibinfo {volume} {D79}},\ \bibinfo {pages} {025015}
  (\bibinfo {year} {2009})},\ \Eprint {http://arxiv.org/abs/0810.5746}
  {arXiv:0810.5746 [hep-lat]} \BibitemShut {NoStop}%
%%CITATION = 0810.5746;%%
\bibitem [{\citenamefont {Elliott}\ \emph {et~al.}(2008)\citenamefont
  {Elliott}, \citenamefont {Giedt},\ and\ \citenamefont
  {Moore}}]{Elliott:2008jp}%
  \BibitemOpen
  \bibfield  {author} {\bibinfo {author} {\bibfnamefont {Joshua~W.}\
  \bibnamefont {Elliott}}, \bibinfo {author} {\bibfnamefont {Joel}\
  \bibnamefont {Giedt}}, \ and\ \bibinfo {author} {\bibfnamefont {Guy~D.}\
  \bibnamefont {Moore}},\ }\bibfield  {title} {\enquote {\bibinfo {title}
  {{Lattice four-dimensional N=4 SYM is practical}},}\ }\href {\doibase
  10.1103/PhysRevD.78.081701} {\bibfield  {journal} {\bibinfo  {journal} {Phys.
  Rev.}\ }\textbf {\bibinfo {volume} {D78}},\ \bibinfo {pages} {081701}
  (\bibinfo {year} {2008})},\ \Eprint {http://arxiv.org/abs/0806.0013}
  {arXiv:0806.0013 [hep-lat]} \BibitemShut {NoStop}%
%%CITATION = 0806.0013;%%
\bibitem [{\citenamefont {Fodor}\ \emph
  {et~al.}(2012{\natexlab{c}})\citenamefont {Fodor}, \citenamefont {Holland},
  \citenamefont {Kuti}, \citenamefont {Nogradi},\ and\ \citenamefont
  {Wong}}]{Fodor:2012td}%
  \BibitemOpen
  \bibfield  {author} {\bibinfo {author} {\bibfnamefont {Zoltan}\ \bibnamefont
  {Fodor}}, \bibinfo {author} {\bibfnamefont {Kieran}\ \bibnamefont {Holland}},
  \bibinfo {author} {\bibfnamefont {Julius}\ \bibnamefont {Kuti}}, \bibinfo
  {author} {\bibfnamefont {Daniel}\ \bibnamefont {Nogradi}}, \ and\ \bibinfo
  {author} {\bibfnamefont {Chik~Him}\ \bibnamefont {Wong}},\ }\bibfield
  {title} {\enquote {\bibinfo {title} {{The Yang-Mills gradient flow in finite
  volume}},}\ }\href {\doibase 10.1007/JHEP11(2012)007} {\bibfield  {journal}
  {\bibinfo  {journal} {JHEP}\ }\textbf {\bibinfo {volume} {1211}},\ \bibinfo
  {pages} {007} (\bibinfo {year} {2012}{\natexlab{c}})},\ \Eprint
  {http://arxiv.org/abs/1208.1051} {arXiv:1208.1051 [hep-lat]} \BibitemShut
  {NoStop}%
%%CITATION = ARXIV:1208.1051;%%
\bibitem [{\citenamefont {Fodor}\ \emph
  {et~al.}(2012{\natexlab{d}})\citenamefont {Fodor}, \citenamefont {Holland},
  \citenamefont {Kuti}, \citenamefont {Nogradi},\ and\ \citenamefont
  {Wong}}]{Fodor:2012qh}%
  \BibitemOpen
  \bibfield  {author} {\bibinfo {author} {\bibfnamefont {Zoltan}\ \bibnamefont
  {Fodor}}, \bibinfo {author} {\bibfnamefont {Kieran}\ \bibnamefont {Holland}},
  \bibinfo {author} {\bibfnamefont {Julius}\ \bibnamefont {Kuti}}, \bibinfo
  {author} {\bibfnamefont {Daniel}\ \bibnamefont {Nogradi}}, \ and\ \bibinfo
  {author} {\bibfnamefont {Chik~Him}\ \bibnamefont {Wong}},\ }\bibfield
  {title} {\enquote {\bibinfo {title} {{The gradient flow running coupling
  scheme}},}\ }\href@noop {} {\bibfield  {journal} {\bibinfo  {journal} {PoS}\
  }\textbf {\bibinfo {volume} {LATTICE2012}},\ \bibinfo {pages} {050} (\bibinfo
  {year} {2012}{\natexlab{d}})},\ \Eprint {http://arxiv.org/abs/1211.3247}
  {arXiv:1211.3247 [hep-lat]} \BibitemShut {NoStop}%
%%CITATION = ARXIV:1211.3247;%%
\bibitem [{\citenamefont {Petropoulos}\ \emph {et~al.}(2012)\citenamefont
  {Petropoulos}, \citenamefont {Cheng}, \citenamefont {Hasenfratz},\ and\
  \citenamefont {Schaich}}]{Petropoulos:2012mg}%
  \BibitemOpen
  \bibfield  {author} {\bibinfo {author} {\bibfnamefont {Gregory}\ \bibnamefont
  {Petropoulos}}, \bibinfo {author} {\bibfnamefont {Anqi}\ \bibnamefont
  {Cheng}}, \bibinfo {author} {\bibfnamefont {Anna}\ \bibnamefont
  {Hasenfratz}}, \ and\ \bibinfo {author} {\bibfnamefont {David}\ \bibnamefont
  {Schaich}},\ }\bibfield  {title} {\enquote {\bibinfo {title} {{MCRG study of
  8 and 12 fundamental flavors}},}\ }\href@noop {} {\bibfield  {journal}
  {\bibinfo  {journal} {PoS}\ }\textbf {\bibinfo {volume} {Lattice 2012}},\
  \bibinfo {pages} {051} (\bibinfo {year} {2012})},\ \Eprint
  {http://arxiv.org/abs/1212.0053} {arXiv:1212.0053} \BibitemShut {NoStop}%
\bibitem [{\citenamefont {Appelquist}\ \emph {et~al.}(2009)\citenamefont
  {Appelquist}, \citenamefont {Fleming},\ and\ \citenamefont
  {Neil}}]{Appelquist:2009ty}%
  \BibitemOpen
  \bibfield  {author} {\bibinfo {author} {\bibfnamefont {Thomas}\ \bibnamefont
  {Appelquist}}, \bibinfo {author} {\bibfnamefont {George~T.}\ \bibnamefont
  {Fleming}}, \ and\ \bibinfo {author} {\bibfnamefont {Ethan~T.}\ \bibnamefont
  {Neil}},\ }\bibfield  {title} {\enquote {\bibinfo {title} {{Lattice Study of
  Conformal Behavior in SU(3) Yang-Mills Theories}},}\ }\href {\doibase
  10.1103/PhysRevD.79.076010} {\bibfield  {journal} {\bibinfo  {journal} {Phys.
  Rev.}\ }\textbf {\bibinfo {volume} {D79}},\ \bibinfo {pages} {076010}
  (\bibinfo {year} {2009})},\ \Eprint {http://arxiv.org/abs/0901.3766}
  {arXiv:0901.3766} \BibitemShut {NoStop}%
\bibitem [{\citenamefont {DeGrand}\ \emph {et~al.}(2010)\citenamefont
  {DeGrand}, \citenamefont {Shamir},\ and\ \citenamefont
  {Svetitsky}}]{DeGrand:2010na}%
  \BibitemOpen
  \bibfield  {author} {\bibinfo {author} {\bibfnamefont {Thomas}\ \bibnamefont
  {DeGrand}}, \bibinfo {author} {\bibfnamefont {Yigal}\ \bibnamefont {Shamir}},
  \ and\ \bibinfo {author} {\bibfnamefont {Benjamin}\ \bibnamefont
  {Svetitsky}},\ }\bibfield  {title} {\enquote {\bibinfo {title} {{Running
  coupling and mass anomalous dimension of SU(3) gauge theory with two flavors
  of symmetric-representation fermions}},}\ }\href {\doibase
  10.1103/PhysRevD.82.054503} {\bibfield  {journal} {\bibinfo  {journal}
  {Phys.Rev.}\ }\textbf {\bibinfo {volume} {D82}},\ \bibinfo {pages} {054503}
  (\bibinfo {year} {2010})},\ \Eprint {http://arxiv.org/abs/1006.0707}
  {arXiv:1006.0707 [hep-lat]} \BibitemShut {NoStop}%
%%CITATION = ARXIV:1006.0707;%%
\bibitem [{\citenamefont {Fodor}\ \emph
  {et~al.}(2012{\natexlab{e}})\citenamefont {Fodor}, \citenamefont {Holland},
  \citenamefont {Kuti}, \citenamefont {Nogradi}, \citenamefont {Schroeder},\
  and\ \citenamefont {Wong}}]{Fodor:2012uw}%
  \BibitemOpen
  \bibfield  {author} {\bibinfo {author} {\bibfnamefont {Zoltan}\ \bibnamefont
  {Fodor}}, \bibinfo {author} {\bibfnamefont {Kieran}\ \bibnamefont {Holland}},
  \bibinfo {author} {\bibfnamefont {Julius}\ \bibnamefont {Kuti}}, \bibinfo
  {author} {\bibfnamefont {Daniel}\ \bibnamefont {Nogradi}}, \bibinfo {author}
  {\bibfnamefont {Chris}\ \bibnamefont {Schroeder}}, \ and\ \bibinfo {author}
  {\bibfnamefont {Chik~Him}\ \bibnamefont {Wong}},\ }\bibfield  {title}
  {\enquote {\bibinfo {title} {{Confining force and running coupling with
  twelve fundamental and two sextet fermions}},}\ }\href
  {http://pos.sissa.it/archive/conferences/164/025/Lattice 2012_025.pdf}
  {\bibfield  {journal} {\bibinfo  {journal} {PoS}\ }\textbf {\bibinfo {volume}
  {Lattice 2012}},\ \bibinfo {pages} {025} (\bibinfo {year}
  {2012}{\natexlab{e}})},\ \Eprint {http://arxiv.org/abs/1211.3548}
  {arXiv:1211.3548} \BibitemShut {NoStop}%
\bibitem [{\citenamefont {Hasenfratz}(2012)}]{Hasenfratz:2011xn}%
  \BibitemOpen
  \bibfield  {author} {\bibinfo {author} {\bibfnamefont {Anna}\ \bibnamefont
  {Hasenfratz}},\ }\bibfield  {title} {\enquote {\bibinfo {title} {{Infrared
  fixed point of the 12-fermion SU(3) gauge model based on 2-lattice MCRG
  matching}},}\ }\href {\doibase 10.1103/PhysRevLett.108.061601} {\bibfield
  {journal} {\bibinfo  {journal} {Phys. Rev. Lett.}\ }\textbf {\bibinfo
  {volume} {108}},\ \bibinfo {pages} {061601} (\bibinfo {year} {2012})},\
  \Eprint {http://arxiv.org/abs/1106.5293} {arXiv:1106.5293} \BibitemShut
  {NoStop}%
\bibitem [{\citenamefont {Catterall}\ \emph
  {et~al.}(2012{\natexlab{c}})\citenamefont {Catterall}, \citenamefont
  {Del~Debbio}, \citenamefont {Giedt},\ and\ \citenamefont
  {Keegan}}]{Catterall:2011zf}%
  \BibitemOpen
  \bibfield  {author} {\bibinfo {author} {\bibfnamefont {Simon}\ \bibnamefont
  {Catterall}}, \bibinfo {author} {\bibfnamefont {Luigi}\ \bibnamefont
  {Del~Debbio}}, \bibinfo {author} {\bibfnamefont {Joel}\ \bibnamefont
  {Giedt}}, \ and\ \bibinfo {author} {\bibfnamefont {Liam}\ \bibnamefont
  {Keegan}},\ }\bibfield  {title} {\enquote {\bibinfo {title} {{MCRG Minimal
  Walking Technicolor}},}\ }\href {\doibase 10.1103/PhysRevD.85.094501}
  {\bibfield  {journal} {\bibinfo  {journal} {Phys.Rev.}\ }\textbf {\bibinfo
  {volume} {D85}},\ \bibinfo {pages} {094501} (\bibinfo {year}
  {2012}{\natexlab{c}})},\ \Eprint {http://arxiv.org/abs/1108.3794}
  {arXiv:1108.3794 [hep-ph]} \BibitemShut {NoStop}%
%%CITATION = ARXIV:1108.3794;%%
\bibitem [{\citenamefont {Banks}\ and\ \citenamefont
  {Casher}(1980)}]{Banks:1979yr}%
  \BibitemOpen
  \bibfield  {author} {\bibinfo {author} {\bibfnamefont {Tom}\ \bibnamefont
  {Banks}}\ and\ \bibinfo {author} {\bibfnamefont {A.}~\bibnamefont {Casher}},\
  }\bibfield  {title} {\enquote {\bibinfo {title} {{Chiral Symmetry Breaking in
  Confining Theories}},}\ }\href {\doibase 10.1016/0550-3213(80)90255-2}
  {\bibfield  {journal} {\bibinfo  {journal} {Nucl.Phys.}\ }\textbf {\bibinfo
  {volume} {B169}},\ \bibinfo {pages} {103} (\bibinfo {year}
  {1980})}\BibitemShut {NoStop}%
%%CITATION = NUPHA,B169,103;%%
\bibitem [{\citenamefont {Giusti}\ and\ \citenamefont
  {Luscher}(2009)}]{Giusti:2008vb}%
  \BibitemOpen
  \bibfield  {author} {\bibinfo {author} {\bibfnamefont {Leonardo}\
  \bibnamefont {Giusti}}\ and\ \bibinfo {author} {\bibfnamefont {Martin}\
  \bibnamefont {Luscher}},\ }\bibfield  {title} {\enquote {\bibinfo {title}
  {{Chiral symmetry breaking and the Banks-Casher relation in lattice QCD with
  Wilson quarks}},}\ }\href {\doibase 10.1088/1126-6708/2009/03/013} {\bibfield
   {journal} {\bibinfo  {journal} {JHEP}\ }\textbf {\bibinfo {volume} {0903}},\
  \bibinfo {pages} {013} (\bibinfo {year} {2009})},\ \Eprint
  {http://arxiv.org/abs/0812.3638} {arXiv:0812.3638} \BibitemShut {NoStop}%
\bibitem [{\citenamefont {Del~Debbio}\ and\ \citenamefont
  {Zwicky}(2010)}]{DelDebbio:2010ze}%
  \BibitemOpen
  \bibfield  {author} {\bibinfo {author} {\bibfnamefont {Luigi}\ \bibnamefont
  {Del~Debbio}}\ and\ \bibinfo {author} {\bibfnamefont {Roman}\ \bibnamefont
  {Zwicky}},\ }\bibfield  {title} {\enquote {\bibinfo {title} {{Hyperscaling
  relations in mass-deformed conformal gauge theories}},}\ }\href {\doibase
  10.1103/PhysRevD.82.014502} {\bibfield  {journal} {\bibinfo  {journal} {Phys.
  Rev.}\ }\textbf {\bibinfo {volume} {D82}},\ \bibinfo {pages} {014502}
  (\bibinfo {year} {2010})},\ \Eprint {http://arxiv.org/abs/1005.2371}
  {arXiv:1005.2371} \BibitemShut {NoStop}%
\bibitem [{\citenamefont {Patella}(2012)}]{Patella:2012da}%
  \BibitemOpen
  \bibfield  {author} {\bibinfo {author} {\bibfnamefont {Agostino}\
  \bibnamefont {Patella}},\ }\bibfield  {title} {\enquote {\bibinfo {title} {{A
  precise determination of the psibar-psi anomalous dimension in conformal
  gauge theories}},}\ }\href {\doibase 10.1103/PhysRevD.86.025006} {\bibfield
  {journal} {\bibinfo  {journal} {Phys. Rev.}\ }\textbf {\bibinfo {volume}
  {D86}},\ \bibinfo {pages} {025006} (\bibinfo {year} {2012})},\ \Eprint
  {http://arxiv.org/abs/1204.4432} {arXiv:1204.4432} \BibitemShut {NoStop}%
\bibitem [{\citenamefont {Cheng}\ \emph {et~al.}(2013)\citenamefont {Cheng},
  \citenamefont {Hasenfratz}, \citenamefont {Petropoulos},\ and\ \citenamefont
  {Schaich}}]{Cheng:2013eu}%
  \BibitemOpen
  \bibfield  {author} {\bibinfo {author} {\bibfnamefont {Anqi}\ \bibnamefont
  {Cheng}}, \bibinfo {author} {\bibfnamefont {Anna}\ \bibnamefont
  {Hasenfratz}}, \bibinfo {author} {\bibfnamefont {Gregory}\ \bibnamefont
  {Petropoulos}}, \ and\ \bibinfo {author} {\bibfnamefont {David}\ \bibnamefont
  {Schaich}},\ }\bibfield  {title} {\enquote {\bibinfo {title}
  {{Scale-dependent mass anomalous dimension from Dirac eigenmodes}},}\
  }\href@noop {} {\  (\bibinfo {year} {2013})},\ \Eprint
  {http://arxiv.org/abs/1301.1355} {arXiv:1301.1355 [hep-lat]} \BibitemShut
  {NoStop}%
%%CITATION = ARXIV:1301.1355;%%
\bibitem [{\citenamefont {Gell-Mann}\ \emph {et~al.}(1968)\citenamefont
  {Gell-Mann}, \citenamefont {Oakes},\ and\ \citenamefont
  {Renner}}]{GellMann:1968rz}%
  \BibitemOpen
  \bibfield  {author} {\bibinfo {author} {\bibfnamefont {Murray}\ \bibnamefont
  {Gell-Mann}}, \bibinfo {author} {\bibfnamefont {R.J.}\ \bibnamefont {Oakes}},
  \ and\ \bibinfo {author} {\bibfnamefont {B.}~\bibnamefont {Renner}},\
  }\bibfield  {title} {\enquote {\bibinfo {title} {{Behavior of current
  divergences under SU(3) x SU(3)}},}\ }\href {\doibase
  10.1103/PhysRev.175.2195} {\bibfield  {journal} {\bibinfo  {journal}
  {Phys.Rev.}\ }\textbf {\bibinfo {volume} {175}},\ \bibinfo {pages}
  {2195--2199} (\bibinfo {year} {1968})}\BibitemShut {NoStop}%
%%CITATION = PHRVA,175,2195;%%
\bibitem [{\citenamefont {Appelquist}\ \emph {et~al.}(2010)\citenamefont
  {Appelquist}, \citenamefont {Avarkian}, \citenamefont {Babich}, \citenamefont
  {Brower}, \citenamefont {Cheng}, \citenamefont {Clark}, \citenamefont
  {Cohen}, \citenamefont {Fleming}, \citenamefont {T.}, \citenamefont {Kiskis},
  \citenamefont {Neil}, \citenamefont {Osborn}, \citenamefont {Rebbi},
  \citenamefont {Schaich},\ and\ \citenamefont {Vranas}}]{Appelquist:2009ka}%
  \BibitemOpen
  \bibfield  {author} {\bibinfo {author} {\bibfnamefont {Thomas}\ \bibnamefont
  {Appelquist}}, \bibinfo {author} {\bibfnamefont {Adam}\ \bibnamefont
  {Avarkian}}, \bibinfo {author} {\bibfnamefont {Ron}\ \bibnamefont {Babich}},
  \bibinfo {author} {\bibfnamefont {Richard~C}\ \bibnamefont {Brower}},
  \bibinfo {author} {\bibfnamefont {Michael}\ \bibnamefont {Cheng}}, \bibinfo
  {author} {\bibfnamefont {Michael}\ \bibnamefont {Clark}}, \bibinfo {author}
  {\bibfnamefont {Saul~D.}\ \bibnamefont {Cohen}}, \bibinfo {author}
  {\bibnamefont {Fleming}}, \bibinfo {author} {\bibfnamefont {George}\
  \bibnamefont {T.}}, \bibinfo {author} {\bibfnamefont {Joseph}\ \bibnamefont
  {Kiskis}}, \bibinfo {author} {\bibfnamefont {Ethan~T.}\ \bibnamefont {Neil}},
  \bibinfo {author} {\bibfnamefont {James}\ \bibnamefont {Osborn}}, \bibinfo
  {author} {\bibfnamefont {Claudio}\ \bibnamefont {Rebbi}}, \bibinfo {author}
  {\bibfnamefont {David}\ \bibnamefont {Schaich}}, \ and\ \bibinfo {author}
  {\bibfnamefont {Pavlos}\ \bibnamefont {Vranas}},\ }\bibfield  {title}
  {\enquote {\bibinfo {title} {{Toward TeV Conformality}},}\ }\href {\doibase
  10.1103/PhysRevLett.104.071601} {\bibfield  {journal} {\bibinfo  {journal}
  {Phys. Rev. Lett.}\ }\textbf {\bibinfo {volume} {104}},\ \bibinfo {pages}
  {071601} (\bibinfo {year} {2010})},\ \Eprint {http://arxiv.org/abs/0910.2224}
  {arXiv:0910.2224 [hep-ph]} \BibitemShut {NoStop}%
%%CITATION = 0910.2224;%%
\bibitem [{\citenamefont {Peskin}\ and\ \citenamefont
  {Takeuchi}(1992)}]{Peskin:1991sw}%
  \BibitemOpen
  \bibfield  {author} {\bibinfo {author} {\bibfnamefont {Michael~E.}\
  \bibnamefont {Peskin}}\ and\ \bibinfo {author} {\bibfnamefont {Tatsu}\
  \bibnamefont {Takeuchi}},\ }\bibfield  {title} {\enquote {\bibinfo {title}
  {{Estimation of oblique electroweak corrections}},}\ }\href {\doibase
  10.1103/PhysRevD.46.381} {\bibfield  {journal} {\bibinfo  {journal} {Phys.
  Rev.}\ }\textbf {\bibinfo {volume} {D46}},\ \bibinfo {pages} {381--409}
  (\bibinfo {year} {1992})}\BibitemShut {NoStop}%
\bibitem [{\citenamefont {Baak}\ \emph {et~al.}(2012)\citenamefont {Baak},
  \citenamefont {Goebel}, \citenamefont {Haller}, \citenamefont {Hoecker},
  \citenamefont {Kennedy} \emph {et~al.}}]{Baak:2012kk}%
  \BibitemOpen
  \bibfield  {author} {\bibinfo {author} {\bibfnamefont {M.}~\bibnamefont
  {Baak}}, \bibinfo {author} {\bibfnamefont {M.}~\bibnamefont {Goebel}},
  \bibinfo {author} {\bibfnamefont {J.}~\bibnamefont {Haller}}, \bibinfo
  {author} {\bibfnamefont {A.}~\bibnamefont {Hoecker}}, \bibinfo {author}
  {\bibfnamefont {D.}~\bibnamefont {Kennedy}},  \emph {et~al.},\ }\bibfield
  {title} {\enquote {\bibinfo {title} {{The Electroweak Fit of the Standard
  Model after the Discovery of a New Boson at the LHC}},}\ }\href {\doibase
  10.1140/epjc/s10052-012-2205-9} {\bibfield  {journal} {\bibinfo  {journal}
  {Eur.Phys.J.}\ }\textbf {\bibinfo {volume} {C72}},\ \bibinfo {pages} {2205}
  (\bibinfo {year} {2012})},\ \Eprint {http://arxiv.org/abs/1209.2716}
  {arXiv:1209.2716 [hep-ph]} \BibitemShut {NoStop}%
%%CITATION = ARXIV:1209.2716;%%
\bibitem [{\citenamefont {Shintani}\ \emph {et~al.}(2008)\citenamefont
  {Shintani}, \citenamefont {Aoki}, \citenamefont {Fukaya}, \citenamefont
  {Hashimoto}, \citenamefont {Kaneko}, \citenamefont {Matsufuru}, \citenamefont
  {Onogi},\ and\ \citenamefont {Yamada}}]{Shintani:2008qe}%
  \BibitemOpen
  \bibfield  {author} {\bibinfo {author} {\bibfnamefont {E.}~\bibnamefont
  {Shintani}}, \bibinfo {author} {\bibfnamefont {S.}~\bibnamefont {Aoki}},
  \bibinfo {author} {\bibfnamefont {H.}~\bibnamefont {Fukaya}}, \bibinfo
  {author} {\bibfnamefont {S.}~\bibnamefont {Hashimoto}}, \bibinfo {author}
  {\bibfnamefont {T.}~\bibnamefont {Kaneko}}, \bibinfo {author} {\bibfnamefont
  {H.}~\bibnamefont {Matsufuru}}, \bibinfo {author} {\bibfnamefont
  {T.}~\bibnamefont {Onogi}}, \ and\ \bibinfo {author} {\bibfnamefont
  {N.}~\bibnamefont {Yamada}},\ }\bibfield  {title} {\enquote {\bibinfo {title}
  {{S Parameter and Pseudo Nambu-Goldstone Boson Mass from Lattice QCD}},}\
  }\href {\doibase 10.1103/PhysRevLett.101.242001} {\bibfield  {journal}
  {\bibinfo  {journal} {Phys. Rev. Lett.}\ }\textbf {\bibinfo {volume} {101}},\
  \bibinfo {pages} {242001} (\bibinfo {year} {2008})},\ \Eprint
  {http://arxiv.org/abs/0806.4222} {arXiv:0806.4222} \BibitemShut {NoStop}%
\bibitem [{\citenamefont {Boyle}\ \emph {et~al.}(2010)\citenamefont {Boyle},
  \citenamefont {Del~Debbio}, \citenamefont {Wennekers},\ and\ \citenamefont
  {Zanotti}}]{Boyle:2009xi}%
  \BibitemOpen
  \bibfield  {author} {\bibinfo {author} {\bibfnamefont {Peter~A.}\
  \bibnamefont {Boyle}}, \bibinfo {author} {\bibfnamefont {Luigi}\ \bibnamefont
  {Del~Debbio}}, \bibinfo {author} {\bibfnamefont {Jan}\ \bibnamefont
  {Wennekers}}, \ and\ \bibinfo {author} {\bibfnamefont {James~M.}\
  \bibnamefont {Zanotti}},\ }\bibfield  {title} {\enquote {\bibinfo {title}
  {{The S Parameter in QCD from Domain Wall Fermions}},}\ }\href {\doibase
  10.1103/PhysRevD.81.014504} {\bibfield  {journal} {\bibinfo  {journal} {Phys.
  Rev.}\ }\textbf {\bibinfo {volume} {D81}},\ \bibinfo {pages} {014504}
  (\bibinfo {year} {2010})},\ \Eprint {http://arxiv.org/abs/0909.4931}
  {arXiv:0909.4931} \BibitemShut {NoStop}%
\bibitem [{\citenamefont {Appelquist}\ \emph {et~al.}(2011)\citenamefont
  {Appelquist}, \citenamefont {Babich}, \citenamefont {Brower}, \citenamefont
  {Cheng}, \citenamefont {Clark}, \citenamefont {Cohen}, \citenamefont
  {Fleming}, \citenamefont {Kiskis}, \citenamefont {Lin}, \citenamefont {Neil},
  \citenamefont {Osborn}, \citenamefont {Rebbi}, \citenamefont {Schaich},\ and\
  \citenamefont {Vranas}}]{Appelquist:2010xv}%
  \BibitemOpen
  \bibfield  {author} {\bibinfo {author} {\bibfnamefont {Thomas}\ \bibnamefont
  {Appelquist}}, \bibinfo {author} {\bibfnamefont {Ron}\ \bibnamefont
  {Babich}}, \bibinfo {author} {\bibfnamefont {Richard~C.}\ \bibnamefont
  {Brower}}, \bibinfo {author} {\bibfnamefont {Michael}\ \bibnamefont {Cheng}},
  \bibinfo {author} {\bibfnamefont {Michael~A.}\ \bibnamefont {Clark}},
  \bibinfo {author} {\bibfnamefont {Saul~D.}\ \bibnamefont {Cohen}}, \bibinfo
  {author} {\bibfnamefont {George~T.}\ \bibnamefont {Fleming}}, \bibinfo
  {author} {\bibfnamefont {Joe}\ \bibnamefont {Kiskis}}, \bibinfo {author}
  {\bibfnamefont {Meifeng}\ \bibnamefont {Lin}}, \bibinfo {author}
  {\bibfnamefont {Ethan~T.}\ \bibnamefont {Neil}}, \bibinfo {author}
  {\bibfnamefont {James~C.}\ \bibnamefont {Osborn}}, \bibinfo {author}
  {\bibfnamefont {Claudio}\ \bibnamefont {Rebbi}}, \bibinfo {author}
  {\bibfnamefont {David}\ \bibnamefont {Schaich}}, \ and\ \bibinfo {author}
  {\bibfnamefont {Pavlos}\ \bibnamefont {Vranas}},\ }\bibfield  {title}
  {\enquote {\bibinfo {title} {{Parity Doubling and the S Parameter Below the
  Conformal Window}},}\ }\href {\doibase 10.1103/PhysRevLett.106.231601}
  {\bibfield  {journal} {\bibinfo  {journal} {Phys. Rev. Lett.}\ }\textbf
  {\bibinfo {volume} {106}},\ \bibinfo {pages} {231601} (\bibinfo {year}
  {2011})},\ \Eprint {http://arxiv.org/abs/1009.5967} {arXiv:1009.5967}
  \BibitemShut {NoStop}%
\bibitem [{\citenamefont {Schaich}(2011)}]{Schaich:2011qz}%
  \BibitemOpen
  \bibfield  {author} {\bibinfo {author} {\bibfnamefont {David}\ \bibnamefont
  {Schaich}},\ }\bibfield  {title} {\enquote {\bibinfo {title} {{S parameter
  and parity doubling below the conformal window}},}\ }\href@noop {} {\bibfield
   {journal} {\bibinfo  {journal} {PoS}\ }\textbf {\bibinfo {volume} {Lattice
  2011}},\ \bibinfo {pages} {087} (\bibinfo {year} {2011})},\ \Eprint
  {http://arxiv.org/abs/1111.4993} {arXiv:1111.4993} \BibitemShut {NoStop}%
\bibitem [{\citenamefont {DeGrand}(2010)}]{DeGrand:2010tm}%
  \BibitemOpen
  \bibfield  {author} {\bibinfo {author} {\bibfnamefont {Thomas}\ \bibnamefont
  {DeGrand}},\ }\bibfield  {title} {\enquote {\bibinfo {title} {{Oblique
  correction in a walking lattice theory}},}\ }\href@noop {} {\  (\bibinfo
  {year} {2010})},\ \Eprint {http://arxiv.org/abs/1006.3777} {arXiv:1006.3777}
  \BibitemShut {NoStop}%
\bibitem [{\citenamefont {Appelquist}\ \emph {et~al.}(2012)\citenamefont
  {Appelquist}, \citenamefont {Babich}, \citenamefont {Brower}, \citenamefont
  {Buchoff}, \citenamefont {Cheng} \emph {et~al.}}]{Appelquist:2012sm}%
  \BibitemOpen
  \bibfield  {author} {\bibinfo {author} {\bibfnamefont {Thomas}\ \bibnamefont
  {Appelquist}}, \bibinfo {author} {\bibfnamefont {Ron}\ \bibnamefont
  {Babich}}, \bibinfo {author} {\bibfnamefont {Richard~C.}\ \bibnamefont
  {Brower}}, \bibinfo {author} {\bibfnamefont {Michael~I.}\ \bibnamefont
  {Buchoff}}, \bibinfo {author} {\bibfnamefont {Michael}\ \bibnamefont
  {Cheng}},  \emph {et~al.},\ }\bibfield  {title} {\enquote {\bibinfo {title}
  {{WW Scattering Parameters via Pseudoscalar Phase Shifts}},}\ }\href
  {\doibase 10.1103/PhysRevD.85.074505} {\bibfield  {journal} {\bibinfo
  {journal} {Phys.Rev.}\ }\textbf {\bibinfo {volume} {D85}},\ \bibinfo {pages}
  {074505} (\bibinfo {year} {2012})},\ \Eprint {http://arxiv.org/abs/1201.3977}
  {arXiv:1201.3977 [hep-lat]} \BibitemShut {NoStop}%
%%CITATION = ARXIV:1201.3977;%%
\bibitem [{\citenamefont {Appelquist}\ and\ \citenamefont
  {Bernard}(1980)}]{Appelquist:1980vg}%
  \BibitemOpen
  \bibfield  {author} {\bibinfo {author} {\bibfnamefont {Thomas}\ \bibnamefont
  {Appelquist}}\ and\ \bibinfo {author} {\bibfnamefont {Claude~W.}\
  \bibnamefont {Bernard}},\ }\bibfield  {title} {\enquote {\bibinfo {title}
  {{Strongly Interacting Higgs Bosons}},}\ }\href {\doibase
  10.1103/PhysRevD.22.200} {\bibfield  {journal} {\bibinfo  {journal}
  {Phys.Rev.}\ }\textbf {\bibinfo {volume} {D22}},\ \bibinfo {pages} {200}
  (\bibinfo {year} {1980})}\BibitemShut {NoStop}%
%%CITATION = PHRVA,D22,200;%%
\bibitem [{\citenamefont {Appelquist}\ and\ \citenamefont
  {Wu}(1993)}]{Appelquist:1993ka}%
  \BibitemOpen
  \bibfield  {author} {\bibinfo {author} {\bibfnamefont {Thomas}\ \bibnamefont
  {Appelquist}}\ and\ \bibinfo {author} {\bibfnamefont {Guo-Hong}\ \bibnamefont
  {Wu}},\ }\bibfield  {title} {\enquote {\bibinfo {title} {{The Electroweak
  chiral Lagrangian and new precision measurements}},}\ }\href {\doibase
  10.1103/PhysRevD.48.3235} {\bibfield  {journal} {\bibinfo  {journal}
  {Phys.Rev.}\ }\textbf {\bibinfo {volume} {D48}},\ \bibinfo {pages}
  {3235--3241} (\bibinfo {year} {1993})},\ \Eprint
  {http://arxiv.org/abs/hep-ph/9304240} {arXiv:hep-ph/9304240 [hep-ph]}
  \BibitemShut {NoStop}%
%%CITATION = HEP-PH/9304240;%%
\bibitem [{\citenamefont {Bagger}\ \emph {et~al.}(1994)\citenamefont {Bagger},
  \citenamefont {Barger}, \citenamefont {Cheung}, \citenamefont {Gunion},
  \citenamefont {Han} \emph {et~al.}}]{Bagger:1993zf}%
  \BibitemOpen
  \bibfield  {author} {\bibinfo {author} {\bibfnamefont {J.}~\bibnamefont
  {Bagger}}, \bibinfo {author} {\bibfnamefont {Vernon~D.}\ \bibnamefont
  {Barger}}, \bibinfo {author} {\bibfnamefont {King-man}\ \bibnamefont
  {Cheung}}, \bibinfo {author} {\bibfnamefont {John~F.}\ \bibnamefont
  {Gunion}}, \bibinfo {author} {\bibfnamefont {Tao}\ \bibnamefont {Han}},
  \emph {et~al.},\ }\bibfield  {title} {\enquote {\bibinfo {title} {{The
  Strongly interacting W W system: Gold plated modes}},}\ }\href {\doibase
  10.1103/PhysRevD.49.1246} {\bibfield  {journal} {\bibinfo  {journal}
  {Phys.Rev.}\ }\textbf {\bibinfo {volume} {D49}},\ \bibinfo {pages}
  {1246--1264} (\bibinfo {year} {1994})},\ \Eprint
  {http://arxiv.org/abs/hep-ph/9306256} {arXiv:hep-ph/9306256 [hep-ph]}
  \BibitemShut {NoStop}%
%%CITATION = HEP-PH/9306256;%%
\bibitem [{\citenamefont {Chatrchyan}\ \emph {et~al.}(2013)\citenamefont
  {Chatrchyan} \emph {et~al.}}]{CMS:2012zz}%
  \BibitemOpen
  \bibfield  {author} {\bibinfo {author} {\bibfnamefont {Serguei}\ \bibnamefont
  {Chatrchyan}} \emph {et~al.} (\bibinfo {collaboration} {CMS Collaboration}),\
  }\bibfield  {title} {\enquote {\bibinfo {title} {{Measurement of the $ZZ$
  production cross section and search for anomalous couplings in 2 l2l ' final
  states in $pp$ collisions at $\sqrt{s}=7$ TeV}},}\ }\href {\doibase
  10.1007/JHEP01(2013)063} {\bibfield  {journal} {\bibinfo  {journal} {JHEP}\
  }\textbf {\bibinfo {volume} {1301}},\ \bibinfo {pages} {063} (\bibinfo {year}
  {2013})},\ \Eprint {http://arxiv.org/abs/1211.4890} {arXiv:1211.4890
  [hep-ex]} \BibitemShut {NoStop}%
%%CITATION = ARXIV:1211.4890;%%
\bibitem [{\citenamefont {Aad}\ \emph {et~al.}(2012{\natexlab{b}})\citenamefont
  {Aad} \emph {et~al.}}]{ATLAS:2012zz}%
  \BibitemOpen
  \bibfield  {author} {\bibinfo {author} {\bibfnamefont {Georges}\ \bibnamefont
  {Aad}} \emph {et~al.} (\bibinfo {collaboration} {ATLAS Collaboration}),\
  }\bibfield  {title} {\enquote {\bibinfo {title} {{Measurement of $ZZ$
  production in $pp$ collisions at $\sqrt{s}=7$ TeV and limits on anomalous
  $ZZZ$ and $ZZ\gamma$ couplings with the ATLAS detector}},}\ }\href@noop {} {\
   (\bibinfo {year} {2012}{\natexlab{b}})},\ \bibinfo {note} {submitted to
  JHEP},\ \Eprint {http://arxiv.org/abs/1211.6096} {arXiv:1211.6096 [hep-ex]}
  \BibitemShut {NoStop}%
%%CITATION = ARXIV:1211.6096;%%
\bibitem [{\citenamefont {Chatrchyan}\ \emph
  {et~al.}(2012{\natexlab{b}})\citenamefont {Chatrchyan} \emph
  {et~al.}}]{CMS:2012wv}%
  \BibitemOpen
  \bibfield  {author} {\bibinfo {author} {\bibfnamefont {Serguei}\ \bibnamefont
  {Chatrchyan}} \emph {et~al.} (\bibinfo {collaboration} {CMS Collaboration}),\
  }\bibfield  {title} {\enquote {\bibinfo {title} {{Measurement of the sum of
  $W W$ and $WZ$ production with $W+$dijet events in $pp$ collisions at
  $\sqrt{s}=7$ TeV}},}\ }\href@noop {} {\  (\bibinfo {year}
  {2012}{\natexlab{b}})},\ \bibinfo {note} {submitted to Eur. Phys. J. C},\
  \Eprint {http://arxiv.org/abs/1210.7544} {arXiv:1210.7544 [hep-ex]}
  \BibitemShut {NoStop}%
%%CITATION = ARXIV:1210.7544;%%
\bibitem [{\citenamefont {Kaplan}(1992)}]{Kaplan:1991ah}%
  \BibitemOpen
  \bibfield  {author} {\bibinfo {author} {\bibfnamefont {David~B.}\
  \bibnamefont {Kaplan}},\ }\bibfield  {title} {\enquote {\bibinfo {title} {{A
  Single explanation for both the baryon and dark matter densities}},}\ }\href
  {\doibase 10.1103/PhysRevLett.68.741} {\bibfield  {journal} {\bibinfo
  {journal} {Phys.Rev.Lett.}\ }\textbf {\bibinfo {volume} {68}},\ \bibinfo
  {pages} {741--743} (\bibinfo {year} {1992})}\BibitemShut {NoStop}%
%%CITATION = PRLTA,68,741;%%
\bibitem [{\citenamefont {Kaplan}\ \emph {et~al.}(2009)\citenamefont {Kaplan},
  \citenamefont {Luty},\ and\ \citenamefont {Zurek}}]{Kaplan:2009ag}%
  \BibitemOpen
  \bibfield  {author} {\bibinfo {author} {\bibfnamefont {David~E.}\
  \bibnamefont {Kaplan}}, \bibinfo {author} {\bibfnamefont {Markus~A.}\
  \bibnamefont {Luty}}, \ and\ \bibinfo {author} {\bibfnamefont {Kathryn~M.}\
  \bibnamefont {Zurek}},\ }\bibfield  {title} {\enquote {\bibinfo {title}
  {{Asymmetric Dark Matter}},}\ }\href {\doibase 10.1103/PhysRevD.79.115016}
  {\bibfield  {journal} {\bibinfo  {journal} {Phys.Rev.}\ }\textbf {\bibinfo
  {volume} {D79}},\ \bibinfo {pages} {115016} (\bibinfo {year} {2009})},\
  \Eprint {http://arxiv.org/abs/0901.4117} {arXiv:0901.4117 [hep-ph]}
  \BibitemShut {NoStop}%
%%CITATION = ARXIV:0901.4117;%%
\bibitem [{\citenamefont {Chivukula}\ and\ \citenamefont
  {Walker}(1990)}]{Chivukula:1989qb}%
  \BibitemOpen
  \bibfield  {author} {\bibinfo {author} {\bibfnamefont {R.~Sekhar}\
  \bibnamefont {Chivukula}}\ and\ \bibinfo {author} {\bibfnamefont {Terry~P.}\
  \bibnamefont {Walker}},\ }\bibfield  {title} {\enquote {\bibinfo {title}
  {{TECHNICOLOR COSMOLOGY}},}\ }\href {\doibase 10.1016/0550-3213(90)90151-3}
  {\bibfield  {journal} {\bibinfo  {journal} {Nucl.Phys.}\ }\textbf {\bibinfo
  {volume} {B329}},\ \bibinfo {pages} {445} (\bibinfo {year}
  {1990})}\BibitemShut {NoStop}%
%%CITATION = NUPHA,B329,445;%%
\bibitem [{\citenamefont {Gudnason}\ \emph {et~al.}(2006)\citenamefont
  {Gudnason}, \citenamefont {Kouvaris},\ and\ \citenamefont
  {Sannino}}]{Gudnason:2006yj}%
  \BibitemOpen
  \bibfield  {author} {\bibinfo {author} {\bibfnamefont {Sven~Bjarke}\
  \bibnamefont {Gudnason}}, \bibinfo {author} {\bibfnamefont {Chris}\
  \bibnamefont {Kouvaris}}, \ and\ \bibinfo {author} {\bibfnamefont
  {Francesco}\ \bibnamefont {Sannino}},\ }\bibfield  {title} {\enquote
  {\bibinfo {title} {{Dark Matter from new Technicolor Theories}},}\ }\href
  {\doibase 10.1103/PhysRevD.74.095008} {\bibfield  {journal} {\bibinfo
  {journal} {Phys.Rev.}\ }\textbf {\bibinfo {volume} {D74}},\ \bibinfo {pages}
  {095008} (\bibinfo {year} {2006})},\ \Eprint
  {http://arxiv.org/abs/hep-ph/0608055} {arXiv:hep-ph/0608055 [hep-ph]}
  \BibitemShut {NoStop}%
%%CITATION = HEP-PH/0608055;%%
\bibitem [{\citenamefont {Appelquist}\ \emph {et~al.}(2013)\citenamefont
  {Appelquist}, \citenamefont {Brower}, \citenamefont {Buchoff}, \citenamefont
  {Cheng}, \citenamefont {Cohen} \emph {et~al.}}]{Appelquist:2013ms}%
  \BibitemOpen
  \bibfield  {author} {\bibinfo {author} {\bibfnamefont {T.}~\bibnamefont
  {Appelquist}}, \bibinfo {author} {\bibfnamefont {R.C.}\ \bibnamefont
  {Brower}}, \bibinfo {author} {\bibfnamefont {M.I.}\ \bibnamefont {Buchoff}},
  \bibinfo {author} {\bibfnamefont {M.}~\bibnamefont {Cheng}}, \bibinfo
  {author} {\bibfnamefont {S.D.}\ \bibnamefont {Cohen}},  \emph {et~al.},\
  }\bibfield  {title} {\enquote {\bibinfo {title} {{Lattice calculation of
  composite dark matter form factors}},}\ }\href@noop {} {\  (\bibinfo {year}
  {2013})},\ \Eprint {http://arxiv.org/abs/1301.1693} {arXiv:1301.1693
  [hep-ph]} \BibitemShut {NoStop}%
%%CITATION = ARXIV:1301.1693;%%
\bibitem [{\citenamefont {Aprile}\ \emph {et~al.}(2012)\citenamefont {Aprile}
  \emph {et~al.}}]{Aprile:2012nq}%
  \BibitemOpen
  \bibfield  {author} {\bibinfo {author} {\bibfnamefont {E.}~\bibnamefont
  {Aprile}} \emph {et~al.} (\bibinfo {collaboration} {XENON100
  Collaboration}),\ }\bibfield  {title} {\enquote {\bibinfo {title} {{Dark
  Matter Results from 225 Live Days of XENON100 Data}},}\ }\href@noop {} {\
  (\bibinfo {year} {2012})},\ \Eprint {http://arxiv.org/abs/1207.5988}
  {arXiv:1207.5988 [astro-ph.CO]} \BibitemShut {NoStop}%
%%CITATION = ARXIV:1207.5988;%%
\bibitem [{\citenamefont {Babich}\ \emph {et~al.}(2010)\citenamefont {Babich},
  \citenamefont {Brannick}, \citenamefont {Brower}, \citenamefont {Clark},
  \citenamefont {Manteuffel} \emph {et~al.}}]{Babich:2010qb}%
  \BibitemOpen
  \bibfield  {author} {\bibinfo {author} {\bibfnamefont {R.}~\bibnamefont
  {Babich}}, \bibinfo {author} {\bibfnamefont {J.}~\bibnamefont {Brannick}},
  \bibinfo {author} {\bibfnamefont {R.C.}\ \bibnamefont {Brower}}, \bibinfo
  {author} {\bibfnamefont {M.A.}\ \bibnamefont {Clark}}, \bibinfo {author}
  {\bibfnamefont {T.A.}\ \bibnamefont {Manteuffel}},  \emph {et~al.},\
  }\bibfield  {title} {\enquote {\bibinfo {title} {{Adaptive multigrid
  algorithm for the lattice Wilson-Dirac operator}},}\ }\href {\doibase
  10.1103/PhysRevLett.105.201602} {\bibfield  {journal} {\bibinfo  {journal}
  {Phys.Rev.Lett.}\ }\textbf {\bibinfo {volume} {105}},\ \bibinfo {pages}
  {201602} (\bibinfo {year} {2010})},\ \Eprint {http://arxiv.org/abs/1005.3043}
  {arXiv:1005.3043 [hep-lat]} \BibitemShut {NoStop}%
%%CITATION = ARXIV:1005.3043;%%
\bibitem [{\citenamefont {Cohen}\ \emph {et~al.}(2011)\citenamefont {Cohen},
  \citenamefont {Brower}, \citenamefont {Clark},\ and\ \citenamefont
  {Osborn}}]{Cohen:2012sh}%
  \BibitemOpen
  \bibfield  {author} {\bibinfo {author} {\bibfnamefont {Saul~D.}\ \bibnamefont
  {Cohen}}, \bibinfo {author} {\bibfnamefont {R.C.}\ \bibnamefont {Brower}},
  \bibinfo {author} {\bibfnamefont {M.A.}\ \bibnamefont {Clark}}, \ and\
  \bibinfo {author} {\bibfnamefont {J.C.}\ \bibnamefont {Osborn}},\ }\bibfield
  {title} {\enquote {\bibinfo {title} {{Multigrid Algorithms for Domain-Wall
  Fermions}},}\ }\href@noop {} {\bibfield  {journal} {\bibinfo  {journal}
  {PoS}\ }\textbf {\bibinfo {volume} {LATTICE2011}},\ \bibinfo {pages} {030}
  (\bibinfo {year} {2011})},\ \Eprint {http://arxiv.org/abs/1205.2933}
  {arXiv:1205.2933 [hep-lat]} \BibitemShut {NoStop}%
%%CITATION = ARXIV:1205.2933;%%
\bibitem [{\citenamefont {Babich}\ \emph
  {et~al.}(2011{\natexlab{a}})\citenamefont {Babich}, \citenamefont {Clark},
  \citenamefont {Joo}, \citenamefont {Shi}, \citenamefont {Brower} \emph
  {et~al.}}]{Babich:2011np}%
  \BibitemOpen
  \bibfield  {author} {\bibinfo {author} {\bibfnamefont {R.}~\bibnamefont
  {Babich}}, \bibinfo {author} {\bibfnamefont {M.A.}\ \bibnamefont {Clark}},
  \bibinfo {author} {\bibfnamefont {B.}~\bibnamefont {Joo}}, \bibinfo {author}
  {\bibfnamefont {G.}~\bibnamefont {Shi}}, \bibinfo {author} {\bibfnamefont
  {R.C.}\ \bibnamefont {Brower}},  \emph {et~al.},\ }\bibfield  {title}
  {\enquote {\bibinfo {title} {{Scaling Lattice QCD beyond 100 GPUs}},}\
  }\href@noop {} {\  (\bibinfo {year} {2011}{\natexlab{a}})},\ \Eprint
  {http://arxiv.org/abs/1109.2935} {arXiv:1109.2935 [hep-lat]} \BibitemShut
  {NoStop}%
%%CITATION = ARXIV:1109.2935;%%
\bibitem [{\citenamefont {Babich}\ \emph
  {et~al.}(2011{\natexlab{b}})\citenamefont {Babich}, \citenamefont {Brower},
  \citenamefont {Clark}, \citenamefont {Gottlieb}, \citenamefont {Joo} \emph
  {et~al.}}]{Babich:2011zz}%
  \BibitemOpen
  \bibfield  {author} {\bibinfo {author} {\bibfnamefont {Ron}\ \bibnamefont
  {Babich}}, \bibinfo {author} {\bibfnamefont {Richard}\ \bibnamefont
  {Brower}}, \bibinfo {author} {\bibfnamefont {Mike}\ \bibnamefont {Clark}},
  \bibinfo {author} {\bibfnamefont {Steven}\ \bibnamefont {Gottlieb}}, \bibinfo
  {author} {\bibfnamefont {Balint}\ \bibnamefont {Joo}},  \emph {et~al.},\
  }\bibfield  {title} {\enquote {\bibinfo {title} {{Progress on the QUDA code
  suite}},}\ }\href@noop {} {\bibfield  {journal} {\bibinfo  {journal} {PoS}\
  }\textbf {\bibinfo {volume} {LATTICE2011}},\ \bibinfo {pages} {033} (\bibinfo
  {year} {2011}{\natexlab{b}})}\BibitemShut {NoStop}%
%%CITATION = POSCI,LATTICE2011,033;%%
\bibitem [{\citenamefont {Brower}\ \emph {et~al.}(2012)\citenamefont {Brower},
  \citenamefont {Fleming},\ and\ \citenamefont {Neuberger}}]{Brower:2012vg}%
  \BibitemOpen
  \bibfield  {author} {\bibinfo {author} {\bibfnamefont {Richard}\ \bibnamefont
  {Brower}}, \bibinfo {author} {\bibfnamefont {George}\ \bibnamefont
  {Fleming}}, \ and\ \bibinfo {author} {\bibfnamefont {Herbert}\ \bibnamefont
  {Neuberger}},\ }\bibfield  {title} {\enquote {\bibinfo {title} {{Lattice
  Radial Quantization: 3D Ising}},}\ }\href@noop {} {\  (\bibinfo {year}
  {2012})},\ \Eprint {http://arxiv.org/abs/1212.6190} {arXiv:1212.6190
  [hep-lat]} \BibitemShut {NoStop}%
%%CITATION = ARXIV:1212.6190;%%
\bibitem [{\citenamefont {Fubini}\ \emph {et~al.}(1973)\citenamefont {Fubini},
  \citenamefont {Hanson},\ and\ \citenamefont {Jackiw}}]{Fubini:1972mf}%
  \BibitemOpen
  \bibfield  {author} {\bibinfo {author} {\bibfnamefont {S.}~\bibnamefont
  {Fubini}}, \bibinfo {author} {\bibfnamefont {Andrew~J.}\ \bibnamefont
  {Hanson}}, \ and\ \bibinfo {author} {\bibfnamefont {R.}~\bibnamefont
  {Jackiw}},\ }\bibfield  {title} {\enquote {\bibinfo {title} {{New approach to
  field theory}},}\ }\href {\doibase 10.1103/PhysRevD.7.1732} {\bibfield
  {journal} {\bibinfo  {journal} {Phys.Rev.}\ }\textbf {\bibinfo {volume}
  {D7}},\ \bibinfo {pages} {1732--1760} (\bibinfo {year} {1973})}\BibitemShut
  {NoStop}%
\bibitem [{\citenamefont {Cardy}(1985)}]{Cardy:1985xx}%
  \BibitemOpen
  \bibfield  {author} {\bibinfo {author} {\bibfnamefont {John~L.}\ \bibnamefont
  {Cardy}},\ }\bibfield  {title} {\enquote {\bibinfo {title} {{Universal
  amplitudes in finite-size scaling: generalization to arbitrary
  dimensionality}},}\ }\href@noop {} {\bibfield  {journal} {\bibinfo  {journal}
  {J.Phys.A}\ }\textbf {\bibinfo {volume} {A18}},\ \bibinfo {pages}
  {L757--L760} (\bibinfo {year} {1985})}\BibitemShut {NoStop}%
\bibitem [{\citenamefont {Brower}\ \emph {et~al.}(2005)\citenamefont {Brower},
  \citenamefont {Neff},\ and\ \citenamefont {Orginos}}]{Brower:2004xi}%
  \BibitemOpen
  \bibfield  {author} {\bibinfo {author} {\bibfnamefont {Richard~C.}\
  \bibnamefont {Brower}}, \bibinfo {author} {\bibfnamefont {Hartmut}\
  \bibnamefont {Neff}}, \ and\ \bibinfo {author} {\bibfnamefont {Kostas}\
  \bibnamefont {Orginos}},\ }\bibfield  {title} {\enquote {\bibinfo {title}
  {{Mobius fermions: Improved domain wall chiral fermions}},}\ }\href {\doibase
  10.1016/j.nuclphysbps.2004.11.180} {\bibfield  {journal} {\bibinfo  {journal}
  {Nucl.Phys.Proc.Suppl.}\ }\textbf {\bibinfo {volume} {140}},\ \bibinfo
  {pages} {686--688} (\bibinfo {year} {2005})},\ \Eprint
  {http://arxiv.org/abs/hep-lat/0409118} {arXiv:hep-lat/0409118 [hep-lat]}
  \BibitemShut {NoStop}%
%%CITATION = HEP-LAT/0409118;%%
\end{thebibliography}%

\end{document}